\documentclass[onecolumn,twoside,12pt,journal,draftclsnofoot]{IEEEtran}

\usepackage{ifpdf}
\ifpdf
  \usepackage[pdftex]{graphicx}
  \graphicspath{{pdf/}}
  \DeclareGraphicsExtensions{.pdf}
\else
  \usepackage[dvips]{graphicx}
  \graphicspath{{eps/}}
  \DeclareGraphicsExtensions{.eps}
\fi
\usepackage{cite}
\usepackage{psfrag}
\usepackage[cmex10]{amsmath}
\usepackage{amssymb}
\usepackage{mdwmath}
\usepackage{mdwtab}
\usepackage[tight,footnotesize]{subfigure}
\usepackage[caption=false]{caption}
\usepackage{fixltx2e}
\usepackage{exscale}
\usepackage{bm}
\usepackage{mathrsfs}
\usepackage{upgreek}
\usepackage[T1]{fontenc}
\usepackage{array}
\usepackage{setspace}
\usepackage[nolist,nohyperlinks]{acronym}
\usepackage{ifthen}
\usepackage[normalem]{ulem}

\newcommand{\equidist}{\stackrel{\mathrm{d}}{=}}
\newcommand{\apprdist}{\stackrel{\mathrm{d}}{\approx}}
\newcommand{\mo}{\mathopen{}}
\newcommand{\degreel}{\ensuremath{^\circ}\!}
\newcommand{\degree}{\ensuremath{^\circ}}

\newcommand{\real}{\mathrm{Re}\mo}
\newcommand{\imag}{\mathrm{Im}\mo}
\newcommand{\trace}{\mathrm{Tr}\mo}

\newcommand{\var}{\mathrm{Var}\mo}
\newcommand{\erf}{\mathrm{erf}\mo}
\newcommand{\erfi}{\mathrm{erfi}\mo}
\newcommand{\vol}{\mathrm{vol}\mo}

\newcommand{\rank}{\mathrm{r}\mo}

\newcommand{\pdf}{\mathrm{p}\mo}
\newcommand{\cdf}{\mathrm{F}\mo}
\newcommand{\ce}{\mathrm{p}\mo}

\newcommand{\defas}{\stackrel{\Delta}{=}}

\newcommand{\conv}{\ast}
\newcommand{\kron}{\otimes}
\renewcommand{\angle}{\text{\color{red}use arg instead of angle}}
\newcommand{\Gammaf}{\Gamma\mo}
\renewcommand{\digamma}{\Psi\mo}
\newcommand{\digammader}{\Psi'\mo}
\newcommand{\polygamma}{\psi\mo}

\newcommand{\TX}{\mathrm{T}}
\newcommand{\RX}{\mathrm{R}}
\newcommand{\DX}{\mathrm{d}}
\newcommand{\BB}{\mathrm{bb}}
\newcommand{\FIX}{\mathrm{f}}
\newcommand{\FREE}{\mathrm{free}}
\newcommand{\WHITE}{\mathrm{w}}
\newcommand{\FROB}{\mathrm{F}}
\newcommand{\MT}{M_\TX}
\newcommand{\MR}{M_\RX}

\newcommand{\INTVL}{\mathfrak{I}}

\newcommand{\VLD}{\mathrm{c}}
\newcommand{\ilen}[1]{|#1|}
\newcommand{\pr}{\varphi}
\newcommand{\PP}{\varphi}

\newcommand{\expect}{\mathbb{E}\mo}

\newcommand{\numID}[1]{\mathbb{#1}\mo}
\newcommand{\numVar}[1]{\mathcal{#1}\mo}
\newcommand{\set}[1]{\mathcal{#1}}

\newcommand{\ud}{\mathrm{d}}

\newcommand{\vect}{\mathrm{vec}\mo}

\newcommand{\diag}{\mathrm{diag}\mo}

\newcommand{\enbrace}[2][leftright]{\ifthenelse{\equal{#1}{leftright}}{\left(#2\right)}{%
                                    \ifthenelse{\equal{#1}{big}}{\big(#2\big)}{%
                                    \ifthenelse{\equal{#1}{Big}}{\Big(#2\Big)}{%
                                    \ifthenelse{\equal{#1}{bigg}}{\bigg(#2\bigg)}{%
                                    \ifthenelse{\equal{#1}{Bigg}}{\Bigg(#2\Bigg)}{%
                                    (#2)}}}}}}

\newcommand{\newappss}[4]{\enbrace[#1]{#2#3}^{#4\!}}

\newcommand{\transp}[3][leftright]{\newappss{#1}{#2}{#3}{T}}
\newcommand{\contra}[3][leftright]{\newappss{#1}{#2}{#3}{H}}

\newcommand{\diagH}[2][leftright]{\newappss{#1}{\diag}{#2}{H}}
\newcommand{\realT}[2][leftright]{\newappss{#1}{\real}{#2}{T}}
\newcommand{\imagT}[2][leftright]{\newappss{#1}{\imag}{#2}{T}}
\newcommand{\fsquare}[2][leftright]{\newappss{#1}{}{#2}{2}}

\renewcommand{\det}{\mathrm{det}\mo}
\renewcommand{\log}{\mathrm{log}\mo}
\newcommand{\cov}{\mathrm{Cov}\mo}
\newcommand{\corrc}{\mathrm{Corr}\mo}
\renewcommand{\jmath}{j}
\renewcommand{\div}{ \, \mathrm{div} \, }
\renewcommand{\mod}{ \, \mathrm{mod} \, }
\renewcommand{\Re}{\real\mo}

\renewcommand{\exp}{\mathrm{exp}\mo}
\renewcommand{\log}{\mathrm{log}\mo}

\renewcommand{\cos}{\mathrm{cos}\mo}
\newcommand{\Jacobian}[2]{\big( \UD #1 (#2) \big)}
\renewcommand{\Jacobian}[2]{\mathcal{J}_{#1}\mo(#2)}
\newcommand{\JacobianT}[2]{\mathcal{J}_{#1}^T\mo(#2)}
\newcommand{\Hessian}[2]{\big( \UD ( \UD #1 (#2) )^T \big)}
\renewcommand{\Hessian}[2]{\mathcal{H}_{#1}\mo(#2)}
\newcommand{\divec}{\mathrm{divec}\mo}
\newcommand{\dg}{\mathrm{dg}\mo}

\newtheorem{prop}{Theorem}
\newtheorem{lemma}{Lemma}

\newtheorem{theorem}[prop]{Theorem}

\newcommand{\vi}{i}
\newcommand{\vk}{k}
\newcommand{\vl}{l}
\newcommand{\vn}{n}
\newcommand{\vrow}{m}
\newcommand{\vcol}{n}

\newcommand{\Elchi}{\mathfrak{X}}
\newcommand{\lrv}{\eta}

\newcommand{\cplxGauss}{\numVar{CN}}
\newcommand{\realGauss}{\numVar{N}}
\newcommand{\cwa}{^{\circ}\mo}
\newcommand{\rsq}{\;}
\newcommand{\unit}[1]{\:#1}
\newcommand{\rsp}{\;}
\newcommand{\propref}[1]{Theorem\rsp\ref{#1}}
\newcommand{\propsref}[1]{Theorems\rsp\ref{#1}}
\newcommand{\lemmaref}[1]{Lemma\rsp\ref{#1}}
\newcommand{\thref}[1]{Theorem\rsp\ref{#1}}
\newcommand{\figref}[1]{Fig.\rsp\ref{#1}}
\newcommand{\figsref}[1]{Figs.\rsp\ref{#1}}
\newcommand{\sfigref}[2]{Fig.\rsp\ref{#1}\,\subref{#2}}
\newcommand{\sfigsref}[2]{Figs.\rsp\ref{#1}\,\subref{#2}}
\newcommand{\sref}[2]{\ref{#1}\,\subref{#2}}
\newcommand{\secref}[1]{Section\rsp\ref{#1}}
\newcommand{\secsref}[1]{Sections\rsp\ref{#1}}

\newcommand{\lemmref}[1]{Lemma\rsp\ref{#1}}
\newcommand{\tabref}[1]{Table\rsp\ref{#1}}
\renewcommand{\ln}{\log_e\mo}

\def\hksqrt{\mathpalette\DHLhksqrt}
\def\DHLhksqrt#1#2{\setbox0=\hbox{$#1\sqrt{#2\,}$}\dimen0=\ht0
  \advance\dimen0-0.2\ht0
  \setbox2=\hbox{\vrule height\ht0 depth -\dimen0}%
  {\box0\lower0.4pt\box2}}

\newcommand{\mysqrt}{\hksqrt}

\newlength{\laenge}

\newcommand{\mytitle}{Information-Theoretic Analysis\\of MIMO Channel Sounding}
\newcommand{\mytshrt}{Information-Theoretic Analysis of MIMO Channel Sounding}

\newcommand{\icg}[1]{\includegraphics[scale=0.7272]{#1}}

\newcommand{\startappendix}{\appendices\section{}}

\begin{document}

\setstretch{1.0}

\title{\mytitle}

\author{\normalsize
\IEEEauthorblockN{\emph{Daniel S.~Baum and Helmut B\"olcskei}\\\vspace{0.125cm}}
\IEEEauthorblockA{Communication Technology Laboratory\\
ETH Zurich\\
8092 Zurich, Switzerland\\
\{dsbaum, boelcskei\}@nari.ee.ethz.ch}%
\thanks{This work was supported in part by the Swiss National Science Foundation (SNF) 
under grant No. 200021-100025/1. 
The paper was presented in part at IEEE VTC Fall 2004, Sept.\ 2004, Los Angeles, CA.
}}

\markboth{}{Baum and B\"olcskei: \mytshrt}

\maketitle

\setstretch{1.8}

\begin{abstract}
The large majority of commercially available 
\ac{MIMOs} radio channel measurement devices (sounders) is based on 
\ac{TDMS} of a single transmit/receive 
\acl{RF} chain into the elements of a transmit/receive antenna array. 
While being cost-effective, such a solution can cause significant
measurement errors due to phase noise and frequency offset in the 
\aclp{LO}. In this paper, we systematically analyze the resulting errors
and show that, in practice, 
\emph{overestimation of channel capacity by several hundred percent} can occur. 
Overestimation is caused by phase noise (and 
to a lesser extent frequency offset) leading to an increase of the 
\ac{MIMOs} channel rank.
Our analysis furthermore reveals that the impact of phase errors is, in general, 
most pronounced if the physical channel has low rank (typical for 
\acl{LOS} or poor scattering scenarios). 
The extreme case of a rank-1 physical channel is analyzed in detail.
Finally, we present 
measurement results obtained from a commercially employed
\ac{TDMS}-based \ac{MIMOs} channel sounder.
In the light of the findings of this paper, the results obtained
through \ac{MIMOs} channel measurement campaigns using 
\ac{TDMS}-based channel sounders should be interpreted with great care. 
\end{abstract}


\acresetall
\newpage

\setcounter{totalnumber}{0}

\begin{acronym}
\acro{ADC}{analog-to-digital converter}
\acro{AWGN}{additive white Gaussian noise}
\acro{AGN}{additive Gaussian noise}
\acro{BPSK}{binary phase-shift keying}
\acro{CW}{continuous wave}
\acro{CDF}[cdf]{cumulative distribution function}
\acro{CDMA}{code division multiple access}
\acro{CSI}{channel state information}
\acro{DAC}{digital-to-analog converter}
\acro{DOA}{direction of arrival}
\acro{IID}[i.i.d.]{independent and identically distributed}
\acro{JG}[JG]{jointly Gaussian}
\acro{LHS}{left-hand side}
\acro{LO}{local oscillator}
\acro{LOS}{line-of-sight}
\acro{LTI}{linear time-invariant}
\acro{MGF}{moment-generating function}
\acro{MI}{mutual information}
\acro{MIMOs}[MIMO]{multiple-input multiple-output}
\acro{MIMO}{Multiple-input multiple-output}
\acro{MISO}{multiple-input single-output}
\acro{OFDM}{orthogonal frequency division multiplexing}
\acro{PAS}{power-angular spectrum}
\acro{PDF}[pdf]{probability density function}
\acro{PDP}{power-delay profile}
\acro{PLL}{phase-locked loop}
\acro{PNS}{pseudo-(random) noise sequence}
\acro{PSD}{power spectral density}
\acro{RF}{radio frequency}
\acro{RHS}{right-hand side}
\acro{RMS}[rms]{root mean-square}
\acro{RV}{random variable}
\acro{SIMO}{single-input multiple-output}
\acro{SISO}{single-input single-output}
\acro{SNR}{signal-to-noise ratio}
\acro{SVD}{singular-value decomposition}
\acro{TDMS}{time-division multiplexed switching}
\acro{VNA}{vector network analyzer}
\acro{WILOG}[w.l.o.g.]{without loss of generality}
\acro{WIP}[w.p.]{with probability}
\acro{WRT}[w.r.t.]{with respect to}
\acro{WSS}{wide-sense stationary}
\end{acronym}

\section{Introduction}

\ac{MIMO} wireless communication promises significant
improvements over existing wireless systems both in terms of spectral efficiency and link
reliability. Obtaining accurate measurements of \ac{MIMO} radio channels is 
of key importance to devising 
accurate \ac{MIMO} radio channel models, which in turn are vital for
system design, simulation, and performance analysis.

A common and widespread \ac{MIMO} channel 
measurement device (a.k.a.\ sounder) design is based on 
time-division multiplexing with synchronous 
switching, or \ac{TDMS} for short,
of a single \ac{RF} chain into the 
individual elements of an antenna array. 
\ac{TDMS} can be used at either the transmitter or the receiver 
(\emph{one-sided \ac{TDMS}}) or at both sides 
of the link (\emph{double-sided \ac{TDMS}}). 
For the latter case, which is practically the most relevant one, 
such an architecture is depicted in \figref{fig:TDDsounder}.
\ac{TDMS} constitutes a natural extension of \ac{SISO}
channel sounders and leads to very cost-effective solutions
as only a single \ac{RF} chain is required 
at either the transmitter or the receiver (one-sided \ac{TDMS})
or at both sides of the link (double-sided \ac{TDMS}). 
To the best of our knowledge, the large majority of 
commercially available \ac{MIMO} channel sounders is
based on the \ac{TDMS} principle.
A major drawback of \ac{TDMS}-based sounder architectures
results from \emph{temporal phase deviations}
between the outputs of the \acp{LO} in the \ac{RF} chains 
at transmitter and receiver being \emph{translated into the
spatial domain} due to switching across antenna elements. 
This can cause an increase of the \ac{MIMO} channel rank
and corresponding measurement errors, 
in terms of estimated \ac{MIMO} channel capacity, that can 
be on the order of several hundred percent.
It is therefore immediately clear that understanding the impact of
phase errors\footnote{For brevity, in the remainder of this paper, 
we use the terminology \emph{phase errors} whenever we refer to 
phase deviations due to phase noise, or frequency offset, or both.}
in \ac{TDMS}-based sounding is of fundamental importance.

One may argue that in a wireless communication link the impact of phase fluctuations 
in the transmitter and/or the receiver can
simply be absorbed into an effective channel consisting of the physical propagation
channel combined with \ac{LO}-related (and potentially other) impairments. 
Channel estimation at the
receiver for demodulation and decoding or for precoding at the transmitter (through
feedback) would then simply work on the effective channel. 
This point of view can certainly be sensible in 
a data transmission setup if the 
frequency dispersion caused by phase fluctuations
is small compared to that induced by the physical channel.
In a channel sounding setup, however, it is 
crucial to separate the physical 
propagation channel (i.e., the object to be measured) from transmitter/receiver
impairments, in order to obtain measurement results that depend as little as
possible on the measurement device (sounder) used.
Furthermore, as already pointed out,
the measurement procedure employed in \ac{TDMS}-based
\ac{MIMO} channel sounding results in very high sensitivity 
of the estimated channel capacity with respect to phase errors.

\subsubsection*{Contributions}

The goal of this paper is to systematically analyze the impact of phase noise
and frequency offset (between transmitter and receiver \ac{LO})
on estimated \ac{MIMO} channel 
capacity when \ac{TDMS}-based channel sounders are used.
In particular, we show that the presence of phase
errors can lead to significant overestimation of \ac{MIMO} channel capacity.
A sensitivity analysis reveals that, in certain cases,
underestimation is possible as well, albeit
typically resulting in significantly smaller errors.

We start by devising a signal model, applicable to the wide class of correlation-based
(as defined in \cite[Sec.{\rsp}III]{MMHS02j}) \ac{MIMO} channel sounders
and taking into account phase noise and frequency offset.
We then systematically identify situations
where phase errors have no (or little) impact on \ac{MIMO}
(ergodic and outage) capacity estimates and where they lead to significant estimation errors.
As an extreme case in the latter category,
we demonstrate that even moderate phase noise can turn a rank-1 physical channel
(e.g., a pin-hole channel \cite{GBGP02,CFGV02}) into a
full-rank effective channel; analytic expressions for
the corresponding (ergodic and outage) capacity estimates
are provided.
Our analytic results are supported by measurement results
obtained from a commercially employed \ac{TDMS}-based \ac{MIMO} channel sounder.

\subsubsection*{Previous related work}

For linear time-varying 
\ac{SISO} physical channels, 
an analysis of the systematic measurement errors 
incurred by correlative channel sounders 
(due to the time-varying nature of the physical channel)
is reported in \cite{MMHS02j}.
Models for the phase noise \ac{PSD} 
of a 5\unit{GHz} and a 50\unit{GHz} frequency synthesizer 
as well as an expression for
the phase-noise induced reduction of the dynamic range of 
m-sequences can be found in \cite{KiVa97}.
A subsequent paper by the same authors \cite{KiVa00} discusses the impact of 
frequency offset on \ac{DOA} estimates in \ac{SIMO} \ac{TDMS}-based measurements
and proposes corresponding mitigation methods. 
The effect of random-walk phase noise on the \ac{RMS}
error of SAGE\footnote{Space-alternating generalized expectation 
maximization (EM) algorithm
\cite{FeHe94j}.
}-based \ac{DOA} estimates is analyzed in \cite{AWTM05c}.

For fully parallel \ac{MIMO} channel sounders, i.e., channel sounders employing a
separate \ac{RF} chain for each transmit and each receive antenna element, 
the impact of gain imbalance in parallel \ac{RF} chains 
and of thermal noise on the estimated capacity of a physical
rank-1 \ac{MIMO} channel is analyzed in \cite{Gans02}. 
The variance of an approximation of the error in the \ac{MI}
of an effective channel resulting from 
a deterministic \ac{MIMO} channel subject to 
additive white complex Gaussian distributed perturbations
is derived in \cite{Kyritsi02}. 
For physical 
pin-hole \cite{GBGP02}\/, a.k.a.\ key-hole \cite{CFGV02,LoKo02j,AlTM03}
(i.e., rank-1), \ac{MIMO} channels
in a controlled indoor environment, the impact of measurement
imperfections such as thermal noise and 
``multi-path leakage'' (i.e., multi-path components propagating between 
the transmitter and the receiver via paths other than through the pin-hole)
on the channel eigenvalue distribution 
and the resulting outage capacity 
are analyzed numerically, based on
measurements and simulations, in \cite{AlTM03}.

\subsubsection*{Organization of the paper}

The remainder of this paper is organized as follows. 
In \secref{sec:TDMS}, the architecture of a \ac{TDMS}-based \ac{MIMO} channel sounder is described and 
the corresponding channel and signal model are provided. 
In \secref{sec:err_stat}, we analyze the effect of phase errors on \ac{MIMO} channel statistics. 
The corresponding impact on estimated \ac{MI} is studied in \secref{sec:MI}.
A framework for analyzing the sensitivity of the \ac{MIMO} channel \ac{MI}
to phase errors is developed in \secref{sec:sens_ana}.
\secref{sec:rank1} is devoted to the special (but practically relevant) case of rank-1 physical channels.
Measurement results performed on a typical, commercially 
employed \ac{TDMS}-based \ac{MIMO} channel sounder 
corroborating our analysis are presented in \secref{sec:meas}.
We conclude in \secref{sec:conc}.

\subsubsection*{Notation}

$\expect\{\cdot\}$ 
denotes the expectation operator.
$f(t)\conv g(t)=\int_{\tau}f(\tau)g(t-\tau)\,\ud\tau$ stands for the convolution 
of the functions $f(t)$ and $g(t)$.
The Dirac delta function is denoted as $\delta(t)$, and $\delta_i=1$ for $i=0$ and $0$ otherwise.
The superscripts $^T$, $^H$, and $^*$
stand for transposition, conjugate transposition, and elementwise conjugation, 
respectively. 
An $\vrow\times\vcol$ matrix is a matrix with $\vrow$ rows and $\vcol$ columns.
$\mathbf{1}$ and $\mathbf{0}$ denote an all-ones and all-zeros 
matrix, respectively, of appropriate size.
If required, the size of a matrix is specified
through subscripts, e.g., $\mathbf{1}_{\vrow,\vcol}$. 
$\mathbf{I}_\vrow$ stands for the $\vrow\times\vrow$ identity matrix.
$\mathbf{A}\circ\mathbf{B}$ and $\mathbf{A}\otimes\mathbf{B}$ 
denote the Hadamard (pointwise) product and the Kronecker product, respectively, 
of the matrices $\mathbf{A}$ and $\mathbf{B}$,
and $f\cwa(\mathbf{A})$ stands for
the matrix resulting from entry-wise application of the function $f(\cdot)$ 
to $\mathbf{A}$.
$\|\mathbf{A}\|_\FROB$ is the Frobenius norm of $\mathbf{A}$ and
$\trace(\mathbf{A})$ is the trace of $\mathbf{A}$. 
$\diag(\mathbf{x})$ denotes the diagonal matrix with the elements 
of the vector $\mathbf{x}$ on its main diagonal,
and $\dg(\mathbf{A}) = \mathbf{A}\circ\mathbf{I}$ zeros out 
all but the diagonal elements of $\mathbf{A}$. 
The element in the $\vrow$th row and $\vcol$th column of $\mathbf{A}$ 
is denoted as $[\mathbf{A}]_{\vrow,\vcol}$.
$\rank(\mathbf{A})$ and $\lambda_i(\mathbf{A})$ stand for the rank and 
the $i$th eigenvalue of $\mathbf{A}$, respectively.
Unless explicitly stated otherwise, eigenvalues are sorted in decreasing order,
i.e., $\lambda_{1}(\mathbf{A})\ge\lambda_2(\mathbf{A})\ge\cdots\ge
\lambda_{n}(\mathbf{A})$.
For an $\vrow\times\vcol$ matrix
$\mathbf{A} = [\, \mathbf{a}_1 \,\;\, \mathbf{a}_2 \,\;\, \cdots \,\;\, \mathbf{a}_\vcol \,]$ 
with columns $\mathbf{a}_i$,
we define the $\vrow\vcol\times1$ vector $\vect(\mathbf{A}) =
[\, \mathbf{a}_1^T \,\;\, \mathbf{a}_2^T \,\;\, \cdots \,\;\, \mathbf{a}_\vcol^T \,]^T$.
The commutation matrix $\mathbf{K}_{(\vrow,\vcol)}$ \cite[Sec.\rsp3.7]{MaNe88b}
is a permutation matrix of size $\vrow\vcol\times\vrow\vcol$ uniquely defined through 
\begin{align}
\mathbf{K}_{(\vrow,\vcol)} \vect(\mathbf{A}) = \vect(\mathbf{A}^T)
\label{eq:comm_mat}
\end{align}
where $\mathbf{A}$ is an $\vrow\times\vcol$ matrix.
For brevity, we define $\divec(\mathbf{A}) = \diag\big( \vect( \mathbf{A} ) \big)$.
For two \acp{RV} $X$ and $Y$, 
$X\equidist Y$ and $X\apprdist Y$ stands for 
equivalence and approximate equivalence in distribution, respectively.
For a \ac{RV} $X$, the \ac{PDF} and \ac{CDF} are denoted by
$\pdf_X(x)$ and $\cdf_X(x)$, respectively,
and the \ac{MGF} is defined as $M_X(s) = \expect\{\exp(sX)\}$.
The variance of a \ac{RV} $X$ is denoted as 
$\var\{X\}$. 
The covariance matrix of a complex random matrix $\mathbf{X}$ is defined as
$\cov\{\mathbf{X}\} 
= \expect\big\{ \big( \vect(\mathbf{X})-\expect\{\vect(\mathbf{X})\} \big) 
                \big( \vect(\mathbf{X})-\expect\{\vect(\mathbf{X})\} \big)^H \big\}$
and the corresponding
matrix of correlation coefficients is the matrix with entries
$[\corrc\{\mathbf{X}\}]_{\vrow,\vcol} = [ \cov\{\mathbf{X}\} ]_{\vrow,\vcol} / 
\left( [ \cov\{\mathbf{X}\} ]_{\vrow,\vrow} [ \cov\{\mathbf{X}\} ]_{\vcol,\vcol} \right)^{1/2}$.
The pseudo-covariance matrix \cite{NeMa93j}
of a complex random matrix $\mathbf{X}$ is defined as
$\cov_\mathrm{p}\mo\{\mathbf{X}\} 
= \expect\big\{ \big( \vect(\mathbf{X})-\expect\{\vect(\mathbf{X})\} \big) 
                \big( \vect(\mathbf{X})-\expect\{\vect(\mathbf{X})\} \big)^T \big\}$.
For the integers $a$ and $b$, 
$a\div b = \lfloor a/b \rfloor$ denotes the integer division of $a$ by $b$, and 
$a\mod b$ stands for the remainder, on division of $a$ by $b$.
For a complex scalar $z\in\numID{C}$, 
the function $\arg(z)$ returns the argument (angle) of $z$ in the 
interval $[0,2\pi)$, 
and $\real(z)$ and $\imag(z)$ stand for the real and imaginary part of $z$, respectively.
All logarithms are to the base 2 unless explicitly stated otherwise.
$\ilen{\INTVL}$ denotes the length of an interval $\INTVL$.
Throughout the paper, the number of transmit and receive antenna 
elements in a \ac{MIMO} channel
is denoted as $\MT$ and $\MR$, respectively,
and we will refer to antenna elements simply as antennas.

A real Gaussian random vector is defined as a vector with \ac{JG} elements
and denoted by $\realGauss(\mathbf{m}, \mathbf{C})$, where  
$\mathbf{m}$ is the mean and $\mathbf{C}$ is the covariance matrix.
A complex Gaussian random vector is defined as a vector 
with \ac{JG} real and imaginary parts.
A complex random vector will be called proper 
if its pseudo-covariance matrix vanishes.
$\cplxGauss(\mathbf{m}, \mathbf{C})$ stands for a proper 
complex Gaussian random vector 
with mean $\mathbf{m}$ and covariance matrix $\mathbf{C}$. 
The complex random vectors $\mathbf{x}$ and $\mathbf{y}$ will
be called jointly proper if the composite random vector
having $\mathbf{x}$ and $\mathbf{y}$ as subvectors is proper.

For a chi-square distributed \ac{RV} 
with $n$ degrees of freedom and variance 
$2n\sigma^4$
we write $\chi^2_{n,\sigma^2}$, 
where $\chi^2_{n,\sigma^2} \equidist \| \mathbf{x} \|^2$
with $\mathbf{x} \equidist \realGauss(\mathbf{0}, \sigma^2\mathbf{I}_n)$.
$[\, x_1 \,\;\, x_2 \,\;\, \cdots \,\;\, x_{n} \,] 
\equidist D_n(a_1, a_2, \ldots, a_n)$, 
$x_i \ge 0$ $(i = 1,2,\ldots,n)$ 
denotes a Dirichlet distributed random vector
with parameters $a_1, a_2, \ldots, a_n$.
The corresponding subvector 
$\mathbf{x} = [\, x_1 \,\;\, x_2 \,\;\, \cdots \,\;\, x_{n-1} \,]$
satisfies 
$\mathbf{x} \equidist D_{n-1}(a_1, a_2, \ldots, a_{n-1}; a_n)$
and has joint \ac{PDF} \cite[Th.\rsp1.2]{FaKN90b}
\begin{align*}
  \pdf_{D_{n-1}(a_1, a_2, \ldots, a_{n-1}; a_n)}(\mathbf{x}) 
&=
  \frac{\Gammaf\left( a \right)}{\prod_{i=1}^{n} \Gammaf(a_i)} 
  \left( \prod_{i=1}^{n-1} x_i^{a_i-1} \right) \left( 1 - \sum_{i=1}^{n-1} x_i \right)^{a_n-1}
\end{align*}
where $a = \sum_{i=1}^{n} a_i$ and
$\Gammaf(z)$ is the Gamma function \cite[Sec.\rsp6.1]{AbSt72b}.
We denote the beta distribution with parameters $a$ and $b$
as $\beta(a,b)$ with \ac{PDF} given by
\begin{align*}
  \pdf_{\beta(a,b)}(x) 
= 
  \frac{\Gammaf(a+b)}{\Gammaf(a)\Gammaf(b)}(1-x)^{b-1}x^{a-1}, 
\quad 
  a>0, b>0, x\ge0.
\end{align*}
The digamma function $\digamma(z)$ is defined as 
\cite[Eq.\rsp6.3.1, Eq.\rsp6.3.16]{AbSt72b}
\begin{align}
\begin{split}
  \digamma(z) 
&= 
  \frac{\ud}{\ud z} \, \ln \Gammaf(z) 
=
  \frac{\Gammaf'(z)}{\Gammaf(z)}
= 
  \polygamma_0(z),
\quad z\in\numID{C}
\\
  \digamma(z)
&=
  -\gamma + \sum_{n=1}^\infty \frac{z}{n(n+z-1)},
\quad 
  z\ne 0, -1, -2, \ldots
\end{split}
\label{eq:dig_def} 
\end{align}
where $\gamma\approx0.5772$ is Euler's
constant, and 
$\polygamma_n(z)$ is the polygamma function \cite[Eq.\rsp6.4.1]{AbSt72b}, 
defined as the $n$th derivative of $\digamma(z)$.
We will also need the following representation of 
the digamma function at positive integer multiples of $1/2$ 
given by \cite[Eq.\rsp6.3.2, Eq.\rsp6.3.4]{AbSt72b}
\begin{align}
  \digamma(k) = -\gamma + \sum_{n=1}^{k-1} \frac{1}{n},
\quad 
  \digamma\left( k - \frac{1}{2} \right) = -\gamma -2\ln(2) + \sum_{n=1}^{k-1} \frac{2}{2n-1},
\quad 
  k\in\numID{N}, k \ge 1.
\label{eq:dig_simpl}
\end{align}
Finally, we note that the first derivative of the digamma function can be 
written as an infinite series as \cite[Eq.\rsp6.4.10]{AbSt72b}
\begin{align}
  \digammader(z) = \polygamma_1(z)
= 
  \frac{\ud^2}{\ud z^2} \, \ln \Gammaf(z) 
=
  \sum_{n=1}^\infty \frac{1}{(n+z-1)^2},
\quad z\ne0,-1,-2, \ldots.
\label{eq:dig_der}
\end{align}

\section{MIMO Channel Sounding Based On Time-Division Multiplexed Switching}
\label{sec:TDMS}

In this section, we describe the system architecture of a correlation-based
\ac{MIMO} channel sounder employing \ac{TDMS}
and we present the corresponding signal model, 
taking into account the presence of phase errors.

\subsection{Channel Sounder Architecture}

\begin{figure}[htbp]
\centering
\psfrag{DAC}{\scriptsize{DAC}}
\psfrag{MIX}{\scriptsize{MIX}}
\psfrag{toDAC}{\scriptsize{chip rate}}
\psfrag{toMIXtx}{\scriptsize{$o_\TX(t)$}}
\psfrag{x(t)}{\scriptsize{$x(t)$}}
\psfrag{MPU}{\scriptsize{MPU}}
\psfrag{CLK}{\scriptsize{CLK}}
\psfrag{ch}{\scriptsize{$\mathbf{H}$}}
\psfrag{u(t)}{\scriptsize{$u(t)$}}
\psfrag{BPF}{\scriptsize{BPF}}
\psfrag{toADC}{\scriptsize{chip rate}}
\psfrag{toMIXrx}{\scriptsize{$o_\RX(t)$}}
\psfrag{ADC}{\scriptsize{ADC}}
\includegraphics[width=\textwidth]{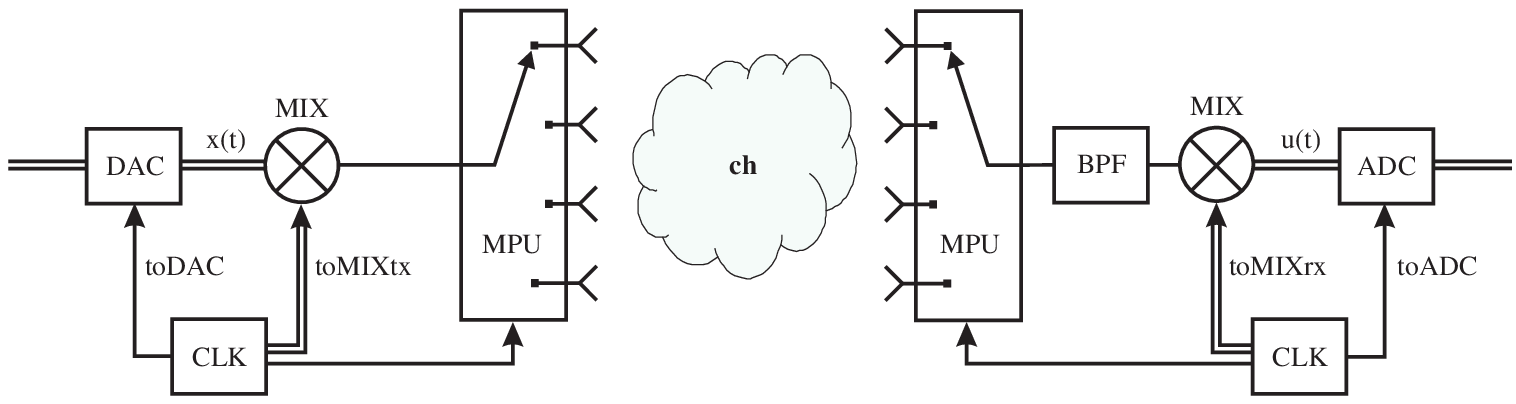}
\caption{\label{fig:TDDsounder}
Architecture of a \ac{TDMS}-based \ac{MIMO} channel sounder. 
ADC, DAC, MIX, MPU, BPF, and CLK stand for 
analog-to-digital converter, digital-to-analog converter, mixer, 
multiplexing unit, bandpass filter, and (reference) clock, respectively.}
\end{figure}

The basic architecture of a \ac{TDMS}-based \ac{MIMO} channel 
sounder is depicted in \figref{fig:TDDsounder}.
A (possibly complex) sounding signal $x(t)$ is generated in baseband and modulated to the
propagation channel center frequency. At the receiver, bandpass filtering
(to remove out-of-band thermal noise)
and downconversion to baseband, resulting in the signal $u(t)$, 
is followed by extraction of the \ac{MIMO} channel estimate.
Both the transmitter and the receiver employ  
a multiplexing unit, which steps a single \ac{RF} chain through 
all transmit/receive antennas sequentially in time following a
prescribed switching pattern. 
Clocks at transmitter and receiver serve as reference for the 
\ac{ADC} and \ac{DAC} sampling rates, 
antenna multiplexing timing,
and \ac{RF} mixing (i.e., \ac{LO}) frequencies.

Another frequently used \ac{MIMO} channel sounder setup 
employs a single antenna (either at the transmitter or the receiver
or at both sides of the link),
which is physically moved (in an automated fashion) to form a ``virtual'' antenna array. 
This setup also fits into the framework described in the paper. 
The impact of phase errors on \ac{MIMO} channel capacity estimates in 
such a time-division-based
``virtual'' antenna array sounder architecture will, 
in general, be significantly more pronounced than in 
the \ac{TDMS}-based case where a single RF chain is switched electronically
into different physical antenna elements. 
This is because 
in the ``virtual'' antenna array case, 
the time that passes when moving the single antenna 
from a given physical position to the next one
is much longer than
the time it takes to switch electronically between
different antenna elements.

\subsection{SISO Signal Model}
\label{sec:SISOsig}

We start by presenting the signal model for a correlative 
\ac{SISO} channel sounder
in the presence of phase errors. This constitutes the basis for the \ac{MIMO}
signal model in \ac{TDMS}-based sounders introduced in 
\secref{sec:MIMOSigMod}.

Apart from frequency offset, a difference between the reference clocks
at transmitter and receiver also causes 
a difference in the corresponding baseband sampling rates.
The effect of this sampling rate offset, however, 
can be neglected 
compared to the frequency offset incurred by up- and down-conversion with 
respect to the \ac{LO} frequencies,
which are significantly higher than the sampling rates.
The signal model presented below will therefore
not account for sampling rate offset.

\subsubsection{Correlation-based sounding}
The baseband sounding signal is given by $x(t)=\sum_k s(t-t_k)$ 
where $s(t)$, supported on an inverval $\INTVL_\TX$, denotes the 
convolution of the (time-continuous) transmit sounding sequence\footnote{The sounding sequence is a weighted
chip-spaced Dirac train with the weights determined by the time-discrete sounding signal, e.g. a \ac{BPSK}-modulated m-sequence.}
with the impulse response of the transmit frontend filter. Further,
$t_k$ is the \ac{SISO} snapshot (measurement) time instant,
referred to as snapshot time in the following, 
and $k$ denotes the \ac{SISO} snapshot index.
Because the sounder can have a measurement duty cycle less than one,
the time intervals $[t_{k-1}, t_{k}]$ generally contain 
measurement time and void time.
In the following, we will refer to the quantities 
$t_{k}-t_{k-1}$ as the SISO snapshot time distances.
After bandpass filtering and downconversion, the receiver
applies a \ac{LTI} filter with impulse response $r(t)$, 
supported on an inverval $\INTVL_\RX$, and consisting of the receive sounding sequence convolved with the receive
frontend filter. The signals
$s(t)$ and $r(t)$ are chosen such that the function
\begin{align*}
  c(t) 
= 
  s(t)\conv r(t)
\end{align*}
is peaky around $t=0$, where it takes on its maximum value
\begin{align}
  c(0) 
= 
  \int_{\INTVL_{\TX\RX}}
  s(t) r(-t) \, \ud t = 1
\label{eq:w_norm}
\end{align}
with $\INTVL_{\TX\RX} = \INTVL_\TX \cap (-\INTVL_\RX)$.
In the following, we will be interested in the behavior of
the function $c(t)$ in an interval $\INTVL_\VLD$ around $t=0$.
The \emph{sequence \ac{SNR}} quantifies the peakiness of $c(t)$ (in an interval $\INTVL_\VLD$).

In the sequel, we denote channel sounders satisfying 
the conditions described in the previous paragraph as
correlation(-based) sounders. 
In the literature, the term correlation sounder
is usually reserved for a more specific setup where $r(t)=s^{\ast}(-t)$ with $s(t)$ resulting from a
\ac{BPSK} modulated \ac{PNS} (typically an m-sequence).
The widely used class of channel sounders employing
chirp or multisine sounding sequences \cite{MMHS02j,MMSH99c,FlJo96c,CuFM93j,PaDT91c}
satisfies the conditions stated in the previous paragraph
and hence fits into our framework.

\subsubsection{Modeling phase noise and frequency offset}
\label{sec:instabilities}
The \acp{LO} in the transmitter and the receiver generate signals that have 
the analytic signal representations
\begin{align*}
o_\TX(t) = e^{ \jmath ( 2 \pi f_\TX t + \pr_\TX(t) ) } \quad \text{and} \quad
o_\RX(t) = e^{ \jmath ( 2 \pi f_\RX t + \pr_\RX(t) ) }
\end{align*}
respectively,
where $f_\TX$ and $f_\RX$ are the corresponding desired \ac{LO} frequencies,
and $\pr_\TX(t)$ and $\pr_\RX(t)$ represent the corresponding additive \ac{LO} phase.
In the remainder of this paper, we assume that $\pr_\TX(t)$ 
and $\pr_\RX(t)$ are real-valued zero-mean 
\ac{WSS} Gaussian random processes.
We shall next justify the modeling assumptions made in this paragraph.

The characterization of phase and frequency instabilities in precision frequency sources has
been a major area of research for many years 
\cite{Rutm78,Rutm91j,GeBa85,DeMR00j,Demi02j,Mehr02j,UFFCe}.
In the ensuing discussion, we distinguish between
the directly observable random process 
$o(t) = \exp\big( \jmath \big(2 \pi f t + \pr(t) \big)\big)$ and the 
underlying phase process $\pr(t)$. 
There are two fundamentally different models for $\pr(t)$ motivated
by the corresponding different methods of frequency generation, 
namely free-running and closed-loop. 
In the case of free-running 
oscillators, the phase process
is often modeled as a continuous-time Wiener
(a.k.a.\ Brownian motion) process, which is Gaussian and nonstationary,
with autocovariance function
$\expect\{ \pr(t) \pr(t+\tau) \} \propto \min(t, t+\tau)$ \cite[Corollary\rsq7.1]{DeMR00j}. 
It is important to note that even though the phase process is nonstationary 
and has a variance that grows without bounds over time, 
the corresponding observable process is stationary with finite power and
has an approximately Lorentzian (i.e., one-pole lowpass) 
\ac{PSD} function around the first harmonic \cite[Eq.\rsp(41)]{DeMR00j}.
In the case 
of a closed-loop (a.k.a.\ phase-feedback) system, 
such as a \ac{PLL}, in the locked and steady state, 
the phase difference between its input and output signal
can be modeled as a Gaussian \ac{WSS} process with bounded variance. 
In practice, frequency generation is often performed by locking
one or more \acp{PLL} to the output of a free-running oscillator,
which results in an overall phase process $\pr(t)$ that is the sum 
of a Brownian motion process and an asymptotically (in time) 
Gaussian \ac{WSS} process \cite{Mehr02j}. 
Generally speaking, the components in $\pr(t)$ corresponding to the free-running oscillator 
and to the \acp{PLL} dominate the long-term and the short-term 
phase process behavior, respectively.
In \ac{TDMS}-based \ac{MIMO} channel sounding, the time-scale of interest, determined by 
the duration of one \ac{MIMO} snapshot (i.e., the duration it takes to measure all scalar
subchannels of the \ac{MIMO} channel), is such that the 
behavior of the phase process
is typically dominated by the component due to the \ac{PLL}(s).
The nonstationary (Brownian motion) component in $\varphi(t)$ can therefore be neglected. 
We shall support this statement in \secref{sec:meas} through
measurements\footnote{While 
we did not observe the Brownian motion component 
within the duration of a \ac{MIMO} snapshot in our measurements, 
we do acknowledge that it may occur for the channel sounder under consideration
if the \ac{MIMO} snapshot duration is sufficiently long.}
performed on a commercially employed \ac{TDMS}-based \ac{MIMO} channel sounder.

The absolute frequencies generated by the independent clock sources at 
transmitter and receiver can differ significantly, which is accounted for
by allowing a nonzero frequency offset $\Delta \omega = 2\pi(f_\TX - f_\RX)$.
The resulting overall model, specified below,
captures all relevant effects in a simple and mathematically tractable manner.

\subsubsection{\acs{SISO} signal model}
\label{sec:narrowband}
We are now ready to state the \ac{SISO} signal model including the effects of
phase errors. Throughout the paper, we shall work in complex
baseband. We assume that the (time-varying) physical channel to be measured 
is \ac{LTI} during each snapshot interval and can change from snapshot to snapshot.
The impulse response during the $k$th snapshot interval 
$t_k+\INTVL_\TX$ $(k\in\numID{Z})$ is given by 
\begin{align}
  h_k(t) 
&= 
  \sum_{l} h^{(l)}_k \, \delta(t-\tau^{(l)}_k),
\quad
  \tau^{(m)}_k\ne\tau^{(n)}_k\;\forall\:m\ne n
\label{eq:CIR}
\end{align} 
and the individual snapshot intervals do not overlap. 

The overall signal model (in complex baseband notation) 
for the $k$th snapshot
is hence given as
\begin{align*}
u_k(t) = r(t) \conv \bigg( 
    o^*_{\RX,\BB}(t) \Big( h_k(t) 
    \conv \big( o_{\TX,\BB}(t) \, s(t-t_k) \big) \Big) + n'(t) 
    \bigg)
\end{align*}
where we set
$o_{\TX,\BB}(t) = o_{\TX}(t) \, \exp(-\jmath2\pi f_\TX t) 
= \exp\big(\jmath\pr_\TX(t)\big)$ and
$o_{\RX,\BB}(t) = o_{\RX}(t) \, \exp(-\jmath2\pi f_\TX t)$
$= \exp\big(-\jmath\big( \Delta\omega t - \pr_\RX(t)\big)\big)$,
and $n'(t)$ represents thermal noise at the receiver input.
Straightforward manipulations yield
\begin{align}
u_k(t) = 
    \sum_{l} h^{(l)}_k \, r(t) \conv 
\Big( \underbrace{
    e^{ \jmath \Delta\omega t} \,
    e^{-\jmath \pr_\RX(t) } \, 
    e^{ \jmath \pr_\TX(t-\tau^{(l)}_k) } 
}_{\theta_{k,l}(t-t_k-\tau^{(l)}_k)}
\,
    s(t-t_k-\tau^{(l)}_k) \Big) 
+ n(t) 
\label{eq:OctB}
\end{align}
where $n(t) = r(t) \conv n'(t)$. 
In the absence of phase errors, i.e., 
$\pr_\TX(t) = \pr_\RX(t) = \Delta\omega = 0$,
we have
\begin{align}
  u_k(t) 
= 
  \sum_{l} h^{(l)}_k \, c(t-t_k-\tau^{(l)}_k) + n(t).
\label{eq:ut_clean}
\end{align}
Now, if the interval $\INTVL_\VLD$ is such that 
$\tau_k^{(l)} \in \INTVL_\VLD \;\forall\:k,l$,
the peakiness of $c(t) = s(t) \conv r(t)$ 
(within $\INTVL_\VLD$) implies that
the channel coefficients $h^{(l)}_k$ can be retrieved by
sampling $u_k(t)$ at the time instants
$t_k+\tau^{(l)}_k$.
The measurement \ac{SNR} in extracting the coefficients
$h^{(l)}_k$ from $u_k(t)$ is clearly limited,
among other factors, by the sequence \ac{SNR}.
From \eqref{eq:OctB}, we can see that in the presence of
phase errors, one has to deal with the quantity
\begin{align*}
  c'_{k,l}(t)
&= 
  \big( \theta_{k,l}(t) \, s(t) \big) \conv r(t) 
\end{align*}
instead of $c(t)$ as in \eqref{eq:ut_clean}.
Phase errors,
accounted for by the term $\theta_{k,l}(t)$,
therefore can lead to a shifting of the peak of 
$c(t)$ (which is at $t=0$ in the phase-error free case)
and to a degradation of the correlation properties,
or equivalently, the peakiness of $c(t)$ 
as quantified by the sequence SNR.
We shall next show that, under quite general conditions,
we can write 
\begin{align}
  c'_{k,l}(t)
&=
  c_{k,l} \, p_{k,l}(t)
\label{eq:zerlegung}
\end{align}
where $p_{k,l}(0) = 1 \;\forall\:k,l$, and 
the $p_{k,l}(t)$ are functions that have their peaks at $t=0$.
As a consequence of \eqref{eq:zerlegung}, we then obtain
\begin{align}
  u_k(t) 
= 
  \sum_{l} h^{(l)}_k \, c_{k,l} \, p_{k,l}(t-t_k-\tau^{(l)}_k) + n(t)
\label{eq:sametime}
\end{align}
which implies that the channel coefficients in the presence of
phase errors are given by $h^{(l)}_k \, c_{k,l}$ 
and can still
be obtained by sampling $u_k(t)$ at the time instants 
$t_k+\tau^{(l)}_k$. Moreover, the functions $p_{k,l}(t)$ 
will be less peaky than $c(t)$ (within the interval $\INTVL_\VLD$), which results in a reduction
of the sequence \ac{SNR} and hence the
measurement SNR in terms of extracting the channel
coefficients by sampling the function $u_k(t)$.

A general but restrictive condition for the 
$p_{k,l}(t)$, $\forall\:k,l$ to have their peaks
at $t=0$ is that $\theta_{k,l}(t)$, $\forall\:k,l$ 
be narrowband relative to $s(t)$ which would imply
$c'_{k,l}(t) \propto c(t)$.
This ``slow phase noise'' condition,
however, is in practice hardly satisfied.
A more systematic approach to assessing the degradation
of sequence SNR due to phase errors consists of decomposing
$\theta_{k,l}(t)$ into the harmonics $\exp(\jmath\omega_d t)$
and evaluating the behavior of the quantity
\begin{align*}
  \big( s(t) \, e^{\jmath\omega_d t} \big) \conv r(t) 
=
  \int_{-\infty}^{\infty} s(-\xi) \, r(t+\xi) \, e^{-\jmath\omega_d\xi} \, \ud\xi 
\end{align*}
which for $r(t) = s^*(-t)$ 
becomes Woodward's correlation function \cite[Ch.~7, Eq.\rsp(17)]{Wood53b}
\begin{align*}
  A_{s^{-}}(t,\omega_d)=
  \int_{-\infty}^{\infty} s(-\xi) \, s^*(-\xi-t) \, e^{-\jmath\omega_d\xi} \, \ud\xi
\end{align*}
where $s^{-}$ stands for $s(-t)$. In the remainder of this paragraph, we restrict ourselves
to $r(t)=s^{\ast}(-t)$, for simplicity. In the general case Woodward's correlation function
is replaced by the cross-correlation function between $s(t)$ and $r(t)$. 
Denoting the Fourier transform of $\theta_{k,l}(t)$ 
by $\Theta_{k,l}(\jmath\omega)=\int_{-\infty}^{\infty}\theta_{k,l}(t)e^{-j\omega t}\,dt$,
it follows that 
\begin{align*}
  c'_{k,l}(t) 
= 
  \frac{1}{2\pi} 
  \int_{-\infty}^{\infty} \Theta_{k,l}(\jmath\omega) \, A_{s^{-}}(t,\omega) \, \ud\omega.
\end{align*}
Analyzing the quantity $\expect\{|c'_{k,l}(t)|^{2}\}$, it can be verified that
for m-sequences (used, e.g., in the channel sounder analyzed in 
\secref{sec:meas}) $A_{s^{-}}(t,\omega)$ is such that, under quite general conditions on the phase noise PSD,
peak-shifting does not occur, i.e.,
$c'_{k,l}(t)$ has its peak at $t=0$, $\forall\:k,l$,
and hence \eqref{eq:zerlegung} is satisfied. However, for sounders employing, for example, chirp sequences
peak-shifting does occur.

We shall next show that, under assumptions validated by 
our measurements in \secref{sec:meas}, \eqref{eq:zerlegung}
can be simplified further in the sense that the 
$c_{k,l}$ do not depend on $l$. 
Straightforward manipulations reveal that
\begin{align}
  c_{k,l} 
&= 
  e^{ \jmath {\Delta\omega} (t_k+\tau^{(l)}_k) } \,
  \int_{\INTVL_{\TX\RX}}
  e^{ \jmath {\Delta\omega} \xi } \,
  e^{-\jmath \pr_\RX(t_k+\tau^{(l)}_k+\xi) } \,
  e^{ \jmath \pr_\TX(t_k+\xi) } \,
  w(\xi) \, 
  \ud\xi
\label{eq:eff_lin_ct}
\end{align}
where $w(t) = s(t) \, r(-t)$ is a window function. 
For \ac{BPSK} modulated m-sequences 
with rectangular chip pulses (used in the sounder analyzed in \secref{sec:meas}) 
and $r(t)=s^*(-t)$,
for example, we have $w(t) = const.$ for 
$t\in\INTVL_{\TX\RX}$.
Next, assuming that
\begin{itemize}
\item[i)] $\pr_{\RX}(t)$, $\pr_{\TX}(t)$, and $\Delta \omega t$ 
do not change appreciably during an interval of length 
$\ilen{\INTVL_{\TX\RX}}$, 
and thus, recalling the normalization of $w(t)$ according to \eqref{eq:w_norm},
\begin{align}
  \left| \int_{\INTVL_{\TX\RX}}
  e^{ \jmath {\Delta\omega} \xi } \,
  e^{-\jmath \pr_\RX(t_k+\tau^{(l)}_k+\xi) } \,
  e^{ \jmath \pr_\TX(t_k+\xi) } \,
  w(\xi) \, 
  \ud\xi \right| 
&\approx 
  1
\label{eq:smallChange}
\end{align}

\item[ii)] the delay spread is small enough for 
\begin{align}
  \Delta\omega\tau^{(l)}_k 
\approx 
  0
\qquad \text{and}& \qquad
  \pr_{\RX}(t_{k}+\tau^{(l)}_k+\xi) 
\approx 
  \pr_{\RX}(t_{k}+\xi), \quad \forall\: k,l
\label{eq:assDS}
\end{align}
to hold,
\end{itemize}
and further assuming \ac{WILOG} that the interval $\INTVL_{\TX\RX}$ is symmetric around $t=0$,
we obtain
\begin{align}
c_{k,l}
\approx
 e^{\jmath \Delta\omega t_k} \, e^{\jmath\pr_k}
\label{eq:simpl_c}
\end{align}
which, obviously, does not depend on $l$. Here,
$\pr_k$ is a real-valued \ac{WSS} Gaussian random process with zero-mean
and variance $\sigma_{\PP}^2$. We note that formally the condition that $\pr_{\RX}(t)$ does not change
appreciably during an interval of length $|\INTVL_{\TX\RX}|$ implies the second condition in \eqref{eq:assDS}
as the delay spread is significantly smaller than $|\INTVL_{\TX\RX}|$. We decided, however, to state the second condition in \eqref{eq:assDS}
separately in order to stress that it is more critical that the approximation error in the second condition in \eqref{eq:assDS} be small.
Inserting \eqref{eq:simpl_c} into \eqref{eq:sametime} finally yields
\begin{align}
 u_k(t) 
=
 \sum_l h_k^{(l)} \, e^{ \jmath {\Delta\omega} t_k } \, 
                     e^{ \jmath \pr_{k} } \, 
p_{k,l}(t-t_k-\tau_k^{(l)}) 
+ n_k
\label{final_sys_mod}
\end{align}
which implies that the impact of phase noise and frequency offset is to modulate the (potentially
frequency-selective) physical channel process according to 
\begin{align*}
 \hat{h}_k^{(l)} 
=
 h_k^{(l)} \, e^{\jmath {\Delta\omega} t_k} \, e^{\jmath\pr_{k}}, \quad \forall\: l.
\end{align*}
For flat-fading physical channels, considered in the remainder of the paper, we simply get
\begin{align}
 \hat{h}_k
=
 h_k \, e^{\jmath\mu_k} \, e^{\jmath\pr_{k}}
\label{eq:OctBeta}
\end{align}
where we set $\mu_k= \Delta\omega t_k$. We conclude by noting that even though the model \eqref{eq:OctBeta} is standard and well-known
in the literature, we decided to provide a detailed derivation in order to exhibit the underlying key
assumptions.
In particular, as already mentioned, these assumptions will be validated in our
measurements in \secref{sec:meas}. 

In practice, processing on the receive side
is often implemented through periodic convolution rather than linear convolution, as described above. 
The corresponding class of sounders will be called circular convolution-based in the following and
entails extending $s(t)$ and $r(t)$ periodically to result in the 
$|\INTVL_{\mathrm{P}}|$-periodic signals $\tilde{s}(t)$ and $\tilde{r}(t)$, where $|\INTVL_{\mathrm{P}}|$ is given
by the length of the sounding (transmit and receive) sequence. Consequently, $\INTVL_{\mathrm{P}}$ denotes one
period of $\tilde{s}(t)$ and $\tilde{r}(t)$.
The signals $u_k(t)$, obtained by convolving, for each $l$, with $\tilde{r}(t-\tau_{k}^{(l)})$ rather than $r(t)$, 
are then sampled at the time instants $t_{k}$. 
As compared to the case of linear convolution, this leads to a slightly modified
expression for the coefficients $c_{k,l}$ given by
\begin{align}
c_{k,l} 
&= 
  e^{ \jmath {\Delta\omega} t_k } \,
  \int_{\INTVL_{\mathrm{P}}}
  e^{ \jmath {\Delta\omega} \xi } \,
  e^{-\jmath \pr_\RX(t_k+\xi) } \,
  e^{ \jmath \pr_\TX(t_k-\tau^{(l)}_k+\xi) } \,
  \widetilde{w}(\xi-\tau^{(l)}_k) \,
  \ud\xi
\label{eq:eff_lin_ct_circ}
\end{align}
with
$\widetilde{w}(t) = \tilde{s}(t) \, \tilde{r}(-t)$.
Now, in order to arrive at the effective input-output relation \eqref{eq:OctBeta}, we need to slightly revise
the assumptions made in the case of a linear convolution-based system.
Instead of Condition i) leading to \eqref{eq:smallChange}, we need to assume that
$\pr_{\RX}(t)$, $\pr_{\TX}(t)$, and $\Delta \omega t$ 
do not change appreciably during an interval of length $|\INTVL_{\mathrm{P}}|$
and thus
\begin{align}
  \left| 
  \int_{\INTVL_{\mathrm{P}}}
  e^{ \jmath {\Delta\omega} \xi } \,
  e^{-\jmath \pr_\RX(t_k+\xi) } \,
  e^{ \jmath \pr_\TX(t_k-\tau^{(l)}_k+\xi) } \,
  \widetilde{w}(\xi-\tau^{(l)}_k) \,
  \ud\xi \right| 
&\approx 
  1.
\label{eq:smallChangeCirc}
\end{align}
Instead of the second condition in \eqref{eq:assDS}, we need to assume
that the delay spread is small enough for
\begin{align}
  \pr_{\TX}(t_{k}-\tau^{(l)}_k+\xi) 
&\approx
  \pr_{\TX}(t_{k}+\xi), \quad \forall\: k,l
\label{eq:assDSCirc}
\end{align}
to hold. The condition $\Delta\omega\tau^{(l)}_k \approx 0$
is not needed. Circular convolution-based sounders have the advantage of reducing the required per SISO snapshot recording
interval length\footnote{The signal recorded is the signal at the output of the receive frontend filter.} from $|\INTVL_{\TX\RX}|+|\INTVL_{c}|$ to $|\INTVL_{\mathrm{P}}|$.
The inequality $|\INTVL_{\TX\RX}|+|\INTVL_{c}|\,>\,|\INTVL_{\mathrm{P}}|$ is obtained by noting that
in the case of circular convolution $|\INTVL_{\TX\RX}|+|\INTVL_{c}|=|\INTVL_{\TX}|+|\INTVL_{c}|\,\ge\,|\INTVL_{\mathrm{P}}|+|\INTVL_{c}|$, where
we used $r(t)=s^{\ast}(-t)$ and the fact that the length of $\INTVL_{\TX}$ is given by the length of the transmit sounding sequence plus the length of the transmit frontend filter impulse response.
We close the discussion by noting that the channel sounder
investigated in \secref{sec:meas} is a circular convolution-based sounder.

\subsection{MIMO Signal Model}
\label{sec:MIMOSigMod}
The basic principle of \ac{TDMS}-based \ac{MIMO} channel sounding is to sequentially measure
the $\MT\MR$ scalar subchannels of the \ac{MIMO} channel. 
Since the individual \ac{SISO} subchannels (note that we 
continue to use the term \emph{\ac{SISO} snapshot} to refer to the measurement
of a subchannel of the \ac{MIMO} matrix)
are band-limited stochastic processes 
(due to finite Doppler spread), it is not necessary to assume that the subchannels are
static during the entire \ac{MIMO} measurement period. Rather, it suffices to choose the sampling rate
in compliance with the 
sampling theorem and to properly align the measurements in time \cite{SRTW02c}.
In the remainder of the paper, we assume that this alignment has already been performed.

Denoting the \emph{effective} (i.e, the physical channel including 
the effect of phase errors) scalar subchannel 
between the $\vcol$th ($\vcol = 1,2,\ldots,\MT$) transmit 
and the $\vrow$th ($\vrow = 1,2,\ldots,\MR$) receive antenna as
$\hat{h}_{\vrow,\vcol} = h_{\vrow,\vcol} \, \exp(\jmath(\mu_{\vrow,\vcol}+\pr_{\vrow,\vcol}))$,
the corresponding effective \ac{MIMO} channel matrix can be expressed as
\begin{align}
\widehat{\mathbf{H}} 
= \mathbf{H} \circ 
\underbrace{ \exp\cwa\big( \jmath (\mathbf{M} + \mathbf{\Phi}) \big) }_{\mathbf{\Theta}}
\label{eq:channelhat}
\end{align}
where $[\mathbf{M}]_{\vrow,\vcol} = \mu_{\vrow,\vcol}$ and $[\mathbf{\Phi}]_{\vrow,\vcol} = \pr_{\vrow,\vcol}$.
What we would like to measure is the \emph{physical} channel matrix $\mathbf{H}$. 
However, due to phase errors, the sounder has access to the effective channel 
matrix $\widehat{\mathbf{H}}$ only. 
The entries in $\mathbf{M}$ and $\mathbf{\Phi}$ depend 
on the switching pattern, i.e., the order in which the individual scalar subchannels are measured,
and the \ac{SISO} snapshot time distances. 
In the following, we denote a switching pattern as an ordered sequence
of pairs $\big( (\vrow_1, \vcol_1), (\vrow_2, \vcol_2), \ldots, (\vrow_{\MT\MR},
\vcol_{\MT\MR}) \big)$ where $(\vrow_k, \vcol_k)$ means that the scalar subchannel
$h_{\vrow_k, \vcol_k}$ is being measured at time $t_k$. 
Let us consider a simple example with $\MT = \MR = 2$. The switching pattern
$\big( (1,1), (2,1), (2,2), (1,2) \big)$ leads to 
\begin{align*}
\mathbf{\Theta}_1 = \left[ 
\begin{array}{ll}
e^{\jmath(\mu_1 + \pr_1)}  &  e^{\jmath(\mu_4 + \pr_4)}  \\
e^{\jmath(\mu_2 + \pr_2)}  &  e^{\jmath(\mu_3 + \pr_3)}
\end{array} \right]
\end{align*}
whereas the switching pattern $\big( (1,1), (2,1), (1,2), (2,2) \big)$ results in 
\begin{align*}
\mathbf{\Theta}_2 = \left[ 
\begin{array}{ll}
e^{\jmath(\mu_1 + \pr_1)}  &  e^{\jmath(\mu_3 + \pr_3)}  \\
e^{\jmath(\mu_2 + \pr_2)}  &  e^{\jmath(\mu_4 + \pr_4)}
\end{array} \right].
\end{align*}
The physical channel matrix $\mathbf{H}$ is, of course, unaffected by the switching pattern.
The dependence of $\mathbf{\Theta}$ on the switching pattern and 
on the \ac{SISO} snapshot time distances
is highly problematic since different switching patterns and/or
\ac{SISO} snapshot times yield different (incorrect) measurement results for
the same physical \ac{MIMO} channel. 
The following simple example for $\MT=\MR=2$ illustrates this undesirable effect
and its implications:
Assume that the physical channel is given by $\mathbf{H}=\mathbf{1}$,
and $\mu_1 + \pr_1 = \mu_4 + \pr_4 = 0$, 
$\mu_2 + \pr_2 = -\pi/2$, and $\mu_3 + \pr_3 = \pi/2$.
It is then easily seen that 
$\lambda_1\mo\big((\mathbf{H}\circ\mathbf{\Theta}_1)(\mathbf{H}\circ\mathbf{\Theta}_1)^H\big)
=\lambda_2\mo\big((\mathbf{H}\circ\mathbf{\Theta}_1)(\mathbf{H}\circ\mathbf{\Theta}_1)^H\big)=2$ whereas
$\lambda_1\mo\big((\mathbf{H}\circ\mathbf{\Theta}_2)(\mathbf{H}\circ\mathbf{\Theta}_2)^H\big)
=4$ and
$\lambda_2\mo\big((\mathbf{H}\circ\mathbf{\Theta}_2)(\mathbf{H}\circ\mathbf{\Theta}_2)^H\big)
=0$.
In summary, starting from a rank-1 physical channel, depending on the switching pattern,
we can get a rank-1 or a rank-2 effective channel.

We conclude this section by introducing an approximation that will frequently be
used throughout the paper. 
For small phase errors, we use the standard first-order Taylor-series approximation
(see, e.g., \cite[Eq.\rsp(4.12)]{HaLe99b})
\begin{align}
\exp\cwa( \jmath \mathbf{\Phi} ) 
\approx \mathbf{1} + \jmath \mathbf{\Phi}.
\label{eq:linApprox}
\end{align}
In the remainder of the paper, whenever referring to $\widehat{\mathbf{H}}$, unless explicitly stated
otherwise, we shall use the exact expression for $\mathbf{\Theta}$ according to \eqref{eq:channelhat}.
We conclude this section by noting that, throughout the paper, whenever we deal with random 
physical channels, $\mathbf{\Theta}$ will be assumed to be statistically independent of $\mathbf{H}$.

\section{Effect of Phase Errors on MIMO Channel Statistics}
\label{sec:err_stat}
In this section, we describe the impact of phase errors
on \ac{MIMO} channel statistics (i.e., the statistics of 
the effective \ac{MIMO} channel $\widehat{\mathbf{H}}$
vs.\ the statistics of the physical \ac{MIMO} channel $\mathbf{H}$)
thereby laying the foundations for the results in 
\secsref{sec:MI}, \ref{sec:rank1}, and \ref{sec:meas}.

In the following, we consider both deterministic and stochastic physical channels $\mathbf{H}$. 
For deterministic $\mathbf{H}$, phase noise
induces randomness and hence makes the static channel appear fading.
In the case of stochastic $\mathbf{H}$, phase errors alter the channel statistics.

\subsection{Mean and Covariance of Effective MIMO Channel Matrix}
\label{sec:MaCoEMCM}

We start by investigating the impact of phase errors on the mean 
$\mathbf{H}_\FIX = \expect\{\mathbf{H}\}$ and on the covariance
$\cov\{\mathbf{H}\}$ of the physical channel. 
The developments in the sequel apply to both deterministic physical channels, where
$\mathbf{H}_\FIX = \mathbf{H}$ and $\cov\{\mathbf{H}\}=\mathbf{0}$,
and random physical channels.
Using \eqref{eq:channelhat},
a straightforward calculation reveals that
\begin{align}
\expect\{\widehat{\mathbf{H}}\} 
& =
\mysqrt{\kappa} \, \exp\cwa(\jmath \mathbf{M}) \circ \mathbf{H}_\FIX
\nonumber
\\
\label{eq:cov_prop}
\cov\{\widehat{\mathbf{H}}\} 
& = \kappa \, ( \mathbf{m} \mathbf{m}^H  ) \circ \Big[
      \exp\cwa\big( \cov\{\mathbf{\Phi}\} \big) 
      \circ \Big( 
      \vect(\mathbf{H}_\FIX) \, \contra[big]{\vect}{(\mathbf{H}_\FIX)}
      + \cov\{\mathbf{H}\} \Big) 
\\
&   \qquad\qquad\qquad
      - \vect(\mathbf{H}_\FIX) \, \contra[big]{\vect}{(\mathbf{H}_\FIX)}
      \Big]
\nonumber
\end{align}
where $\kappa = \exp(-\sigma_{\pr}^2)$, 
$\mathbf{m} = \vect\big(\exp\cwa(\jmath\mathbf{M})\big)$,
and we made use of the relations 
$\vect(\mathbf{A}\circ\mathbf{B}) \, \contra[big]{\vect}{(\mathbf{A}\circ\mathbf{B})} 
=\big( \vect(\mathbf{A})\,\contra[big]{\vect}{(\mathbf{A})} \big)
\circ \big( \vect(\mathbf{B})\,\contra{\vect}{(\mathbf{B})} \big)$ and
$\expect\{\exp(\jmath X)\}=\exp(\jmath m_X-\sigma^2_X/2)$ 
for $X\sim\realGauss(m_X,\sigma^2_X)$.

We observe that phase noise leads to an attenuation of the first moment of the
physical channel by a factor of $\mysqrt{\kappa}$. The presence of a frequency offset
(reflected by the matrix $\mathbf{M}$)
can result in $\expect\{\widehat{\mathbf{H}}\}$ having a higher rank than $\mathbf{H}_\FIX$
which in turn implies that the spatial 
multiplexing gain of the effective channel can be higher than that 
of the underlying physical channel. Take for example a deterministic
rank-1 physical channel with $\mathbf{H}=\mathbf{H}_\FIX=\mathbf{1}$. In the
absence of phase noise, we have $\expect\{\widehat{\mathbf{H}}\}
=\exp\cwa(\jmath\mathbf{M})$ which, depending on the frequency offset characteristics, 
can even have full rank.
As a simple example, consider the switching pattern 
$\big( (1,1), (2,1), (2,2), (1,2) \big)$
with $\mu_k = \exp(\jmath\pi k/2)$, which results in the full-rank matrix
\begin{align*}
\expect\{\widehat{\mathbf{H}}\} =
\exp\cwa(\jmath\mathbf{M}) &= \left[ 
\begin{array}{ll}
\jmath  &  1 \\
-1      &  -\jmath 
\end{array} \right].
\end{align*}
The conditions for frequency offset (in terms of the measurement setup)
to have a significant impact on the measurement error in terms of \ac{MI}
will be discussed in \secref{sec:freq_offs}. 

The impact of phase errors on the channel's second-order statistics is more
involved. Consider, for example, the case of no frequency offset
($\mathbf{M}=\mathbf{0}$ and hence $\mathbf{m}=\mathbf{1}$) 
and fully correlated phase noise, i.e., 
$\cov\{\mathbf{\Phi}\}=\sigma^2_{\PP}\mathbf{1}$,
representative of a high-quality \ac{LO}.
In this case, \eqref{eq:cov_prop} yields
\begin{align*}
\cov\{\widehat{\mathbf{H}}\} 
= \cov\{\mathbf{H}\} 
+ \vect(\mathbf{H}_\FIX) \contra{\vect}{(\mathbf{H}_\FIX)} \, ( 1 - \kappa )
\end{align*}
which shows that the 
mere presence of phase noise, even if it is fully correlated, alters the covariance
matrix of the physical channel by adding a rank-1 component to $\cov\{\mathbf{H}\}$. 
For a deterministic physical channel, where 
$\cov\{\mathbf{H}\}=\mathbf{0}$, we can see that the presence of phase noise randomizes
the channel and yields an effective channel with the rank-1 covariance matrix $\cov\{\widehat{\mathbf{H}}\} 
=\vect(\mathbf{H}_\FIX) \contra{\vect}{(\mathbf{H}_\FIX)} \, (1-\kappa)$. 
We conclude this discussion by investigating the case of fully uncorrelated phase noise,
representative of a very poor \ac{LO}, still assuming $\mathbf{M}=\mathbf{0}$.
In this case, we have $\mathbf{\Phi}=\sigma^2_{\PP}\mathbf{I}$ and consequently 
\begin{align}
  \cov\{\widehat{\mathbf{H}}\} = (1-\kappa) \, 
  \dg\Big( \vect(\mathbf{H}_\FIX) \contra[big]{\vect}{(\mathbf{H}_\FIX)} \Big)
+ \left( \begin{array}{ccccc}
  1 & \kappa & \kappa & \cdots & \kappa \\
  \kappa & 1 & \kappa & \cdots & \kappa \\
  \vdots & \ddots &        & \ddots & \vdots \\
  \kappa & \kappa & \kappa & \cdots & 1 
  \end{array} \right) \circ \cov\{\mathbf{H}\}.
\label{eq:eqZ}
\end{align}
We can see that again for deterministic physical channels (where $\cov\{\mathbf{H}\}=\mathbf{0}$)
the effective channel has a
nonzero second moment and is hence randomized by phase noise. An important effect 
is brought out by 
starting with a purely Rayleigh fading physical channel (i.e., 
$\mathbf{H}_\FIX=\mathbf{0}$) with fully correlated entries (due to insufficient
antenna spacing for example) so that $\cov\{\mathbf{H}\}=\mathbf{1}$. 
If the phase noise
variance is high so that $\kappa=\exp(-\sigma^2_{\PP})$ is small, 
\eqref{eq:eqZ} implies that $\cov\{\widehat{\mathbf{H}}\}\approx\mathbf{I}$ 
which amounts to a full decorrelation of the
channel entries. Consequently, the effective \ac{MIMO} channel 
``looks like'' an \ac{IID} channel. There is one subtle difference, though,
to the widely used \ac{IID} Rayleigh fading channel model, 
namely that the effective \ac{MIMO} channel matrix 
will not have \ac{JG} entries (see \secref{sec:lJG}).
We can therefore not conclude that the entries in the effective channel matrix
are statistically independent. Nevertheless, the statistics of the effective channel will cause
significant overestimation of the channel's \ac{MI} and capacity
(see, for example, the numerical results in \secref{sec:num_res}).
In the case of fully correlated phase noise, 
the effective \ac{MIMO} channel matrix will have JG entries and, 
as discussed above, we have (assuming 
$\mathbf{H}_\FIX=\mathbf{0}$) 
$\cov\{\widehat{\mathbf{H}}\}=\cov\{\mathbf{H}\}$ 
so that overestimation does not occur. 

We close this discussion by noting that we have identified a number
of possible scenarios where the presence of phase errors
can significantly alter the
\ac{MIMO} channel statistics and, in particular, can lead to substantial rank increase of the
effective channel's covariance matrix \ac{WRT} the underlying physical 
channel's covariance matrix. From a channel measurement point
of view, the consequences are significant measurement errors in terms of \ac{MI}
and capacity. Corresponding quantitative results (analytic and numerical) will be
provided in \secsref{sec:MI} and \ref{sec:rank1}. 
Finally, we recall that the situation is exacerbated by the fact that 
measurement errors depend significantly on the antenna switching pattern 
as well as the \ac{SISO} snapshot time distances.

\subsection{Loss of (Joint and Individual) Gaussianity}
\label{sec:lJG}

The impact of phase noise on the \ac{MIMO} channel statistics is not restricted
to the first and second moments as discussed above. 
Rather, as shown below, it affects the joint and individual
distributions of the scalar subchannels
in a profound way. In the following, for the sake of clarity of exposition, 
we set $\mathbf{M}=\mathbf{0}$.

\subsubsection{Effect on the joint distribution of the Rayleigh fading
physical channel elements}
\label{sec:ejl}
We start by assuming that the physical channel is Rayleigh distributed, 
i.e., $\mathbf{H}_\FIX = \expect\{\mathbf{H}\} = \mathbf{0}$ 
and $\mathbf{H}$ is a zero-mean proper 
(or equivalently ``circularly symmetric'')
complex Gaussian random matrix.
We shall next show that even though the individual entries of $\widehat{\mathbf{H}}$
continue to be circularly-symmetric complex Gaussian, 
they will, in general, not be \ac{JG}. 
The first part of this statement is made precise in the following
Lemma for which, even though it is straightforward, we could not find
a reference in the literature. 

\begin{lemma}
Let $h\equidist\cplxGauss(0,\sigma^2_h)$ and take $\PP$ to be a continuous 
\ac{RV} with \ac{PDF} $\pdf_{\PP}(x)$. The \ac{RV} $h \, \exp(\jmath\PP)$ is 
$\cplxGauss(0,\sigma^2_h)$, irrespective of $\pdf_{\PP}(x)$.

\begin{IEEEproof} 
Since $| \exp(\jmath\PP) | = 1$, it follows immediately that the \ac{PDF} of
$|h \, \exp(\jmath\PP)|$ is equal to the \ac{PDF} of $|h|$. The statistics of the 
phase of $h \, \exp(\jmath\PP)$ are obtained by noting that
$\arg(h \, \exp(\jmath\PP)) 
= ( \arg(h)+\PP ) \mod 2\pi
= \tilde{z}$ 
with 
$\arg(h)$ being uniformly distributed in $[0,2\pi)$.
Our task is therefore reduced to finding the \ac{PDF} $\pdf_{\tilde{z}}(x)$.
Next, denote the $2\pi$-periodic 
continuation of the \ac{PDF} of $\arg(h)$ and of $\PP$ as 
$\pdf_{\tilde{h}}(x) = 1/(2\pi)$ and 
$\pdf_{\tilde{\PP}}(x) = \sum_{\vl=-\infty}^{\infty} \pdf_{\PP}(x-2\pi\vl)$, respectively. 
Noting that $\arg(h)$ and $\PP$ are statistically independent, 
we obtain $\pdf_{\tilde{z}}(x) 
= \int_{-\pi}^{\pi} \pdf_{\tilde{h}}(x) \, \pdf_{\tilde{\PP}}(y-x) \, \ud x 
= 1/(2\pi)\int_{-\pi}^{\pi} \pdf_{\tilde{\PP}}(y-x) \, \ud x 
= 1/(2\pi)$. 
\end{IEEEproof}
\end{lemma}

It remains to show that the elements in
$\widehat{\mathbf{H}}$ will,
in general, not be \ac{JG}. 
This will be done by considering a simple example. 
Since two complex Gaussian \acp{RV} $x_1$ and $x_2$ are 
\ac{JG} if and only if the linear combination
$ax_1+bx_2$ is (complex) Gaussian $\forall\:\{a,b\}\in\numID{C}$, 
it suffices to show that
$z = h_1\,\exp(\jmath\pr_1)+h_2\,\exp(\jmath\pr_2)$ with 
$h_1,h_2\equidist\cplxGauss(0,1)$
will not be Gaussian 
if $h_1$ and $h_2$ are fully correlated, i.e., 
$h_1=h_2$ and hence $z = h_1 y$ where 
$y=\exp(\jmath\pr_1)+\exp(\jmath\pr_2)$. 
A simple proof is obtained by
writing the \ac{MGF} of $z$ as $M_z(s) = \expect\{\exp(sz)\} 
= \expect_y \expect_{h_1|y}\{\exp(sh_1y)\} = \expect_y\{\exp(|y|^2|s|^2/4)\}$ 
which is the \ac{MGF} of a complex Gaussian \ac{RV} 
only if $|y|$ is deterministic (which is satisfied for
$\pr_1$ and $\pr_2$ fully correlated or both being deterministic, but not in general).

\subsubsection{Effect on the joint distribution of the 
Ricean fading physical channel elements}
\label{sec:eil}
Next, we consider a Ricean fading physical channel according to 
$\mathbf{H} \equidist \cplxGauss(\mathbf{H}_\FIX, \mathbf{\Sigma}_\mathbf{H})$,
where $\mathbf{H}_\FIX \ne 0$.
We shall show that 
the presence of phase noise results not only in a loss of joint 
Gaussianity of the elements of $\widehat{\mathbf{H}}$, as in the
Rayleigh fading case, but also in a loss of 
properness and Gaussianity
of the individual entries in $\widehat{\mathbf{H}}$.
The loss of properness follows by noting that
\begin{align*}
  \cov_\mathrm{p}\mo\{ \hat{h} \} 
&=
  \expect\big\{ ( \hat{h} - \expect\{ \hat{h} \} )^2 \big\} 
=
  \expect\{ \hat{h}^2 \} - (\expect\{ \hat{h} \})^2
=
  \expect\big\{ h^2 \, e^{\jmath2\pr} \big\}  - (\mysqrt{\kappa} \, h_f)^2
\\
&=
  h_f^2 \kappa^2 - \kappa \, h_f^2 = h_f^2(\kappa^2-\kappa)
\ne 
  0
\end{align*}
for $\sigma^2_\PP > 0$ and hence $\kappa < 1$.
It follows immediately that the elements of $\widehat{\mathbf{H}}$
are no more jointly proper as well.
The loss of Gaussianity is a direct consequence
of $h_\FIX \, e^{\jmath \pr}$ not being Gaussian 
distributed. Since the marginals are not Gaussian,
we can immediately conclude that the elements of 
$\mathbf{H}$ will not be \ac{JG} either.

\subsubsection{Special cases}
We conclude this discussion by dropping the 
assumption $\mathbf{M}=\mathbf{0}$ and
identifying two interesting special cases where, 
despite the presence of phase errors, 
the effective \ac{MIMO} channel matrix has \ac{JG} entries.

The first case is that of a deterministic physical channel $\mathbf{H} = \mathbf{H}_\FIX$
subject to ``small'' (in the sense of approximation \eqref{eq:linApprox} being appropriate) 
phase noise. 
More specifically, we have 
\begin{align*}
\widehat{\mathbf{H}} = \mathbf{H}_\FIX \circ \exp\cwa(\jmath\mathbf{M}) 
                   \circ ( \mathbf{1} + \jmath\mathbf{\Phi}).
\end{align*}
Since the entries in $\mathbf{\Phi}$ are samples 
of a zero-mean (real) Gaussian process,
any finite set of such samples is zero-mean \ac{JG} 
which results in the entries of $\widehat{\mathbf{H}}$ being \ac{JG}.
The corresponding first and second moments are given by
\begin{align}
\expect\{ \widehat{\mathbf{H}} \} & = \mathbf{H}_\FIX 
                              \circ \exp\cwa(\jmath\mathbf{M}) 
\nonumber
\\
\cov\{ \widehat{\mathbf{H}}\} & = (\mathbf{m}\mathbf{m}^H) \circ 
      \big( \vect(\mathbf{H}_\FIX) \contra{\vect}{(\mathbf{H}_\FIX)} \big)
      \circ \cov\{ \mathbf{\Phi} \}. 
\label{eq:cm2lin}
\end{align}
Note that while the entries in $\widehat{\mathbf{H}}$ are \ac{JG}, 
they will neither be jointly proper
nor individually proper in general
so that the second-order description 
in \eqref{eq:cm2lin} is incomplete 
without specifying the pseudo-covariance matrix 
obtained as
\begin{align*}
\cov_\mathrm{p}\mo\{ \widehat{\mathbf{H}}\} 
&=  - (\mathbf{m}\mathbf{m}^T) \circ 
      \big( \vect(\mathbf{H}_\FIX) \transp{\vect}{(\mathbf{H}_\FIX)} \big)
      \circ \cov\{ \mathbf{\Phi} \}. 
\end{align*}

\label{sec:noeffect}
The second special case is that of an \ac{IID} purely Rayleigh fading 
physical channel $\mathbf{H}$. 
Defining $\mathbf{h}=\vect(\mathbf{H}) \equidist\cplxGauss(\mathbf{0},\mathbf{I})$,
we want to show that $\mathbf{D}\mathbf{h}\equidist\cplxGauss(\mathbf{0},\mathbf{I})$, where 
$\mathbf{D}=\diag\big(
  \big[\, \exp\big(\jmath(\mu_1+\pr_1)\big) \,\;\,
          \exp\big(\jmath(\mu_2+\pr_2)\big) \,\;\, \cdots \,\;\, 
          \exp\big(\jmath(\mu_{\MT\MR}+\pr_{\MT\MR})\big) \,\big] \big)$.
We start by noting that the characteristic function of $\mathbf{h}$
is given by
\cite[Eq.\rsp(20)]{Wood56j}
\begin{align*}
\Psi_\mathbf{\mathbf{h}}(\jmath\bm{\nu}) 
= \expect\{ e^{\jmath \Re(\mathbf{h}^H\bm{\nu}) } \}
= e^{-\bm{\nu}^H\bm{\nu}/4}
, \quad
\bm{\nu}\in\numID{C}^{\MT\MR}.
\end{align*}
The characteristic function of $\mathbf{D}\mathbf{h}$
is obtained as
\begin{align*}
\begin{split}
\Psi_{\mathbf{D}\mathbf{h}}(\jmath\bm{\nu}) 
& =  \expect\{ e^{\jmath \Re( \mathbf{h}^H \mathbf{D}^H \bm{\nu})  } \}
  =  \expect_{\mathbf{D}}\{ \expect_{\mathbf{h}|\mathbf{D}}\{ e^{\jmath \Re( \mathbf{h}^H \mathbf{D}^H \bm{\nu} ) }  \} \} 
\\
& =  \expect_{\mathbf{D}}\{ e^{-\bm{\nu}^H \mathbf{D} \mathbf{D}^H \bm{\nu}/4} \}
  = \expect_{\mathbf{D}}\{ e^{-\bm{\nu}^H \bm{\nu}/4} \}
  = e^{-\bm{\nu}^H \bm{\nu}/4}
  = \Psi_\mathbf{\mathbf{h}}(\jmath\bm{\nu}) 
\end{split}
\end{align*}
where we made use of the fact that $\mathbf{D} \mathbf{D}^H = \mathbf{I}$.
We have therefore shown that \ac{IID} purely Rayleigh fading physical channels
are not affected by phase errors. 

In closing, we would like to remark that the effective channel not 
having \ac{JG} entries, in general, is one of the main 
factors contributing to the 
difficulties in analyzing the impact of phase errors on \ac{MI}.

\section{Effect of Phase Errors on Mutual Information}
\label{sec:MI}

Having seen in the previous section that phase errors can alter the \ac{MIMO} channel
statistics significantly, the purpose of this section is to analyze the corresponding 
impact on \ac{MI} for random physical channels.
Analytic results for the general (arbitrary rank (\ac{WIP}\!\rsp1) of
the physical channel) case seem very difficult to obtain. We shall therefore restrict
our discussion to identifying cases where the \ac{MI} is not affected 
even though the channel statistics are. 
In addition,
representative numerical results bringing out the key consequences of phase errors
will be provided.
Analytic results for deterministic physical channels and for (deterministic or random)
rank-1 physical channels will be provided in \secsref{sec:sens_ana} and \ref{sec:rank1}, respectively.

We analyze a \ac{MIMO} channel with input-output relation
\begin{align*}
\mathbf{r} = \mathbf{H} \mathbf{s} + \mathbf{n}
\end{align*}
where $\mathbf{s}$ is the $\MT \times 1$ transmit vector, $\mathbf{r}$ is the 
$\MR \times 1$ receive vector, and $\mathbf{n}$ is an $\MR \times 1$ 
noise vector distributed as $\cplxGauss(\mathbf{0},\mathbf{I}_{\MR})$.
Assuming no \ac{CSI} at the transmitter and perfect \ac{CSI} at the receiver, the \ac{MI} 
(in bit/s/Hz) of this channel is given by \cite{Tela99j}
\begin{align}
I = \log\,\det\bigg( \mathbf{I}_{\MR} + \frac{\rho}{\MT} \mathbf{H} \mathbf{H}^H \bigg) 
  = \log\,\det\bigg( \mathbf{I}_{\MT} + \frac{\rho}{\MT} \mathbf{H}^H \mathbf{H} \bigg) 
\label{eq:telatar}
\end{align}
where the input signal vector was assumed to be 
circularly-symmetric complex Gaussian 
with covariance matrix $(\rho/\MT)\mathbf{I}_{\MT}$ and
$\rho$ is the average \ac{SNR} at each of the receive antennas. 

The purpose of this section is to study how $I$ changes 
when $\mathbf{H}$ in \eqref{eq:telatar} is replaced
by $\widehat{\mathbf{H}}$ in \eqref{eq:channelhat}, i.e.,
to analyze the statistics of
$\hat{I} 
= \log\,\det\big( \mathbf{I}_{\MR} + (\rho/\MT) \widehat{\mathbf{H}} \widehat{\mathbf{H}}^H \big) 
= \log\,\det\big( \mathbf{I}_{\MT} + (\rho/\MT) \widehat{\mathbf{H}}^H \widehat{\mathbf{H}} \big)$.

\subsection{Cases Where Mutual Information is Not Affected}
\label{sec:MI_not_aff}

In the last paragraph of \secref{sec:err_stat},
we showed that \ac{IID} Rayleigh fading channels
are not affected by phase errors in the sense that 
the effective channel $\widehat{\mathbf{H}}$ is \ac{IID}
Rayleigh fading as well, irrespectively of the statistics of the phase errors. 
We have furthermore seen (\secref{sec:MaCoEMCM})
that for correlated
Rayleigh or Ricean fading physical channels, phase errors can have a significant 
impact on the channel statistics and thus on the corresponding
\ac{MI}.
There are cases, however, where even though the statistics 
of the effective channel $\widehat{\mathbf{H}}$ 
differ from the statistics of $\mathbf{H}$, we still have $\hat{I}=I$. 
Intuitively, this happens because of the quadratic dependence of $\hat{I}$ on $\widehat{\mathbf{H}}$.
In the following, we shall discuss two such practically very relevant cases.

\subsubsection{The low-\ac{SNR} regime} 
Phase errors have no impact on \ac{MI} in the low-\ac{SNR} regime, irrespectively of 
the physical channel's statistics and the phase error statistics.
This can easily be seen by noting that for low \ac{SNR}
\begin{align*}
I \approx \log \bigg( 1 + \frac{\rho}{\MT} \|\mathbf{H}\|_\FROB^2 \bigg)
\end{align*}
which, combined with the fact that 
$\|\mathbf{H}\|_\FROB^2 = \|\widehat{\mathbf{H}}\|_\FROB^2$, proves the statement. 
In general, we can conclude that the impact of phase errors on \ac{MI} is
more pronounced for higher \ac{SNR}. This is because phase errors lead to a
rank increase of the \ac{MIMO} channel and high-\ac{SNR} \ac{MI} depends strongly
on the rank (or multiplexing gain) of the channel, whereas low-\ac{SNR} \ac{MI}
depends only on the Frobenius norm $\|\mathbf{H}\|_\FROB$.

\subsubsection{One-sided switching or fully parallel sounding} 
\label{sec:one_sided}
For \ac{MIMO} channel sounders where either the transmitter or the receiver employs
one \ac{RF} chain per antenna (i.e., parallel sounding)
and hence no switching is necessary on the corresponding side
of the link, the effective channel matrix is given by
$\widehat{\mathbf{H}} = \mathbf{D}_\RX \mathbf{H} d_\TX$ and
$\widehat{\mathbf{H}} = d_\RX \mathbf{H} \mathbf{D}_\TX$
in the case of switching only at the receive and the transmit side, respectively. 
Here, $d_\RX$ and $d_\TX$ as well as the entries
of the diagonal matrices $\mathbf{D}_\RX$ and $\mathbf{D}_\TX$ 
are of the form $\exp\big(\jmath(\mu+\pr)\big)$. Even though
\eqref{eq:cov_prop} implies that the statistics of the effective channel $\widehat{\mathbf{H}}$
are different from the statistics of the physical channel $\mathbf{H}$, 
it is easily seen by direct insertion into \eqref{eq:telatar} that
the \ac{MI} is not affected by ``one-sided'' phase errors. 
Obviously, this is also true for fully parallel 
(both the transmitter and the receiver employ one \ac{RF} chain per antenna)
\ac{MIMO} channel sounders. 

We would like to add a word of caution. In practice, the \ac{MI} of measured \ac{MIMO}
channels is often not evaluated by directly inserting the measured channel realizations
into the \ac{MI} formula \eqref{eq:telatar}. Rather, the measurement results are used to extract
parameters of a statistical \ac{MIMO} channel model, e.g., the
\ac{PAS} or the distribution of the \ac{DOA}. The resulting statistical description
of $\mathbf{H}$ is then used to evaluate the channel's \ac{MI}. Now, one-sided
switching does, in general, entail errors in the estimation of the parameters of a 
statistical \ac{MIMO} channel model so that the procedure described above will lead to, 
in general significant, errors in \ac{MI}. This brings out an interesting and
practically very relevant point. \emph{While one-sided switching does not entail errors
in \ac{MI} when the measurement results are used to directly evaluate the \ac{MI},
significant errors can be expected if one takes a detour via a specific statistical
\ac{MIMO} channel model.}

\subsubsection{Impact of frequency offset for separable timing matrix}
\label{sec:freq_offs}
First, we define the \emph{timing matrix}
$[\mathbf{T}]_{m_k, n_k} = t_k$ of size $\MR\times\MT$, 
which contains the \ac{SISO} snapshot measurement times in matrix form
arranged corresponding to the switching pattern.
We call $\mathbf{T}$ \emph{separable} if it can be written in the form 
$[\mathbf{T}]_{\vrow,\vcol} = [\mathbf{t}_\RX]_\vrow + [\mathbf{t}_\TX]_\vcol$,
where $\mathbf{t}_\TX$ and $\mathbf{t}_\RX$ 
are the transmit and receive timing vectors of size $\MT\times1$ and $\MR\times1$, respectively.
An example of a switching pattern 
and \ac{SISO} snapshot times
leading to a separable timing matrix is given 
by the \emph{regular sounding pattern} 
(where sounding pattern denotes the combination of 
a switching pattern and a set of \ac{SISO} snapshot times)
\begin{align}
\begin{split}
m_k &= (k-1) \mod \MR + 1 \\
n_k &= (k-1) \div \MR + 1 \\
t_k &= T_{\RX} ( m_k-1 ) + T_{\TX} ( n_k-1 )
\end{split}
\label{eq:reg_pat}
\end{align}
with $k = 1,2,\ldots,\MT\MR$, 
characterized by the timing parameters $T_{\TX}, T_{\RX}\in\numID{R}^{+}$. This
corresponds to starting with transmit antenna 1, 
switching through the receive antennas $1,2,\ldots,\MR$ sequentially,
then switching to transmit antenna 2, again switching through the receive 
antennas sequentially with the same \ac{SISO} snapshot
time distances between the receive antennas
as before, and so on.
Likewise, if we start with receive antenna 1,
switch through the transmit antennas 
sequentially and so on, a separable timing matrix will be obtained.

Now, for 
$[\mathbf{T}]_{\vrow,\vcol} = [\mathbf{t}_\RX]_\vrow + [\mathbf{t}_\TX]_\vcol$,
we have $\exp\cwa(\jmath\mathbf{M})
= \exp\cwa(\jmath\Delta\omega\mathbf{t}_\RX) \big( \exp\cwa(\jmath\Delta\omega\mathbf{t}_\TX) \big)^T$
and hence by \propref{prop:svd} in \secref{sec:rank1}, 
it follows that the eigenvalues of 
$\mathbf{H} \circ \exp\cwa\big(\jmath(\mathbf{M}+\mathbf{\Phi})\big)$ 
are equal to the eigenvalues of 
$\mathbf{H} \circ \exp\cwa(\jmath\mathbf{\Phi})$. This implies that
for $\mathbf{T}$ separable, the frequency offset has no effect on \ac{MI}. 
In practice, one typically has control over the sounding pattern. In the light of
what was said above, it is therefore sensible to choose the sounding pattern such that
$\mathbf{T}$ is separable. 
Of course, even if \ac{MI} is not affected, 
frequency offset can still cause significant errors in other parameters,
as discussed in the last paragraph of \secref{sec:one_sided}. 
We note, however, that in contrast to phase noise, frequency offset, 
due to its deterministic nature, can be estimated 
and mitigated with relative ease \cite{KiVa00}.
In the remainder of the paper, we shall, therefore, often neglect 
the frequency offset and consider $\mathbf{H} \circ \exp\cwa(\jmath\mathbf{\Phi})$
only.

\subsection{Numerical Results} 
\label{sec:num_res}
We shall next provide numerical results to quantify 
the impact of phase errors on \ac{MI}.
In particular, we will also quantify 
the impact of the scalar subchannels in the effective \ac{MIMO} channel
not being \ac{JG} distributed.

\subsubsection{Ergodic capacity increase due to phase errors}
\label{sec:Cidtpe}
\begin{figure}[htbp]
\centering
\psfrag{r(R) === 1, 0o RMS}{\scriptsize{$\rank(\mathbf{R})=1$, 0\degreel\ \acs{RMS}}}
\psfrag{r(R) === 1, 3.5o RMS}{\scriptsize{$\rank(\mathbf{R})=1$, 3.5\degreel\ \acs{RMS}}}
\psfrag{r(R) === 1, 7o RMS}{\scriptsize{$\rank(\mathbf{R})=1$, 7\degreel\ \acs{RMS}}}
\psfrag{r(R) === 4, 0o RMS}{\scriptsize{$\rank(\mathbf{R})=4$, 0\degreel\ \acs{RMS}}}
\psfrag{r(R) === 4, 3.5o RMS}{\scriptsize{$\rank(\mathbf{R})=4$, 3.5\degreel\ \acs{RMS}}}
\psfrag{r(R) === 4, 7o RMS}{\scriptsize{$\rank(\mathbf{R})=4$, 7\degreel\ \acs{RMS}}}
\psfrag{r(R) === 8, 0o RMS}{\scriptsize{$\rank(\mathbf{R})=8$, 0\degreel\ \acs{RMS}}}
\psfrag{r(R) === 8, 3.5o RMS}{\scriptsize{$\rank(\mathbf{R})=8$, 3.5\degreel\ \acs{RMS}}}
\psfrag{r(R) === 8, 7o RMS}{\scriptsize{$\rank(\mathbf{R})=8$, 7\degreel\ \acs{RMS}}}
\psfrag{SNR w [dB]}{\scriptsize{\acs{SNR} $\rho$ [dB]}}
\psfrag{Ergodic capacity C [bit/s/Hz]}{\scriptsize{Ergodic capacity $\widehat{C}$ [bit/s/Hz]}}
\icg{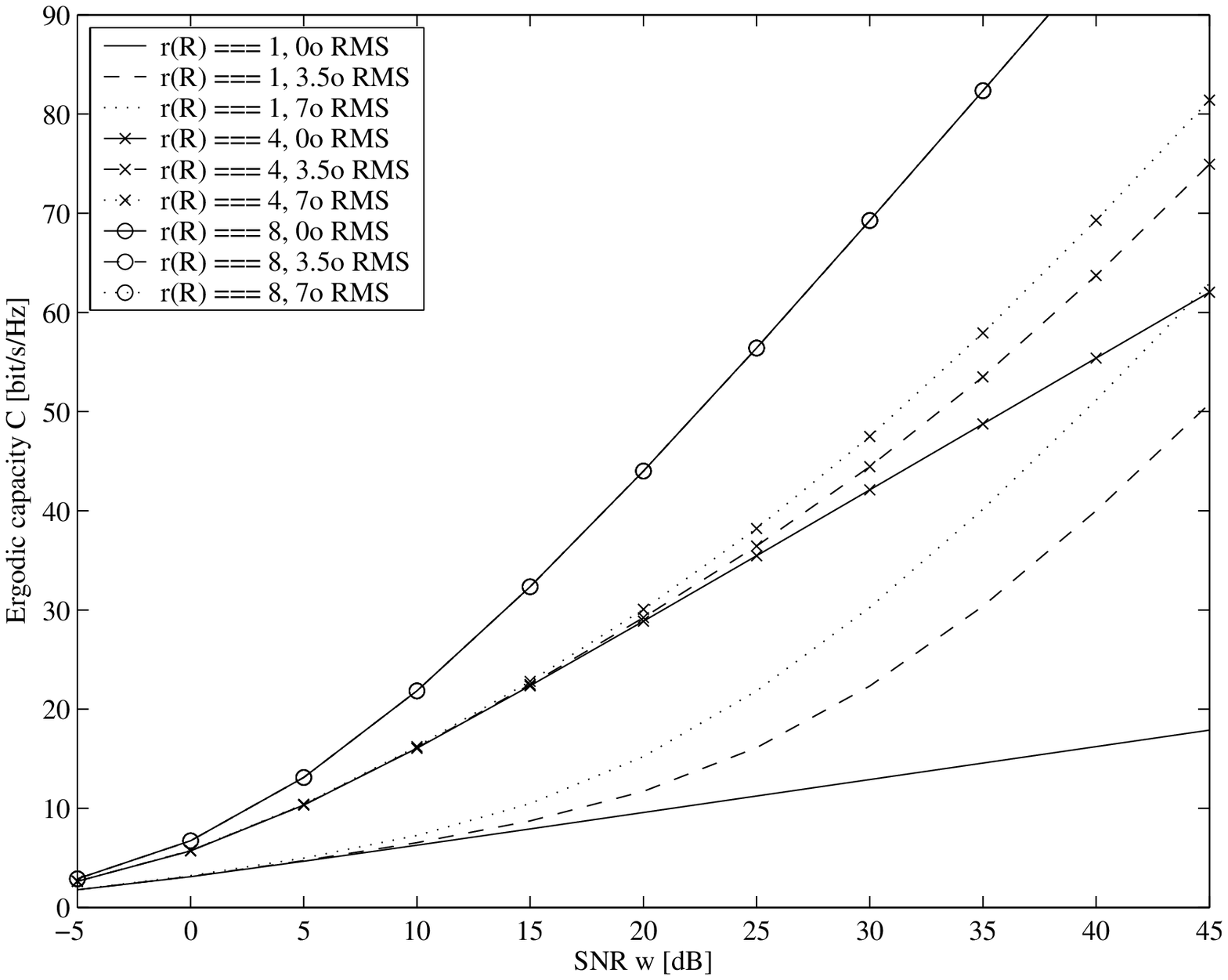}
\vspace{-2mm}
\caption{\label{fig:capPN}
Ergodic capacity of an $8\times8$ effective channel for
fully uncorrelated phase noise with varying \ac{RMS} values
and various $\rank(\mathbf{R})$ of the underlying physical channel.}
\end{figure}

We examine an $\MT = \MR = M = 8$ physical channel 
(with receive antenna correlation)
given by $\mathbf{H}=\mathbf{R}^{1/2}\mathbf{H}_{\WHITE}$, 
where $\mathbf{R}=\mathbf{R}^{1/2}\mathbf{R}^{1/2}$
denotes the receive correlation matrix and the entries of $\mathbf{H}_{\WHITE}$ 
are \ac{IID} $\cplxGauss(0,1)$.
We choose $\mathbf{R}$ such that 
$\lambda_i\mo(\mathbf{R}) = M/\rank(\mathbf{R})$ for $i=1,2,\ldots,\rank(\mathbf{R})$
and $\lambda_i\mo(\mathbf{R})=0$ otherwise. 
This normalization ensures that 
$\trace(\mathbf{R}) = M$, irrespectively of the rank of $\mathbf{R}$.
Employing Monte Carlo simulations, 
\figref{fig:capPN} shows the ergodic capacity\footnote{Throughout
the paper, we tacitly assume that the effective channel is ergodic.} 
$\widehat{C}=\expect\{\hat{I}\}$ of the effective channel $\widehat{\mathbf{H}}$
as a function of \ac{SNR} for varying $\rank(\mathbf{R})$, 
varying $\sigma^2_{\PP}$ with fully uncorrelated 
(i.e., worst-case) phase noise,
i.e., $\cov\{\mathbf{\Phi}\}=\sigma^2_{\PP}\,\mathbf{I}_{\MT\MR}$,
and no frequency offset. 
We have chosen 3.5\degreel\ and 7\degreel\ \ac{RMS} 
phase noise (corresponding to $\sigma^2_{\PP}\approx0.0037$ and 
$\sigma^2_{\PP}\approx0.0149$, respectively)
as typical and worst case values, respectively. 
These values were derived from our measurements on a commercially 
employed \ac{TDMS}-based \ac{MIMO} channel sounder (see \secref{sec:meas}).
In general, phase noise correlation properties quantified by $\cov\{\mathbf{\Phi}\}$
and phase noise variance $\sigma^2_{\PP}$ depend on
the \ac{LO} characteristics, \ac{SISO} snapshot times, 
and the number of transmit and receive antennas.
Fully uncorrelated phase noise, as assumed in this example, corresponds to 
relatively long \ac{SISO} snapshot time distances and represents the worst 
(though not necessarily untypical) conditions. 
We can see in \figref{fig:capPN}, that in none of the considered cases
phase noise results in a reduction of ergodic capacity. 
Moreover, we observe, in agreement with what was 
shown in \secref{sec:noeffect}, that in the case
of \ac{IID} physical channels (i.e., $\rank(\mathbf{R})=8$ 
in this example\footnote{Recall that 
   the nonzero eigenvalues of $\mathbf{R}$
   were chosen to be equal, which together with $\rank(\mathbf{R})=8$ implies that 
   the entries of $\mathbf{H}=\mathbf{R}^{1/2}\mathbf{H}_\WHITE$ are \ac{IID} 
   $\cplxGauss(0,1)$.}) 
phase noise has no impact at all on ergodic capacity. 
In the $\rank(\mathbf{R})=1$ case, at $\rho=35$\unit{dB} 
for the typical \ac{RMS} phase noise value of 3.5\degree, 
the error in ergodic capacity due to phase noise is about 100\%.
For the worst-case value of 7\degreel\ \ac{RMS} phase noise, the error, 
again at $\rho=35$\unit{dB}, is about 175\%. 
Furthermore, for $\rank(\mathbf{R})=1$ and \ac{RMS} phase noise of 3.5\degree, 
the multiplexing gain of the effective channel
(i.e., the ergodic capacity pre-log, obtained by determining the high-\ac{SNR}
slope of capacity as a function of \ac{SNR}) lies between 4 and 8, 
making the effective channel look like a \ac{MIMO} channel with rank between 4 and 8.
Furthermore, as predicted by theory, we can see that phase noise has little impact in
the low-\ac{SNR} regime. 
We conclude this simulation example by noting that the
impact of phase noise is most pronounced for low-rank physical channels at high \ac{SNR}. 
This is one of the motivating factors for 
the detailed analysis of the rank-1 physical channel in \secref{sec:rank1}.

\subsubsection{Impact of loss of joint Gaussianity}

As demonstrated in \secref{sec:lJG}, albeit the individual elements of
the effective \ac{MIMO} channel 
associated with a physical purely Rayleigh fading channel
will be circularly-symmetric complex Gaussian distributed, 
they will, in general, not be \ac{JG} distributed. 
Characterizing the consequences of this loss
of joint Gaussianity analytically seems very difficult
as known techniques (e.g., \cite{SFGK00j,ONBP03j,ALTV04c,ZhCL05j,TuLV05j,KaAl06j-Rayl})
for deriving analytic expressions for ergodic and outage
capacity of \ac{MIMO} channels (or bounds thereon) almost exclusively hinge on the assumption
of the elements in $\mathbf{H}$ being \ac{JG} distributed. 
There are two practically relevant points related to the loss of joint Gaussianity,
which will be brought out through numerical results in the following. 

\begin{figure}[htbp]
\centering
\psfrag{c == 0.95 phys. channel}{\scriptsize{$c=0.95$, phys.\ channel}}
\psfrag{c == 0.95 eff. channel, exact}{\scriptsize{$c=0.95$, eff.\ channel, exact}}
\psfrag{c == 0.95 eff. channel, linear}{\scriptsize{$c=0.95$, eff.\ channel, linear}}
\psfrag{c == 0.95 K, synthetic channel}{\scriptsize{$c=0.95\kappa$, synthetic channel}}
\psfrag{c == 0.7 phys. channel}{\scriptsize{$c=0.7$, phys.\ channel}}
\psfrag{c == 0.7 eff. channel, exact}{\scriptsize{$c=0.7$, eff.\ channel, exact}}
\psfrag{c == 0.7 eff. channel, linear}{\scriptsize{$c=0.7$, eff.\ channel, linear}}
\psfrag{c == 0.7 K, synthetic channel}{\scriptsize{$c=0.7\kappa$, synthetic channel}}
\psfrag{MI I [bit/s/Hz]}{\scriptsize{\acs{MI} [bit/s/Hz]}}
\psfrag{outage probability}{\scriptsize{outage probability}}
\icg{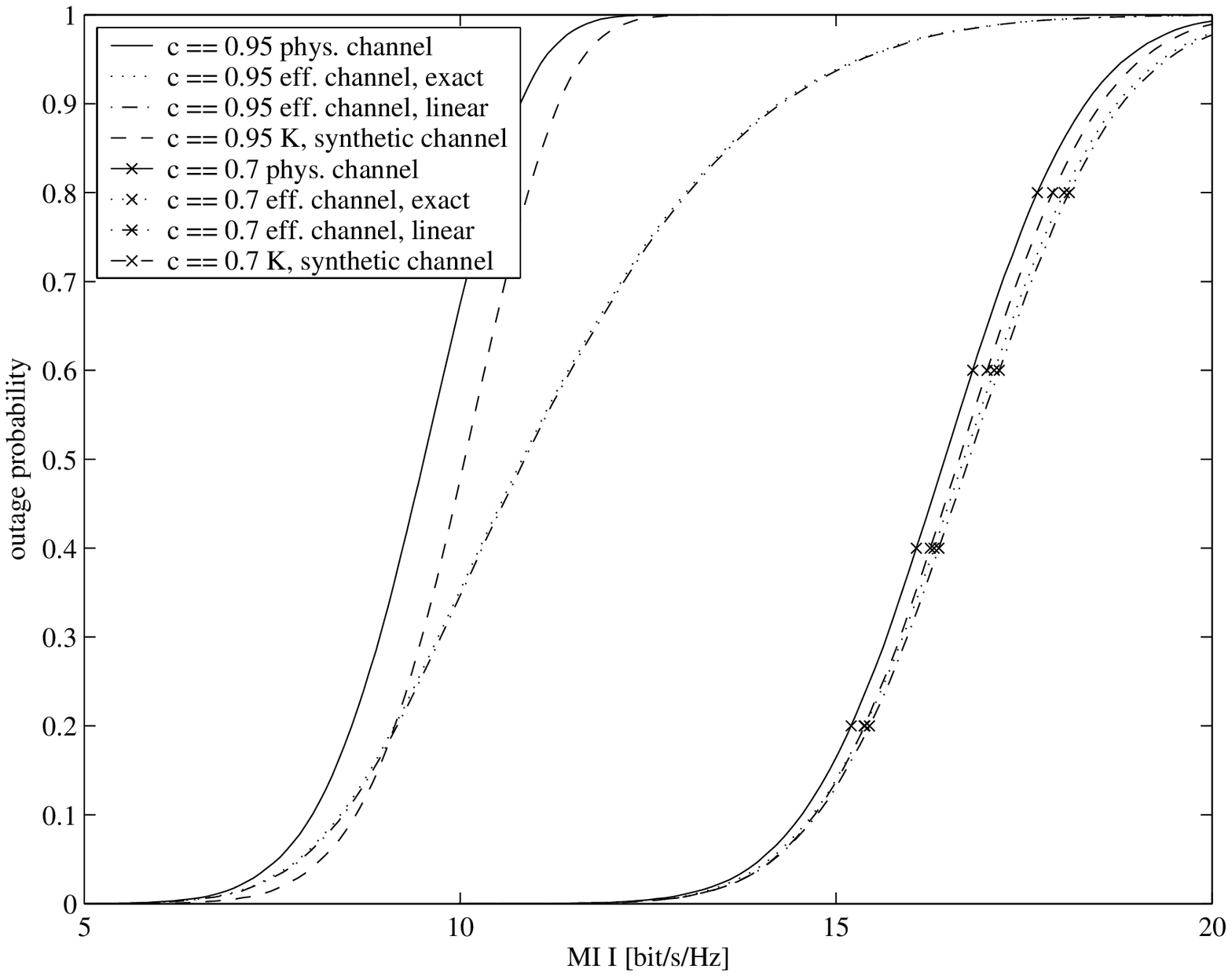}
\vspace{-2mm}
\caption{\label{fig:jga_left}
\acs{CDF} of \acl{MI} at \acs{SNR} $\rho=20$\unit{dB} showing the impact of 
the loss of joint Gaussianity for a $4\times4$ physical channel 
with $[\corrc(\mathbf{H})]_{\vrow,\vcol} = c$, $\forall\:\vrow\ne\vcol$, 
and fully uncorrelated 7\degreel\ \acs{RMS} phase noise.}
\end{figure}

We investigate a $4\times4$ purely Rayleigh fading (i.e., $\mathbf{H}_\FIX =\mathbf{0}$) 
physical channel with two different correlation levels according to 
$[\corrc\{\mathbf{H}\}]_{\vrow,\vcol}= 0.7$ and $0.95$, respectively, for $\vrow\ne\vcol$ 
in the presence of fully uncorrelated 7\degreel\ \ac{RMS} phase noise and no frequency offset. 
For $\rho=20$\unit{dB}, \figref{fig:jga_left} shows the \ac{CDF} of \ac{MI}
for the physical channel $\mathbf{H}$, the resulting effective channel $\widehat{\mathbf{H}}$, 
and a synthetic \ac{MIMO} channel obtained by assuming that 
the channel matrix is circularly-symmetric
complex Gaussian with
covariance matrix according to \eqref{eq:eqZ}, i.e., 
$[\cov\{\widehat{\mathbf{H}}\}]_{\vrow,\vcol} = 0.7\kappa$ and $0.95\kappa$, respectively, for $\vrow\ne\vcol$. 
In addition, the \ac{CDF} of \ac{MI} under the linear phase noise approximation \eqref{eq:linApprox} is shown.
Before proceeding, we note that the synthetic channel has the same first and second order statistics
as the effective channel and should hence exhibit (when compared to the effective channel)
the impact of the loss of joint Gaussianity due to phase errors.

We can now draw the following conclusions for this specific setup:
\begin{itemize}
\item 
The impact of phase noise is significant at the (high) correlation level of 0.95.
In addition, comparing the synthetic channel to the effective channel,
we can see that neglecting the loss of joint Gaussianity would result in 
a slight overestimation and a significant underestimation of outage capacity 
at low and high outage levels, respectively. 
For outage levels of practical interest, i.e., up to 30\% outage probability,
the error in outage capacity (due to the loss of joint Gaussianity) is not more than 8\%.
\item 
The impact of phase noise is negligible at the correlation level of 0.7. The synthetic channel
exhibits a behavior that is very close to that of the effective channel. Interestingly, a 
correlation level of 0.7-0.75 is often quoted in the literature \cite{Jake91j,Lee73j} as
a threshold above which ``the channel starts behaving as highly correlated''. 
\item 
Even for the worst case \ac{RMS} phase noise value of 7\degree,
the linear phase noise approximation \eqref{eq:linApprox} yields 
very accurate results. 
This approximation will be used extensively in \secref{sec:rank1}.
\end{itemize}

With regards to Ricean fading physical channels, we content ourselves with noting that
for high Ricean K-factors, the situation approaches that 
for deterministic physical channels (studied in \secsref{sec:sens_ana} and \ref{sec:rank1}),
whereas for low Ricean K-factors, the behavior will be close to that of 
the purely Rayleigh fading case (treated in this section and in \secref{sec:rank1}).

\section{Sensitivity Analysis}
\label{sec:sens_ana}

Throughout this section, we assume a separable timing matrix
in which case frequency offset has no effect on \ac{MI} and can therefore be
neglected. For the sake of simplicity of exposition, we shall furthermore
assume $\MR\le\MT$. 
The purpose of this section is twofold. First, we introduce a tool for evaluating
the sensitivity of the \ac{MI} of a fixed physical channel to phase noise.
This will be accomplished by computing the first two terms in the Taylor series
expansion of $\hat{I}(\mathbf{\Phi})$ around the phase-noise-free case 
$\mathbf{\Phi}=\mathbf{0}$. Based on this framework, we will then be able to
provide analytic expressions for approximations of the first and second moment of the \ac{MI}
of the effective channel for arbitrary phase noise covariance matrix.

We shall be concerned with computing 
the second-order Taylor series expansion of 
\begin{align*}
\hat{I}(\mathbf{\Phi}) = \log\,\det\left( \mathbf{I}_{\MR} + \frac{\rho}{\MT}
\big( \mathbf{H} \circ \exp\cwa(\jmath\mathbf{\Phi}) \big)
\big( \mathbf{H} \circ \exp\cwa(\jmath\mathbf{\Phi}) \big)^H \right)
\end{align*}
around $\mathbf{\Phi}=\mathbf{0}$ given by
\begin{align}
  \tilde{I}(\mathbf{\Phi}) 
= 
  \hat{I}(\mathbf{0}) 
+ \Jacobian{\hat{I}}{\mathbf{0}} \, \vect(\mathbf{\Phi}) 
+ \frac{1}{2} \, \transp[big]{\vect}{(\mathbf{\Phi})} \, 
                 \Hessian{\hat{I}}{\mathbf{0}} \, 
                 \vect(\mathbf{\Phi})
\label{eq:Taylor}
\end{align}
where $\Jacobian{\hat{I}}{\mathbf{0}}$ denotes the $1\times\MT\MR$ Jacobian matrix
(vector, in this case)
and $\Hessian{\hat{I}}{\mathbf{0}}$ is the $\MT\MR\times\MT\MR$ Hessian matrix of
$\hat{I}=\hat{I}(\mathbf{\Phi})$ at $\mathbf{\Phi}=\mathbf{0}$. 
Clearly, we have $\hat{I}(\mathbf{0}) = I$. We shall see below that 
the second-order Taylor series 
expansion of \ac{MI} is, in general, 
accurate for full-rank physical channels, but tends to yield loose
approximations for rank-deficient physical channels. This problem can be
mitigated by either using more terms in the Taylor series expansion of 
\ac{MI} or by performing the Taylor series expansion on the channel's eigenvalues
rather than on the channel's \ac{MI} itself. 
Both approaches are, in general, cumbersome.
The latter approach will not be
detailed here, for brevity of exposition, but will be outlined briefly in 
\secref{sec:aesoTse}. Even though the second-order Taylor series
expansion of \ac{MI} does not yield accurate approximations in the case
of rank-deficient physical channels, explicit expressions for the Jacobian matrix $\Jacobian{\hat{I}}{\mathbf{0}}$
and the Hessian matrix $\Hessian{\hat{I}}{\mathbf{0}}$
can be used to test whether $\hat{I}(\mathbf{\Phi})$ has an extremum at $\mathbf\Phi={\bf 0}$.

\subsection{Sensitivity Analysis}

Even though the computation of the Jacobian matrix
$\Jacobian{\hat{I}}{\mathbf{0}}$ and the Hessian matrix
$\Hessian{\hat{I}}{\mathbf{0}}$ does not pose any major technical 
difficulties, it still requires the application of tools
that are not completely standard, namely matrix differential calculus
\cite{MaNe88b} and matrix-variate Wirtinger a.k.a.\ $\numID{CR}$ calculus
as described in \cite{Kreu05}.
We shall therefore present the corresponding derivations in some detail.

\begin{prop}
The $1\times\MT\MR$ Jacobian matrix (vector) $\Jacobian{\hat{I}}{\mathbf{0}}$ in
\eqref{eq:Taylor} is given by
\begin{align}
  \Jacobian{\hat{I}}{\mathbf{0}}
&=
  2\, \log(e) \, \imag\mo\bigg( \transp[Big]{\vect}{\big( (\mathbf{Y}^{-1} \mathbf{H}) 
  \circ \mathbf{H}^* \big)} \, \bigg) 
\label{eq:diff1}
\end{align}
where
\begin{align*}
  \mathbf{Y}
&= 
  \frac{\MT}{\rho} \mathbf{I}_{\MR} + \mathbf{H} \mathbf{H}^H.
\end{align*}

The $\MT\MR\times\MT\MR$ Hessian matrix $\Hessian{\hat{I}}{\mathbf{0}}$ in
\eqref{eq:Taylor} is given by
\begin{align}
\begin{split}
  \Hessian{\hat{I}}{\mathbf{0}} 
= 
  2\,\log(e) \, \real \bigg( 
& 
  \divec( \mathbf{H} ) 
  \mathbf{K}_{(\MT,\MR)} 
  \big( (\mathbf{H}^H \mathbf{Y}^{-1})^T \kron \mathbf{H}^H \mathbf{Y}^{-1} \big)
  \divec( \mathbf{H} )
\\
& 
+ \divec( \mathbf{H} )
  \big( (\mathbf{I}_{\MT} - \mathbf{H}^H \mathbf{Y}^{-1} \mathbf{H}) \kron (\mathbf{Y}^{-1})^T \big)
   \divec( \mathbf{H}^* )
\\
& 
- \divec\big((\mathbf{Y}^{-1} \mathbf{H}) \circ \mathbf{H}^* \big)
\bigg).
\end{split}
\label{eq:diff2MI}
\end{align}

\begin{IEEEproof}
We start by defining 
\begin{align*}
\widehat{\mathbf{Y}}
= \frac{\MT}{\rho}\mathbf{I}_{\MR} + 
  \underbrace{ \big( \mathbf{H} \circ \exp\cwa(\jmath\mathbf{\Phi}) \big) 
             }_{\widehat{\mathbf{H}}}
  \underbrace{ \big( \mathbf{H} \circ \exp\cwa(\jmath\mathbf{\Phi}) \big)^H 
             }_{\widehat{\mathbf{H}}^H}
\end{align*}
so that $\mathbf{Y} = \widehat{\mathbf{Y}} |_{\mathbf{\Phi} = \mathbf{0}}$
and hence $\hat{I} = \log\,\det\big( (\rho/\MT) \widehat{\mathbf{Y}} \big)$.
The strategy used in the proof of both parts of the statement is to
compute $\ud \hat{I}$ and $\ud^2 \hat{I}$ and to bring the resulting expressions into
the form
\begin{align}
  \ud \hat{I}    
&=  
  \mathbf{A} \, \vect(\ud\mathbf{\Phi}) 
\nonumber
\\
  \ud^2 \hat{I}  
&=  
  \transp[big]{\vect}{(\ud\mathbf{\Phi})} \, \mathbf{B} \, \vect(\ud\mathbf{\Phi})
\label{eq:sec_ident}
\end{align}
which will then allow us to apply the first \cite[Ch.\rsp5, Th.\rsq6]{MaNe88b} 
and the second \cite[Ch.\rsp6, Th.\rsq6]{MaNe88b} identification theorem
for a real-valued function of real-valued parameters 
to conclude that 
\begin{align*}
\Jacobian{\hat{I}}{\mathbf{0}} = \mathbf{A}
\qquad\text{and}\qquad
\Hessian{\hat{I}}{\mathbf{0}} = \frac{1}{2}(\mathbf{B}+\mathbf{B}^T).
\end{align*} 

\emph{Computing the Jacobian matrix:}
Using the basic rules of differentiation
together with the relation
$\ud\,\ln\,\det(\mathbf{A}) = \trace(\mathbf{A}^{-1} \, \ud\mathbf{A})$ 
\cite[Sec.\rsp8.3, Eq.\rsp(11)]{MaNe88b}, we obtain
\begin{align*}
  \ud \, \log\,\det \big( (\rho/\MT) \widehat{\mathbf{Y}} \big)
&=
  \log(e) \,
  \trace \left( \widehat{\mathbf{Y}}^{-1} \, \ud \widehat{\mathbf{Y}} \right).
\end{align*}
Applying the product rule
$\ud(\mathbf{AB})=(\ud\mathbf{A})\mathbf{B}+\mathbf{A}\ud\mathbf{B}$ 
\cite[Sec.\rsp8.2, Eq.\rsp(15)]{MaNe88b} and 
$\ud(\mathbf{A}^T)=(\ud\mathbf{A})^T$ \cite[Sec.\rsp8.2, Eq.\rsp(18)]{MaNe88b},
we get, using Wirtinger calculus, 
\begin{align*}
  \ud \widehat{\mathbf{Y}} 
&=
  \ud (\widehat{\mathbf{H}} \, \widehat{\mathbf{H}}^H)
 =
  (\ud \widehat{\mathbf{H}}) \, \widehat{\mathbf{H}}^H
+ \widehat{\mathbf{H}} \, \big( \ud (\widehat{\mathbf{H}}^*) \big)^T.
\end{align*}
With $\ud(\mathbf{A}\circ\mathbf{B})=(\ud\mathbf{A})\circ\mathbf{B}+\mathbf{A}\circ(\ud\mathbf{B})$
\cite[Sec.\rsp8.2, Eq.\rsp(17)]{MaNe88b},
we have
$\ud \widehat{\mathbf{H}} = \ud(\mathbf{H}\circ\mathbf{\Theta})
= \mathbf{H}\circ\ud\mathbf{\Theta}$
and $\ud(\widehat{\mathbf{H}}^*) = \mathbf{H}^*\circ\ud(\mathbf{\Theta}^*)$.
Noting that $[\ud\big( f\cwa(\mathbf{X}) \big)]_{\vrow,\vcol}$
$=\ud\big( [f\cwa(\mathbf{X})]_{\vrow,\vcol} \big)$
$=\ud\big( f([\mathbf{X}]_{\vrow,\vcol}) \big)$, 
we obtain
\begin{align}
  \ud \, \log\,\det \big( (\rho/\MT) \widehat{\mathbf{Y}} \big)
&=
  \log(e) \,
  \trace\left( \widehat{\mathbf{Y}}^{-1} \, 
               \Big(
                     \frac{1}{\jmath} \widehat{\mathbf{H}} \, 
                     \big( \widehat{\mathbf{H}} \circ \ud \mathbf{\Phi} \big)^H 
                   - \frac{1}{\jmath} \big( \widehat{\mathbf{H}} \circ \ud \mathbf{\Phi} \big) \, 
                     \widehat{\mathbf{H}}^H 
               \Big) \right)
\nonumber
\\
&=
  \log(e) \,
  \trace\left( \widehat{\mathbf{Y}}^{-1} \, 
               \bigg(
                     \frac{1}{\jmath} \widehat{\mathbf{H}} \, 
                     \big( \widehat{\mathbf{H}} \circ \ud \mathbf{\Phi} \big)^H 
                   - \frac{1}{\jmath} \Big( \widehat{\mathbf{H}}^* \,
                     \big( \widehat{\mathbf{H}} \circ \ud \mathbf{\Phi} \big)^H \Big)^*
               \bigg) \right)
\nonumber
\\
&=
  2\,\log(e) \, \imag\,\left(
  \trace\left( \widehat{\mathbf{Y}}^{-1} \, 
               \widehat{\mathbf{H}} \, 
               (\widehat{\mathbf{H}} \circ \ud \mathbf{\Phi})^H \right)\right)
\nonumber
\\
&=
  2\,\log(e) \, \imag\,\left(
  \trace\left( (\widehat{\mathbf{Y}}^{-1} \, \widehat{\mathbf{H}})^T \, 
               (\widehat{\mathbf{H}}^* \circ \ud \mathbf{\Phi}) \right)\right).
\label{eq:Jlast}
\end{align}
It remains to turn \eqref{eq:Jlast} into the form 
$\ud \hat{I} = \mathbf{A} \, \vect(\ud\mathbf{\Phi})$. This can be done by
first showing that
\begin{align}
  \trace\big( \mathbf{A} ( \mathbf{B} \circ \mathbf{C} ) \big) 
&= 
  \transp[big]{\vect}{( \mathbf{A}^T \circ \mathbf{B} )} \, \vect( \mathbf{C} ) 
\label{eq:trace_special}
\end{align}
and then applying \eqref{eq:trace_special} to \eqref{eq:Jlast}.
In order to prove \eqref{eq:trace_special}, we start by noting that
with $\trace(\mathbf{A}^T\mathbf{B})$
$ = \vect(\mathbf{A})^T \vect(\mathbf{B})$ 
\cite[Sec.\rsp2.4, Eq.\rsp(4)]{MaNe88b}, we have
\begin{align*}
  \trace\big( \mathbf{A} ( \mathbf{B} \circ \mathbf{C} ) \big) 
&=
  \transp[big]{\vect}{(\mathbf{A}^T)} \, \vect( \mathbf{B} \circ \mathbf{C} ) 
\end{align*}
which upon application of 
\begin{align}
  \vect(\mathbf{A}\circ\mathbf{B}) 
&= 
  \divec(\mathbf{A}) \, \vect(\mathbf{B}) 
\label{eq:vect_prod_Had} 
\end{align}
yields the desired result.
Finally, applying \eqref{eq:trace_special} to \eqref{eq:Jlast}, we obtain
\begin{align*}
  \ud \, \log\,\det \big( (\rho/\MT) \widehat{\mathbf{Y}} \big)
&=
2\, \log(e) \, \imag\mo\bigg( \transp[Big]{\vect}{\big( (\mathbf{Y}^{-1} \mathbf{H}) 
  \circ \mathbf{H}^* \big)} \, \bigg) \, \vect( \ud \mathbf{\Phi} )
\end{align*}
which proves \eqref{eq:diff1}.

\emph{Computing the Hessian matrix:}
We start by noting that 
\begin{align}
  \ud^2 \, \log\,\det\big( (\rho/\MT) \widehat{\mathbf{Y}} \big)
&= 
  \log(e) \; \ud \, \trace( \widehat{\mathbf{Y}}^{-1} \, \ud \widehat{\mathbf{Y}} )
= 
  \log(e) \; \trace\big( \widehat{\mathbf{Y}}^{-1} \; (\ud^2 \widehat{\mathbf{Y}} )
- ( \widehat{\mathbf{Y}}^{-1} \, \ud \widehat{\mathbf{Y}})^2 \big)
\label{eq:twotrace} 
\end{align}
where we used $\ud\,\trace(\mathbf{A}) = \trace(\ud\mathbf{A})$ 
\cite[Sec.\rsp8.2, Eq.\rsp(20)]{MaNe88b} 
and $\ud(\mathbf{A}^{-1})=-\mathbf{A}^{-1}(\ud\mathbf{A})\mathbf{A}^{-1}$ 
\cite[Sec.\rsp8.4, Eq.\rsp(1)]{MaNe88b} along with the product rule.
In order to keep the following exposition simple, we set
$\ud\widehat{\mathbf{H}} = \jmath \widehat{\mathbf{H}} \circ \ud \mathbf{\Phi} = \jmath\dot{\mathbf{H}}$
so that 
$\ud \widehat{\mathbf{Y}} = \ud ( \widehat{\mathbf{H}} \widehat{\mathbf{H}}^H )
= \jmath\dot{\mathbf{H}} \, \widehat{\mathbf{H}}^H -\jmath \widehat{\mathbf{H}} \, \dot{\mathbf{H}}^H
$.
Expanding the second term on the \ac{RHS} of \eqref{eq:twotrace} 
through similar manipulations as in the derivation of the Jacobian matrix, 
and using $(\widehat{\mathbf{Y}}^{-1})^T = (\widehat{\mathbf{Y}}^{-1})^*$, we get
\begin{align}
-\trace\big( & (\widehat{\mathbf{Y}}^{-1} \ud \widehat{\mathbf{Y}} )^2 \big)
\nonumber 
\\
&= 
\trace\big( \widehat{\mathbf{Y}}^{-1} (\dot{\mathbf{H}} \, \widehat{\mathbf{H}}^H 
                                           - \widehat{\mathbf{H}} \, \dot{\mathbf{H}}^H) 
        \widehat{\mathbf{Y}}^{-1} (\dot{\mathbf{H}} \, \widehat{\mathbf{H}}^H 
                                           - \widehat{\mathbf{H}} \, \dot{\mathbf{H}}^H) \big)
\nonumber
\\
&= 
 \trace( \widehat{\mathbf{Y}}^{-1} \dot{\mathbf{H}} \, \widehat{\mathbf{H}}^H 
             \widehat{\mathbf{Y}}^{-1} \dot{\mathbf{H}} \, \widehat{\mathbf{H}}^H )
-\trace( \widehat{\mathbf{Y}}^{-1} \dot{\mathbf{H}} \, \widehat{\mathbf{H}}^H 
             \widehat{\mathbf{Y}}^{-1} \widehat{\mathbf{H}} \, \dot{\mathbf{H}}^H )
\nonumber
\\
&+
 \trace\big( \dot{\mathbf{H}}^* \, \widehat{\mathbf{H}}^T \, (\widehat{\mathbf{Y}}^{-1})^* 
             \dot{\mathbf{H}}^* \, \widehat{\mathbf{H}}^T \, (\widehat{\mathbf{Y}}^{-1})^* \big)
-\trace\big( \widehat{\mathbf{H}}^* \, \dot{\mathbf{H}}^T \, (\widehat{\mathbf{Y}}^{-1})^* 
             \dot{\mathbf{H}}^* \, \widehat{\mathbf{H}}^T \, (\widehat{\mathbf{Y}}^{-1})^* \big)
\nonumber
\\
&= 
 \trace( \widehat{\mathbf{Y}}^{-1} \dot{\mathbf{H}} \, \widehat{\mathbf{H}}^H 
             \widehat{\mathbf{Y}}^{-1} \dot{\mathbf{H}} \, \widehat{\mathbf{H}}^H )
-\trace( \widehat{\mathbf{Y}}^{-1} \dot{\mathbf{H}} \, \widehat{\mathbf{H}}^H 
             \widehat{\mathbf{Y}}^{-1} \widehat{\mathbf{H}} \, \dot{\mathbf{H}}^H )
\nonumber
\\
&+
 \trace\big( (\widehat{\mathbf{Y}}^{-1})^* \dot{\mathbf{H}}^* \, \widehat{\mathbf{H}}^T \, 
             (\widehat{\mathbf{Y}}^{-1})^* \dot{\mathbf{H}}^* \, \widehat{\mathbf{H}}^T \big)
-\trace\big( (\widehat{\mathbf{Y}}^{-1})^* \dot{\mathbf{H}}^* \, \widehat{\mathbf{H}}^T \, 
             (\widehat{\mathbf{Y}}^{-1})^* \widehat{\mathbf{H}}^* \, \dot{\mathbf{H}}^T \big)
\nonumber
\\
&=
 2 \, \real \, \left(\trace(
                 \widehat{\mathbf{Y}}^{-1} \, \dot{\mathbf{H}} \widehat{\mathbf{H}}^H \,
                 \widehat{\mathbf{Y}}^{-1} \, \dot{\mathbf{H}} \widehat{\mathbf{H}}^H ) \right)
-2 \, \real \, \left( \trace(
                 \widehat{\mathbf{Y}}^{-1} \, \dot{\mathbf{H}} \widehat{\mathbf{H}}^H \, 
                 \widehat{\mathbf{Y}}^{-1} \, \widehat{\mathbf{H}} \dot{\mathbf{H}}^H ) \right)
\nonumber
\\
\begin{split}
&=
  2 \, \real \, \left( \trace(
  \widehat{\mathbf{H}}^H
  \widehat{\mathbf{Y}}^{-1} \dot{\mathbf{H}} \widehat{\mathbf{H}}^H \,
  \widehat{\mathbf{Y}}^{-1} \dot{\mathbf{H}} ) \right)
- 2 \, \real \, \left( \trace\big(
  (\widehat{\mathbf{Y}}^{-1})^T \dot{\mathbf{H}}^* \widehat{\mathbf{H}}^T \, 
  (\widehat{\mathbf{Y}}^{-1})^* \widehat{\mathbf{H}}^* \dot{\mathbf{H}}^T \big) \right).
\end{split}
\label{eq:sec_term}
\end{align}
Next, applying \cite[Ch.\rsp2, Th.\rsp3]{MaNe88b}
\begin{align}
\begin{split}
  \trace(\mathbf{ABCD}) 
&= 
  \transp[big]{\vect}{(\mathbf{D}^T)} \, (\mathbf{C}^T \kron \mathbf{A}) \, \vect(\mathbf{B})
 = 
  \transp[big]{\vect}{(\mathbf{D})} \, (\mathbf{A} \kron \mathbf{C}^T) \, \vect(\mathbf{B}^T)  
\end{split}
\label{eq:trABCD}
\end{align}
with 
$\mathbf{A} = \widehat{\mathbf{H}}^H \widehat{\mathbf{Y}}^{-1}$,
$\mathbf{B} = \dot{\mathbf{H}}$,
$\mathbf{C} = \widehat{\mathbf{H}}^H \, \widehat{\mathbf{Y}}^{-1}$, and 
$\mathbf{D} = \dot{\mathbf{H}}$ 
to the first term on the \ac{RHS} of \eqref{eq:sec_term}
and with 
$\mathbf{A} = (\widehat{\mathbf{Y}}^{-1})^T$,
$\mathbf{B} = \dot{\mathbf{H}}^*$,
$\mathbf{C} = \widehat{\mathbf{H}}^T \, (\widehat{\mathbf{Y}}^{-1})^* \widehat{\mathbf{H}}^*$, and 
$\mathbf{D} = \dot{\mathbf{H}}^T$ 
to the second term on the \ac{RHS} of \eqref{eq:sec_term},
we obtain
\begin{align}
  -\trace\big( (\widehat{\mathbf{Y}}^{-1} \ud \widehat{\mathbf{Y}} )^2 \big) \nonumber
&= 
  2 \, \real\Big(
  \transp[big]{\vect}{(\dot{\mathbf{H}}^T)} \, 
  \widehat{\mathbf{R}}_1 \, \vect( \dot{\mathbf{H}} ) \Big)
  -2 \, \real\Big(
  \transp[big]{\vect}{(\dot{\mathbf{H}}^T)} \, \widehat{\mathbf{R}}_2 \, \vect( \dot{\mathbf{H}}^H ) \Big)
\\ 
\begin{split}
&\stackrel{\text{(a)}}{=} 
  2\,\transp[Big]{\vect}{\big((\ud \mathbf{\Phi})^T\big)} 
  \real\big( \divec(\widehat{\mathbf{H}}^T) 
  \, \widehat{\mathbf{R}}_1 \, 
  \divec( \widehat{\mathbf{H}} ) \big) 
  \, \vect(\ud\mathbf{\Phi})
\\
& 
  -2\,\transp[Big]{\vect}{((\ud \mathbf{\Phi})^T)} \real\big( \divec(\widehat{\mathbf{H}}^T) 
  \, \widehat{\mathbf{R}}_2 \, 
  \divec( \widehat{\mathbf{H}}^H ) \big) 
  \, \vect\big( (\ud \mathbf{\Phi})^T \big)
\end{split}
\label{eq:PhiIsT}
\end{align}
where we set 
$\widehat{\mathbf{R}}_1 
= (\widehat{\mathbf{H}}^H \, \widehat{\mathbf{Y}}^{-1})^T \kron \widehat{\mathbf{H}}^H \, \widehat{\mathbf{Y}}^{-1}
= (\widehat{\mathbf{Y}}^{-1} \kron \widehat{\mathbf{H}}^*)^T (\widehat{\mathbf{H}}^* \kron \widehat{\mathbf{Y}}^{-1})
$
and
$\widehat{\mathbf{R}}_2 = (\widehat{\mathbf{Y}}^{-1})^T \kron \widehat{\mathbf{H}}^H \, \widehat{\mathbf{Y}}^{-1} \, \widehat{\mathbf{H}}$,
and we used 
$\dot{\mathbf{H}} = \widehat{\mathbf{H}} \circ \ud \mathbf{\Phi}$
and \eqref{eq:vect_prod_Had} in (a).
Next, we need to rewrite \eqref{eq:PhiIsT} in terms of 
$\transp[big]{\vect}{(\ud\mathbf{\Phi})}$ 
and $\vect(\ud\mathbf{\Phi})$ only, 
which requires getting rid of the terms 
$(\ud\mathbf{\Phi})^T$ inside the $\transp[big]{\vect}{(\cdot)}$.
Upon applying \eqref{eq:comm_mat} in \eqref{eq:PhiIsT}, we obtain
\begin{align*}
\begin{split}
  -\trace\big( (\widehat{\mathbf{Y}}^{-1} \ud \widehat{\mathbf{Y}} )^2 \big)
&= 
  2\,\transp[big]{\vect}{(\ud\mathbf{\Phi})} \mathbf{K}_{(\MR,\MT)}^T 
  \real\big( \divec(\widehat{\mathbf{H}}^T) 
  \, \widehat{\mathbf{R}}_1 \, 
  \divec( \widehat{\mathbf{H}} ) \big) 
  \, \vect(\ud\mathbf{\Phi})
\\
& 
  -2\,\transp[big]{\vect}{(\ud\mathbf{\Phi})} \mathbf{K}_{(\MR,\MT)}^T 
  \real\big( \divec(\widehat{\mathbf{H}}^T) 
  \, \widehat{\mathbf{R}}_2 \, 
  \divec( \widehat{\mathbf{H}}^H ) \big) \mathbf{K}_{(\MR,\MT)} \vect(\ud\mathbf{\Phi})
\end{split}
\end{align*}
which using 
$\mathbf{K}_{(\vrow,\vcol)}^T = \mathbf{K}_{(\vcol,\vrow)}$ 
\cite[Sec.\rsp3.7, Eq.\rsp(2)]{MaNe88b}
and 
\begin{align}
  \transp[big]{\vect}{(\mathbf{A})} \, \mathbf{K}_{(\vcol,\vrow)} \, \divec(\mathbf{B})
&= 
  \transp[big]{\vect}{(\mathbf{A})} \, \divec(\mathbf{B}^T) \, \mathbf{K}_{(\vcol,\vrow)}
\label{comm_mat_shift}
\end{align}
results in 
\begin{align}
&
  -\trace\big( (\widehat{\mathbf{Y}}^{-1} \ud \widehat{\mathbf{Y}} )^2 \big) 
\nonumber
\\
\begin{split}
&\qquad= 
  2\,\transp[big]{\vect}{(\ud \mathbf{\Phi})} \real\big( \divec(\widehat{\mathbf{H}}) 
  \mathbf{K}_{(\MT,\MR)} 
  \, \widehat{\mathbf{R}}_1 \,
  \divec( \widehat{\mathbf{H}} ) \big) \, \vect(\ud\mathbf{\Phi})
\\
&\qquad 
  -2\,\transp[big]{\vect}{(\ud \mathbf{\Phi})} \real\big( \divec(\widehat{\mathbf{H}}) 
  \mathbf{K}_{(\MT,\MR)} 
  \, \widehat{\mathbf{R}}_2 \,
  \mathbf{K}_{(\MR,\MT)} 
  \divec( \widehat{\mathbf{H}}^* ) \big) \, \vect( \ud \mathbf{\Phi} ).
\end{split}
\label{eq:secndT}
\end{align}
The validity of \eqref{comm_mat_shift} can easily be seen by 
noting that 
$\divec(\mathbf{B}) \, \mathbf{K}_{(\vrow,\vcol)} \, \vect(\mathbf{A})=$\linebreak 
$\divec(\mathbf{B}) \, \vect(\mathbf{A}^T)
=\vect( \mathbf{B} \circ \mathbf{A}^T )
=\vect\big( (\mathbf{B}^T \circ \mathbf{A})^T \big)
=  \mathbf{K}_{(\vrow,\vcol)} \, \vect( \mathbf{B}^T \circ \mathbf{A} )=$\linebreak
$\mathbf{K}_{(\vrow,\vcol)} \, \divec(\mathbf{B}^T) \, \vect(\mathbf{A}).
$

Finally, employing \cite[Sec.\rsp3.7, Eq.\rsp(5)]{MaNe88b}
\begin{align}
  \mathbf{K}_{(p,m)} (\mathbf{A}\kron\mathbf{B}) \mathbf{K}_{(n,q)}
&=
  \mathbf{B}\kron\mathbf{A}
\label{eq:reverseK}
\end{align}
to the $m\times n$ matrix $\mathbf{A}$ and the $p\times q$ matrix $\mathbf{B}$, 
we can simplify the second term on the \ac{RHS} of \eqref{eq:secndT} to obtain
\begin{align}
  -\trace\big( (\widehat{\mathbf{Y}}^{-1} \ud \widehat{\mathbf{Y}} )^2 \big) 
\qquad\qquad& \nonumber\\
\begin{split}
= 
  2\,\transp[big]{\vect}{(\ud \mathbf{\Phi})} \real\Big( &
  \divec(\widehat{\mathbf{H}}) 
  \mathbf{K}_{(\MT,\MR)} \, \big(
  (\widehat{\mathbf{H}}^H \, \widehat{\mathbf{Y}}^{-1})^T 
  \kron (\widehat{\mathbf{H}}^H \, \widehat{\mathbf{Y}}^{-1}) \big) \,
  \divec( \widehat{\mathbf{H}} ) 
\\
- &
  \divec(\widehat{\mathbf{H}}) 
  \big( (\widehat{\mathbf{H}}^H \, \widehat{\mathbf{Y}}^{-1} \, \widehat{\mathbf{H}})
  \kron (\widehat{\mathbf{Y}}^{-1})^T \big)
  \divec( \widehat{\mathbf{H}}^* ) \Big) \vect( \ud \mathbf{\Phi} ).
\end{split}
\label{eq:finalSecT}
\end{align}

It remains to turn the first term on the \ac{RHS} of \eqref{eq:twotrace} 
into the form of the \ac{RHS} of \eqref{eq:sec_ident}.
To this end, we start by noting that, using Wirtinger calculus,
\begin{align}
  \ud^2 \widehat{\mathbf{Y}} 
=
  \ud \, \ud ( \widehat{\mathbf{H}} \widehat{\mathbf{H}}^H )
=
  \ud \big( \jmath\dot{\mathbf{H}} \, \widehat{\mathbf{H}}^H 
           -\jmath \widehat{\mathbf{H}} \, \dot{\mathbf{H}}^H \big)
= 
  - \ddot{\mathbf{H}} \, \widehat{\mathbf{H}}^H 
  + \dot{\mathbf{H}} \, \dot{\mathbf{H}}^H
  + \dot{\mathbf{H}} \, \dot{\mathbf{H}}^H
  - \widehat{\mathbf{H}} \, \ddot{\mathbf{H}}^H
\label{eq:sdH}
\end{align}
where we set $\ud^2 \widehat{\mathbf{H}} 
= - \widehat{\mathbf{H}} \circ \ud\mathbf{\Phi} \circ \ud\mathbf{\Phi}
= - \ddot{\mathbf{H}}$.
Next, inserting \eqref{eq:sdH} into the first term on the \ac{RHS} of \eqref{eq:twotrace}, we get
\begin{align}
  & \trace ( \widehat{\mathbf{Y}}^{-1} \, \ud^2 \widehat{\mathbf{Y}} )\\
& \qquad =
 2\,\trace ( \widehat{\mathbf{Y}}^{-1} \, \dot{\mathbf{H}} \, \dot{\mathbf{H}}^H )
-   \trace ( \widehat{\mathbf{Y}}^{-1} \, \ddot{\mathbf{H}} \, \widehat{\mathbf{H}}^H )
-   \trace ( \widehat{\mathbf{Y}}^{-1} \, \widehat{\mathbf{H}} \, \ddot{\mathbf{H}}^H )
\nonumber
\\
& \qquad =
 2\,\trace ( \widehat{\mathbf{Y}}^{-1} \, \dot{\mathbf{H}} \, \dot{\mathbf{H}}^H )
-   \trace ( \widehat{\mathbf{Y}}^{-1} \, \ddot{\mathbf{H}} \, \widehat{\mathbf{H}}^H )
-   \trace\big( (\widehat{\mathbf{Y}}^{-1})^* \, \ddot{\mathbf{H}}^* \, \widehat{\mathbf{H}}^T \big)
\nonumber
\\
& \qquad = 
 2\,\trace \Big( \widehat{\mathbf{Y}}^{-1} \dot{\mathbf{H}} \dot{\mathbf{H}}^H 
- \real\big( \widehat{\mathbf{Y}}^{-1} \ddot{\mathbf{H}} \widehat{\mathbf{H}}^H \big) \Big)
\label{eq:first_term}
\\
& \qquad =
  2 \, \trace \big( \widehat{\mathbf{Y}}^{-1} \dot{\mathbf{H}} \mathbf{I}_{\MT} \dot{\mathbf{H}}^H \big)
- 2 \, \real\Big( \trace \big( \widehat{\mathbf{H}}^H \widehat{\mathbf{Y}}^{-1} \ddot{\mathbf{H}} \big) \Big)
\nonumber
\\
& \qquad \stackrel{\text{(a)}}{=}
  2 \, \transp[big]{\vect}{( \dot{\mathbf{H}}^T )}
  \big( (\widehat{\mathbf{Y}}^{-1})^T \kron \mathbf{I}_{\MT} \big)
               \vect( \dot{\mathbf{H}}^H )
- 2 \, \real\bigg( 
       \transp[Big]{\vect}{\big( (\widehat{\mathbf{H}}^H \widehat{\mathbf{Y}}^{-1})^T \big)} 
               \vect( \ddot{\mathbf{H}} ) \bigg)
\nonumber
\\
\begin{split}
& \qquad \stackrel{\text{(b)}}{=}
  2 \, \transp[big]{\vect}{( \ud\mathbf{\Phi} )} \, 
  \mathbf{K}_{(\MT,\MR)} \, 
  \divec( \widehat{\mathbf{H}}^T ) 
  \big( (\widehat{\mathbf{Y}}^{-1})^T \kron \mathbf{I}_{\MT} \big) \,
  \divec( \widehat{\mathbf{H}}^H ) \, \mathbf{K}_{(\MR,\MT)} \, \vect( \ud\mathbf{\Phi} )
\\
& \qquad
- 2 \, \real\bigg(
       \transp[Big]{\vect}{\big( (\widehat{\mathbf{H}}^H \widehat{\mathbf{Y}}^{-1} )^T \big)} \, 
  \divec( \widehat{\mathbf{H}} ) \, \divec( \ud\mathbf{\Phi} ) \, \vect( \ud\mathbf{\Phi} ) \bigg)
\end{split}
\nonumber
\\
\begin{split}
& \qquad \stackrel{\text{(c)}}{=} 
  2 \, \transp[big]{\vect}{( \ud\mathbf{\Phi} )} \,
  \divec( \widehat{\mathbf{H}} ) \, 
  \big( \mathbf{I}_{\MT} \kron (\widehat{\mathbf{Y}}^{-1})^T \big) \,
  \divec( \widehat{\mathbf{H}}^* ) \, 
  \vect( \ud\mathbf{\Phi} )
\\
& \qquad
- 2 \, \transp[big]{\vect}{( \ud\mathbf{\Phi} )} \, 
  \real\Big( \divec\big( (\widehat{\mathbf{H}}^H \widehat{\mathbf{Y}}^{-1})^T \circ \widehat{\mathbf{H}} \big) \Big) \, 
  \vect( \ud\mathbf{\Phi} )
\end{split}
\label{eq:firstT}
\end{align}
where 
(a) results from applying \eqref{eq:trABCD} with 
$\mathbf{A} = \widehat{\mathbf{Y}}^{-1}$, 
$\mathbf{B} = \dot{\mathbf{H}}$, 
$\mathbf{C} = \mathbf{I}_{\MT}$, and 
$\mathbf{D} = \dot{\mathbf{H}}^H$ to the first term,
transposing the result,
and, as before, applying
$\trace(\mathbf{A}^T\mathbf{B}) = \vect(\mathbf{A})^T \vect(\mathbf{B})$ 
with
$\mathbf{A}^T = \widehat{\mathbf{H}}^H\widehat{\mathbf{Y}}^{-1}$ 
and $\mathbf{B} = \ddot{\mathbf{H}}$
to the second term.
Step (b) is a consequence of applying \eqref{eq:vect_prod_Had}, 
the commutation relation \eqref{eq:comm_mat},
and
$\mathbf{K}_{(\vrow,\vcol)}^T = \mathbf{K}_{(\vcol,\vrow)}$.
To obtain (c), we used \eqref{comm_mat_shift} and \eqref{eq:reverseK} 
for the first term and 
\begin{align*}
  \transp[big]{\vect}{\big( (\widehat{\mathbf{H}}^H \widehat{\mathbf{Y}}^{-1} )^T \big)} \, 
  \divec( \widehat{\mathbf{H}} ) \, \divec( \ud\mathbf{\Phi} )
&=
  \transp[big]{\vect}{\big( (\widehat{\mathbf{H}}^H \widehat{\mathbf{Y}}^{-1} )^T \big)} \, 
  \divec( \ud\mathbf{\Phi} ) \, \divec( \widehat{\mathbf{H}} )
\\
&=
 \transp[big]{\vect}{( \ud\mathbf{\Phi} )} \, 
 \divec\big( (\widehat{\mathbf{H}}^H \widehat{\mathbf{Y}}^{-1})^T \circ \widehat{\mathbf{H}} \big)
\end{align*}
for the second term.
The final result follows by identifying 
\eqref{eq:firstT} and \eqref{eq:finalSecT} with the \ac{RHS} of \eqref{eq:sec_ident},
noting that $\widehat{\mathbf{H}} = \mathbf{H}$ for $\mathbf{\Phi} = \mathbf{0}$,
and applying the second identification theorem.
All the terms, except for the first term in \eqref{eq:firstT},
can be verified to be real-symmetric\footnote{A matrix 
$\mathbf{X}$ is said to be real-symmetric if $\mathbf{X}=\mathbf{X}^T$.}
so that
$(1/2)(\mathbf{B}+\mathbf{B}^T) = \mathbf{B}$.
The first term in \eqref{eq:firstT} is Hermitian and hence
$(1/2)(\mathbf{B}+\mathbf{B}^T) = \real(\mathbf{B})$.
\end{IEEEproof}
\end{prop}

\subsection{Approximations for First and Second Moment of MI}

Even though the physical channel is deterministic, the effective channel
$\widehat{\mathbf{H}} = \mathbf{H} \circ \exp\cwa(\jmath\mathbf{\Phi})$ will be 
random due to phase noise. We shall next compute approximations
of the effective channel's ergodic capacity 
$\widehat{C} = \expect\{ \hat{I} \}$ and of $\var\{ \hat{I} \}$
based on the second-order Taylor series expansion \eqref{eq:Taylor}. 
For an explanation of the
operational significance of $\var\{ \hat{I} \}$ the reader is referred to 
\secref{sec:rank1}
and to \cite{ONBP03j}. Unlike the 
results in \secref{sec:rank1}, which are restricted to the (extreme,
but not necessarily untypical) case of
fully uncorrelated phase noise, we will allow a general phase noise covariance
matrix, i.e., $\mathbf{\Phi} \equidist \realGauss(\mathbf{0}, \mathbf{\Sigma}_{\PP})$.
Noting that $\Jacobian{\hat{I}}{\mathbf{0}} \, \vect(\mathbf{\Phi})$ is zero-mean
Gaussian and $\mathbf{Q}_{\hat{I}} 
= \mathbf{Q}_{\hat{I}}\big( \vect(\mathbf{\Phi}) \big) 
= \transp[big]{\vect}{(\mathbf{\Phi})} \, \Hessian{\hat{I}}{\mathbf{0}} \, 
\vect(\mathbf{\Phi})$ is a quadratic form in real-valued Gaussian \acp{RV} with distribution
\cite[Eq.\rsp(4.1.1)]{MaPr92b}
\begin{align*}
  \mathbf{Q}_{\hat{I}}
&\equidist
  \sum_{i=1}^{\MT\MR}
  \lambda_i\big( \mathbf{\Sigma}_{\PP}^{1/2} \, 
                 \Hessian{\hat{I}}{\mathbf{0}} \, 
                 \mathbf{\Sigma}_{\PP}^{1/2} \big)
  X_i
\end{align*}
where the $X_i \equidist \chi^2_{1,1}$ $(i=1,2,\ldots,\MT\MR)$
are statistically independent,
straightforward manipulations reveal that
\begin{align}
  \expect\{ \tilde{I}(\mathbf{\Phi}) \}
&= 
  \hat{I}(\mathbf{0}) 
+ \frac{1}{2} \, \trace\big( \Hessian{\hat{I}}{\mathbf{0}} \, \mathbf{\Sigma}_{\PP} \big)
\label{eq:biasTerm}
\end{align}
and
\begin{align}
  \var\{ \tilde{I}(\mathbf{\Phi}) \}
&= 
  \Jacobian{\hat{I}}{\mathbf{0}} \, \mathbf{\Sigma}_{\PP} \, 
  \JacobianT{\hat{I}}{\mathbf{0}}
+ \frac{1}{2} \, \trace\Big( \big( \Hessian{\hat{I}}{\mathbf{0}} \, 
  \mathbf{\Sigma}_{\PP} \big)^2 \Big).
\label{eq:varTerm}
\end{align}

Inserting \eqref{eq:diff1} and \eqref{eq:diff2MI} into \eqref{eq:biasTerm} and \eqref{eq:varTerm}, 
we have analytic approximations of the
ergodic capacity and the variance of the \ac{MI} of the effective
channel as a function of the physical channel and of the phase noise
covariance matrix $\mathbf{\Sigma}_\PP$. 
The following numerical results demonstrate that
these approximations tend to be quite accurate for full-rank physical
channels but rather loose for rank-deficient physical channels.

\begin{figure}[htbp]
\centering
\subfigure[First moment of
$\hat{I}$ and of
second order Taylor series expansion $\tilde{I}$]{
\psfrag{E{V} - V}{\scriptsize{$\expect\{\hat{I}\}-I$}}
\psfrag{E{W} - W}{\scriptsize{$\expect\{\tilde{I}\}-I$}}
\icg{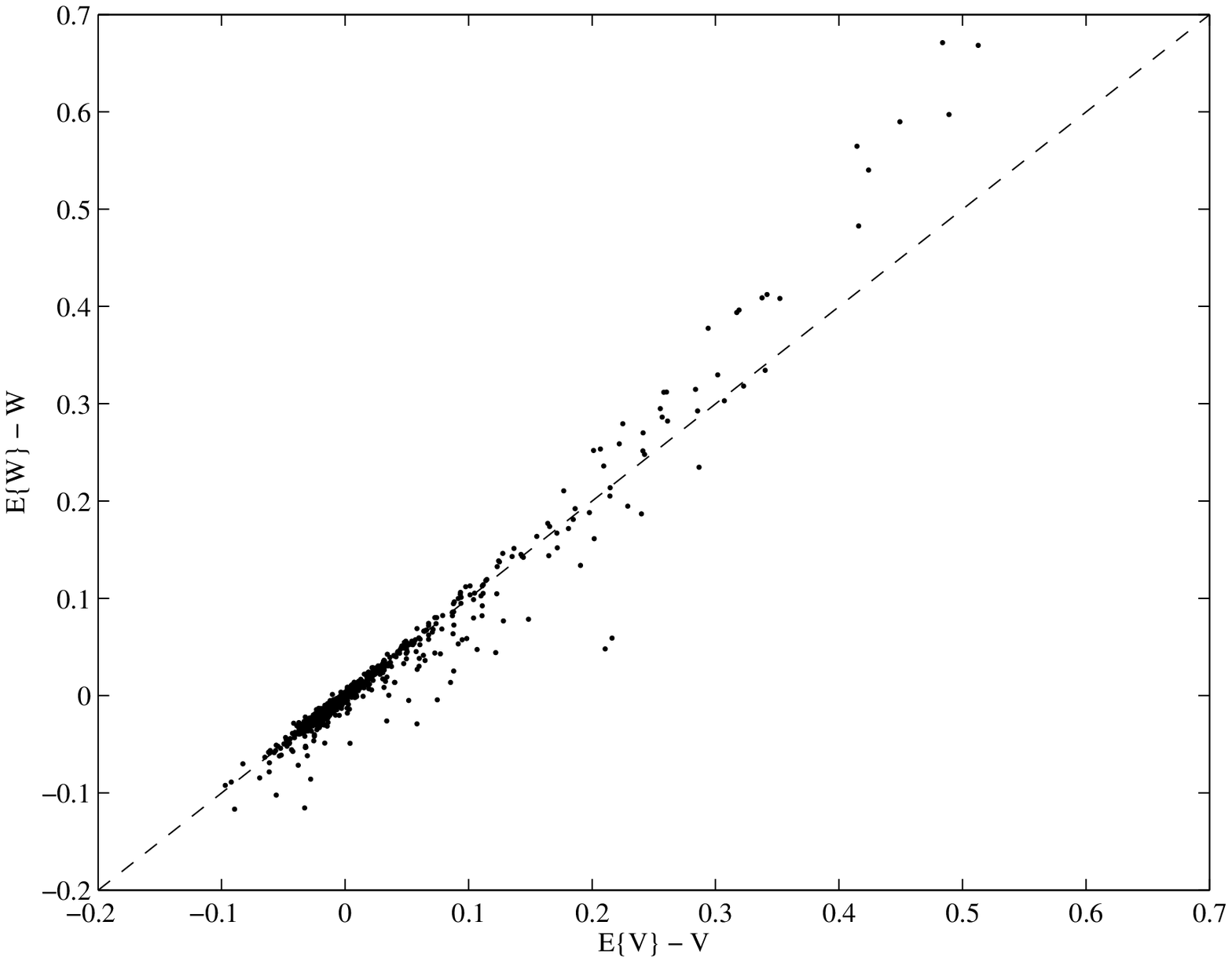}
\label{fig:moments_mean}
}
\subfigure[Second moment of 
$\hat{I}$ and of 
second order Taylor series expansion $\tilde{I}$]{
\psfrag{Stdv{V} - Stdv{V}}{\scriptsize{$(\var\{\hat{I}\})^{1/2}$}}
\psfrag{Stdv{W} - Stdv{W}}{\scriptsize{$(\var\{\tilde{I}\})^{1/2}$}}
\icg{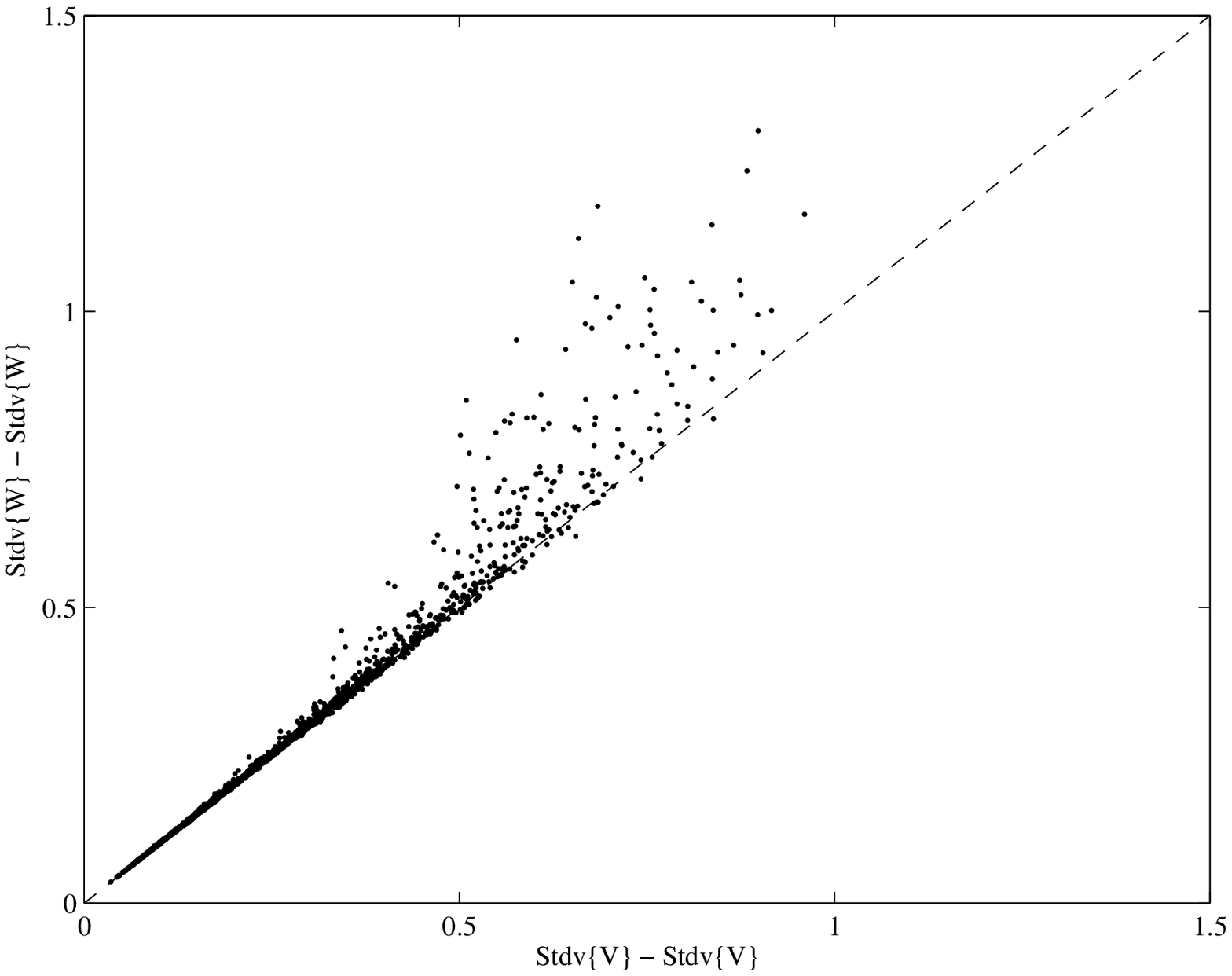}
\label{fig:moments_std}
}
\vspace{-2mm}
\caption{\label{fig:moments}
Moments of \acl{MI} $\hat{I}$ and of \acl{MI} 
second order Taylor series expansion $\tilde{I}$, both 
for 2000 realizations of an \acs{IID} Gaussian physical channel 
subject to fully uncorrelated 3.5\degreel\ \ac{RMS} phase noise 
at \acs{SNR} $\rho=30$\unit{dB}.
Moments of $\hat{I}$ are obtained from Monte Carlo simulation
over 10\,000 
phase noise samples per physical channel realization. 
} 
\end{figure}

{\it Numerical results}\/: In \figref{fig:moments}, we plot
$\expect\{\tilde{I}(\mathbf{\Phi})\}$
and 
$\big( \var\{\tilde{I}(\mathbf{\Phi})\} \big)^{1/2}$
in \eqref{eq:biasTerm} and \eqref{eq:varTerm}, respectively,
versus the exact values
$\expect\{\hat{I}(\mathbf{\Phi})\}$
and 
$\big( \var\{\hat{I}(\mathbf{\Phi})\} \big)^{1/2}$
obtained 
by Monte Carlo simulation over 10\,000 phase noise samples
with 3.5\degreel\ \ac{RMS} fully uncorrelated phase noise.
Each point in the figures 
represents a pair of exact and approximate 
\ac{MI} first and second moments for one of 2000 realizations 
of an \ac{IID} physical channel with $\MT=\MR=4$ at $\rho=30$\unit{dB}.
We observe that the deviation of the first moment of \ac{MI}
can be positive or negative and is essentially 
independent of the exact value of the first moment. 
The deviation of the second moment is predominantly positive, 
and the accuracy shows a strong dependence 
on the exact value of the second moment.
Quantitatively speaking, 
the deviation of the estimates 
$\expect\{\tilde{I}(\mathbf{\Phi})\}$
and 
$\big( \var\{\tilde{I}(\mathbf{\Phi})\} \big)^{1/2}$
is significant for about 5\% of the 
\ac{IID} channel realizations.

\subsection{Taylor Series Expansion of the Channel's Eigenvalues}
\label{sec:aesoTse}

\begin{figure}[htbp]
\centering
\subfigure[\acsp{CDF} corresponding to the second-order Taylor series 
approximation of \acl{MI}]{
\psfrag{Mutual information I [bit/s/Hz]}{\scriptsize{Mutual information $\hat{I}$ [bit/s/Hz]}}
\psfrag{CDF}{\scriptsize{\acs{CDF}}}
\psfrag{exact}{\scriptsize{exact}}
\psfrag{Taylor series approx. of Mutual Information}{\scriptsize{Taylor series approx.\ of \acl{MI}}}
\icg{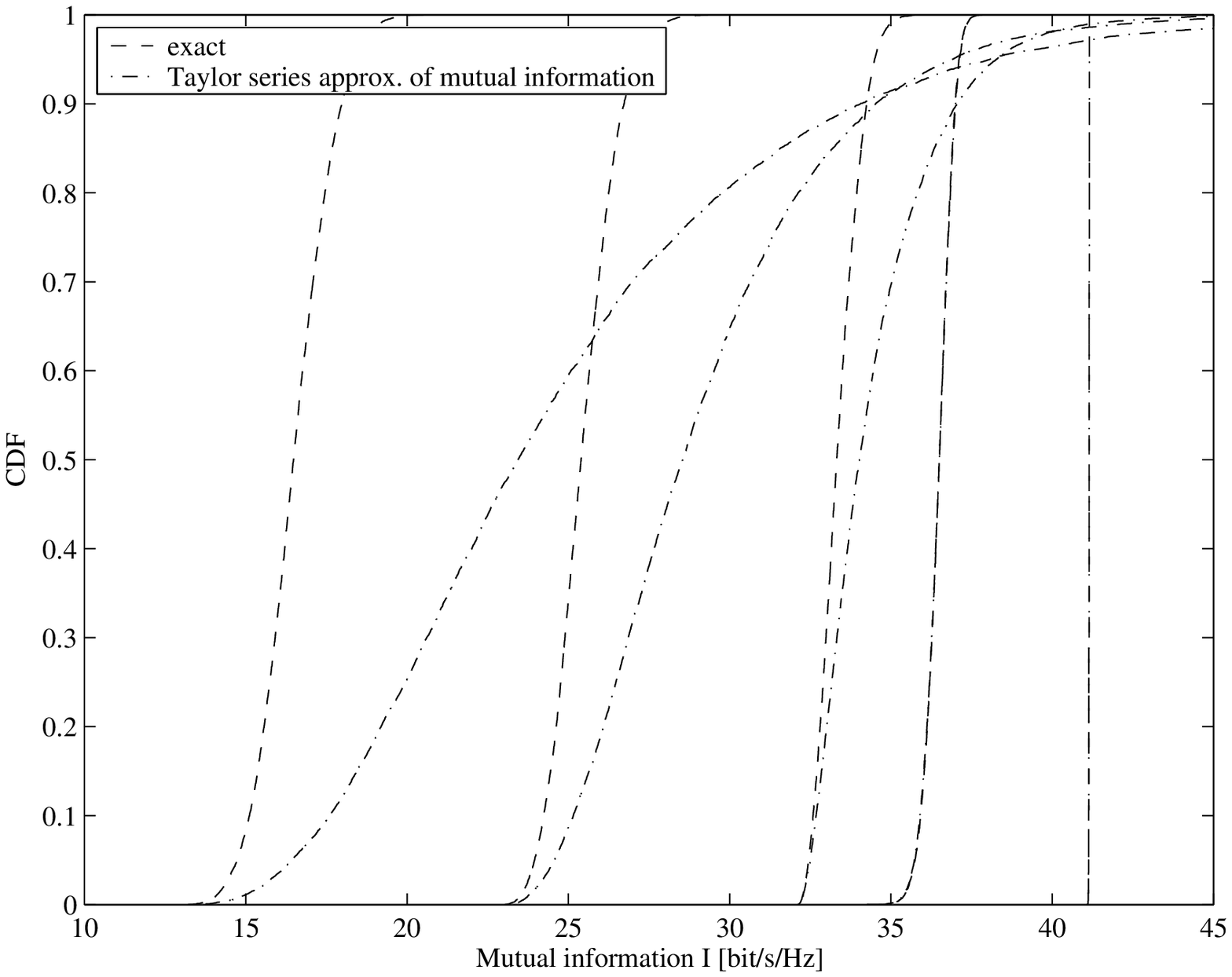}
\label{fig:evapprox} 
}
\subfigure[\acsp{CDF} corresponding to 
second-order Taylor series approximation of channel eigenvalues]{
\psfrag{Mutual information I [bit/s/Hz]}{\scriptsize{Mutual information $\hat{I}$ [bit/s/Hz]}}
\psfrag{CDF}{\scriptsize{\acs{CDF}}}
\psfrag{exact}{\scriptsize{exact}}
\psfrag{Taylor series approx. of eigenvalues}{\scriptsize{Taylor series approx.\ of eigenvalues}}
\icg{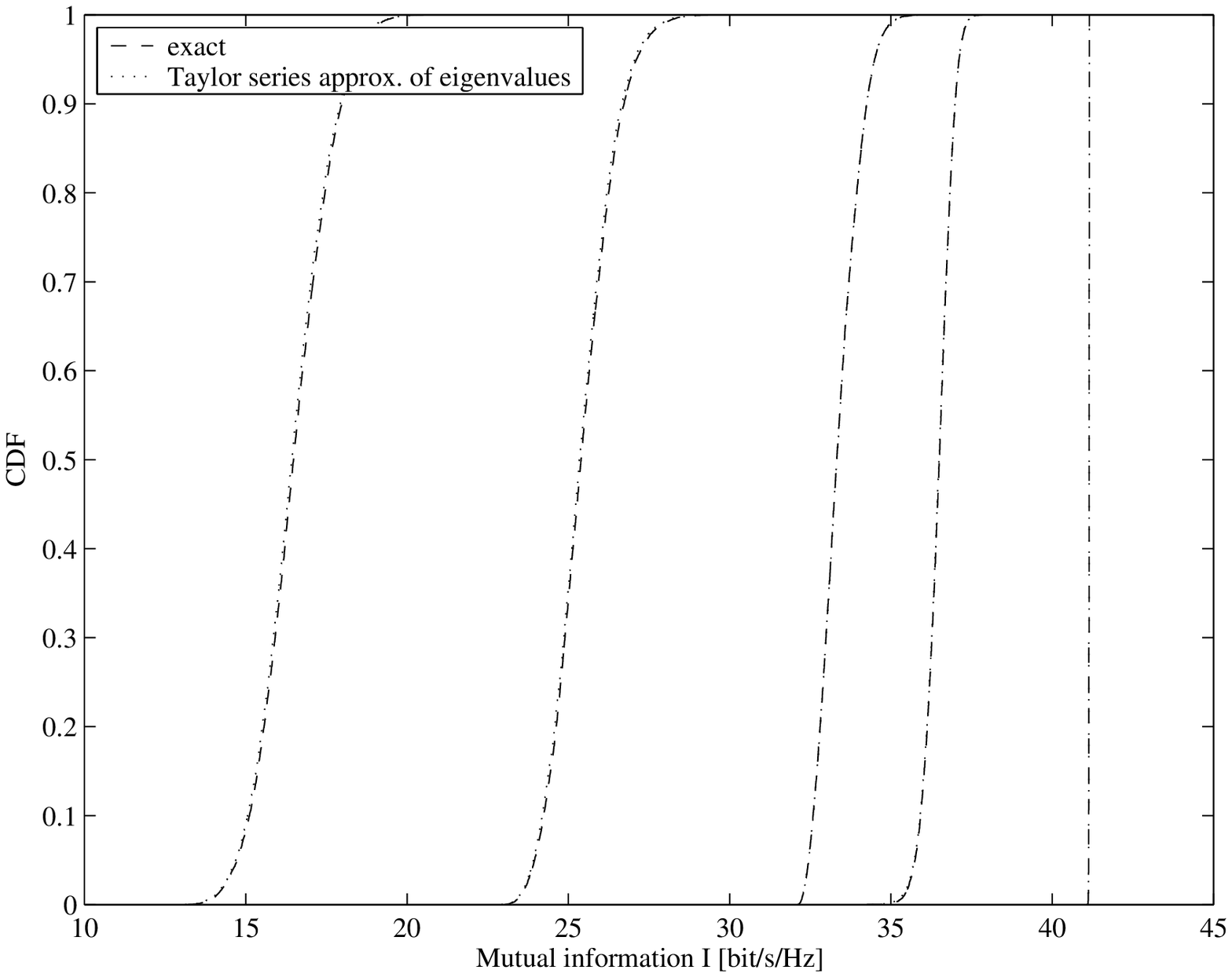}
\label{fig:trueevapprox}
}
\vspace{-2mm}
\caption{\label{fig:Taylorapprox}
\acsp{CDF} of \acl{MI} and of corresponding approximation based on 
second-order Taylor series expansion \eqref{eq:Taylor},
as well as second-order Taylor series approximation of eigenvalues,
for realizations of a (from left to right)
rank-1, balanced rank-2, balanced rank-3, rank-4,
and balanced rank-4 physical $4\times4$ channel 
subject to fully uncorrelated 
3.5\degreel\ \acs{RMS} phase noise 
(10\,000 realizations)
at \acs{SNR} $\rho=30$\unit{dB}.
For all channels, we employed the normalization $\|\mathbf{H}\|_\FROB^2=\MT\MR$.} 
\end{figure}

To obtain more insight into the quality-of-fit
of the second-order Taylor series approximation of \ac{MI},
we show in \sfigref{fig:Taylorapprox}{fig:evapprox} the \acp{CDF} of the exact \ac{MI} 
and the \ac{MI} obtained through the approximation \eqref{eq:Taylor}. 
In all cases, Monte Carlo simulation at $\rho=30$\unit{dB} 
with 3.5\degreel\ \ac{RMS} fully uncorrelated phase noise
was employed.
As (deterministic) physical channels we have chosen 
balanced (i.e., all nonzero singular values of the channel matrix are equal) 
rank-$M$ $4\times4$ channels for $M=1,2,3,4$ and 
an unbalanced full-rank channel. 
All physical channels were normalized to satisfy
$\|\mathbf{H}\|_\FROB^2=\MT\MR=16$.
The figure shows that the Taylor series approximation is
very loose for rank-1 and rank-2 physical channels,
acceptable for the rank-3 physical channel,
and very accurate for the two full-rank physical channels.
This example shows that the second-order Taylor series
expansion of \ac{MI}
tends to yield poor approximations for low-rank or, more generally,
poorly balanced physical channels, i.e., physical channels with
large eigenvalue spread. 
An alternative approach for obtaining approximations 
of $\hat{I}$ is to compute
a second-order Taylor series expansion of the unordered 
(but continuous w.r.t. $\mathbf{\Phi}$) eigenvalues 
$\lambda_{\widehat{\mathbf{H}}\widehat{\mathbf{H}}^H}^{(i)}(\mathbf{\Phi})$ 
where 
\begin{align}
  \hat{I}(\mathbf{\Phi}) 
= 
  \sum_{i=1}^{\MR} 
         \log\left( 1 + \frac{\rho}{\MT} \, 
         \lambda_{\widehat{\mathbf{H}}\widehat{\mathbf{H}}^H}^{(i)}(\mathbf{\Phi})
             \right).
\label{eq:EVexp}
\end{align}

Before briefly outlining how this can be done analytically, 
we show in \sfigref{fig:Taylorapprox}{fig:trueevapprox} the 
result of replacing \eqref{eq:Taylor} in 
\sfigref{fig:Taylorapprox}{fig:evapprox} by the second-order
Taylor series expansion of the eigenvalues 
$\lambda_{\widehat{\mathbf{H}}\widehat{\mathbf{H}}^H}^{(i)}(\mathbf{\Phi})$ 
used in the \ac{RHS} of \eqref{eq:EVexp}.
It is clearly
seen that the Taylor series expansion of the
eigenvalues 
$\lambda_{\widehat{\mathbf{H}}\widehat{\mathbf{H}}^H}^{(i)}(\mathbf{\Phi})$ 
yields outstanding accuracy and 
significantly better results 
than the Taylor series expansion of \ac{MI} directly.
A few comments on how to obtain analytic expressions for the
second-order Taylor series expansion of the 
$\lambda_{\widehat{\mathbf{H}}\widehat{\mathbf{H}}^H}^{(i)}(\mathbf{\Phi})$ 
are in order. For the sake of space
and focus of the paper, we shall not present the details, but rather
refer the interested reader to suitable references. In general, obtaining analytic
expressions for the Taylor series of the 
$\lambda_{\widehat{\mathbf{H}}\widehat{\mathbf{H}}^H}^{(i)}(\mathbf{\Phi})$ 
is difficult and tedious; in fact, in general,
more tedious than computing the Taylor series
expansion of \ac{MI} directly. 
While the case of a physical channel with
nonrepeated eigenvalues can be treated with relative ease by
employing \cite[Ch.\rsp8, Ths.\rsp7, 8, 10, and 11]{MaNe88b}, 
the general case of physical channels that have eigenvalues
of multiplicity larger than 1 (e.g., multiple eigenvalues
equal to zero in the case 
of rank-deficient physical channels) is significantly more involved. 
Results relevant in this context can be found in 
\cite[Ch.\rsp8, Sec.\rsp12]{MaNe88b} and in \cite{Tork01j}. 
The main difficulty in obtaining
analytic expressions in the case of repeated eigenvalues 
is that the results depend on the eigenvectors
of the channel.

\section{The Rank-1 Physical Channel}
\label{sec:rank1}
As already mentioned in \secref{sec:MI}, the impact of phase errors 
is more pronounced for low-rank physical \ac{MIMO} channels. 
In the following, we shall therefore analyze the extreme case of a rank-1 physical channel 
in detail. 
In practice, deterministic rank-1 channels occur in \ac{LOS} scenarios with
small angle-spread \cite{GBGP02} (green-field like propagation conditions). Stochastic rank-1 \ac{MIMO}
channels are channels where the realization of the \ac{MIMO} channel matrix has rank 1 
\ac{WIP}\!\!\rsp1. A prominent member of this class of channels is 
the pin-hole \cite{GBGP02} or key-hole \cite{CFGV02,LoKo02j,AlTM03} channel 
reflecting propagation conditions with significant scattering close to the transmitter
and the receiver and at the same time long distances between transmitter and receiver. 
Finally, the following \ac{MIMO} channel sounder ``calibration procedure'' 
provides a practical motivation for 
studying (and quantifying) the impact of phase errors on rank-1 physical channels.
The main idea underlying this calibration procedure is based on the fact that
connecting transmitter and receiver in a \ac{TDMS}-based sounder by a cable results 
in a deterministic rank-1 physical channel $\mathbf{H} = \alpha \mathbf{1}$, 
where $\alpha\in\numID{C}$ is the gain corresponding to the constant (across frequency)
cable transfer function.
The channel sounder then acquires samples of the effective channel matrix
which contains channel coefficients that (after power normalization) have unit magnitude 
and a phase that varies due to phase errors created by the sounder.
An inspection of the resulting
eigenvalue histogram yields the number of significant eigenvalues and the corresponding
eigenvalue distribution. Since the underlying physical channel has rank 1, it follows that
any additional (\ac{WRT} the one resulting from the physical channel) significant modes in
the effective channel must necessarily be due to phase errors (and/or potentially 
other imperfections in the measurement
equipment). This ``calibration measurement'' can therefore loosely be interpreted as 
revealing the highest possible rank increase due to phase errors.

We shall see that, unlike the general case discussed in \secsref{sec:MI} and \ref{sec:sens_ana},
rank-1 physical channels allow to establish a number of insightful analytic results 
on the impact of phase errors on \ac{MI}.
Throughout this section, unless explicitly stated otherwise, the results are valid for a general
(i.e., not necessarily separable) timing matrix $\mathbf{T}$.
We consider channels given by $\mathbf{H} = \mathbf{g}\mathbf{h}^T$
where the vectors $\mathbf{g}$ and $\mathbf{h}$ can be either deterministic or stochastic
(and with entries that are not necessarily unit modulus).

Let us start with a simple basic result which will be needed later in this section.

\begin{lemma}
\label{lemma:prod}
The Hadamard product of a rank-1 matrix 
$\mathbf{H} = \mathbf{g}\mathbf{h}^T$
and an arbitrary matrix $\mathbf{\Theta}$ 
can be written as a matrix product according to
\begin{align*}
(\mathbf{g} \mathbf{h}^T) \circ \mathbf{\Theta} 
= \diag(\mathbf{g}) \, \mathbf{\Theta} \, \diag(\mathbf{h}).
\end{align*}

\begin{IEEEproof}
The elements of a rank-1 matrix $\mathbf{H}=\mathbf{g}\mathbf{h}^T$ 
are given by $[\mathbf{H}]_{m,n} = [\mathbf{g}]_{m} 
[\mathbf{h}]_{n}$. Consequently, we have
$[\mathbf{H} \circ \mathbf{\Theta}]_{m,n} =
[\mathbf{g}]_{m} [\mathbf{h}]_{n} [\mathbf{\Theta}]_{m,n}$. 
On the other hand, it follows immediately that 
$[\diag(\mathbf{g}) \, \mathbf{\Theta} \, \diag(\mathbf{h}) ]_{m,n} 
= [\mathbf{g}]_{m} \, [\mathbf{\Theta}]_{m,n} \, [\mathbf{h}]_{n}$ 
which concludes the proof. 
\end{IEEEproof}
\end{lemma}

The following three Theorems state that a physical
rank-1 channel subject to severe enough phase errors results in a full-rank
effective channel.

\begin{prop}
\label{prop:sep}
For a rank-1 physical channel $\mathbf{H} = \mathbf{g} \mathbf{h}^T$ subject to 
phase errors with $\mathbf{\Theta}$ in \eqref{eq:channelhat} having full rank \ac{WIP}\!\!\rsp1,
we have
\begin{align*}
\det( \widehat{\mathbf{H}}\widehat{\mathbf{H}}^H ) 
& = 
\left( \prod_{i=1}^{\MR} \big| [\mathbf{g}]_i \big|^2 \varepsilon_i \right) \,
\det( \mathbf{\Theta}\mathbf{\Theta}^H ), \quad \MR\le \MT
\\
\det( \widehat{\mathbf{H}}^H\widehat{\mathbf{H}} ) 
& = 
\left( \prod_{i=1}^{\MT} \big| [\mathbf{h}]_i \big|^2 \nu_i \right) \,
\det( \mathbf{\Theta}^H\mathbf{\Theta} ), \quad \MR\ge\MT
\end{align*}
where 
\begin{align*}
&\min_i \big| [\mathbf{h}]_i \big|^2 \le \varepsilon_i \le \max_i \big| [\mathbf{h}]_i \big|^2,&
&\min_i \big| [\mathbf{g}]_i \big|^2 \le \nu_i \le \max_i \big| [\mathbf{g}]_i \big|^2.
\end{align*}

\begin{IEEEproof}
We provide the proof for the case $\MR\le \MT$ only. 
The proof for $\MR>\MT$ follows exactly the same line of reasoning. 
We start by noting that \lemmaref{lemma:prod} implies
\begin{align*}
  \det(\widehat{\mathbf{H}}\widehat{\mathbf{H}}^H) 
&= 
  \det\big( \diag(\mathbf{g}) \, \mathbf{\Theta} \, \diag(\mathbf{h}) \, 
  \diagH[big]{(\mathbf{h})} \, \mathbf{\Theta}^H \, \diagH[big]{(\mathbf{g})} \big) \\
&= 
  \left( \prod_{i=1}^{\MR} \big| [\mathbf{g}]_i \big|^2 \right)
  \prod_{i=1}^{\MR} \lambda_i\mo\big( \mathbf{\Theta} \, 
  \diag(\mathbf{h}) \, \diagH[big]{(\mathbf{h})} \, \mathbf{\Theta}^H \big) \\
&\stackrel{\text{(a)}}{=} 
  \left( \prod_{i=1}^{\MR} \big| [\mathbf{g}]_i \big|^2 \right)
  \prod_{i=1}^{\MR} \lambda_i\mo\big( \diagH[big]{(\mathbf{h})} \, 
  \mathbf{\Theta}^H \mathbf{\Theta} \, \diag(\mathbf{h}) \big) \\
&\stackrel{\text{(b)}}{=} 
  \left( \prod_{i=1}^{\MR} \big| [\mathbf{g}]_i \big|^2 \right)
  \prod_{i=1}^{\MR} \varepsilon_i \, \lambda_i\mo\big( 
  \mathbf{\Theta} \mathbf{\Theta}^H \big)
\end{align*}
where the second product on the \ac{RHS} of (a) is taken over the $\MR$ nonzero eigenvalues of 
$\diagH[big]{(\mathbf{h})} \, \mathbf{\Theta}^H \mathbf{\Theta} \, \diag(\mathbf{h})$ 
only (note that the $\lambda_i$ are ordered as defined in the Notations section)
and (b) follows from a Corollary to Ostrowski's Theorem 
\cite[Corollary\rsq4.5.11]{HoJo85b}. 
\end{IEEEproof}
\end{prop}

\propref{prop:sep} thus states
that a physical rank-1 channel 
subject to phase noise such that $\mathbf{\Theta}$ has full rank \ac{WIP}\!\!\rsp1
results in a full-rank effective \ac{MIMO} channel 
(provided that $[\mathbf{g}]_m\ne0$, $\forall\,m$, 
and $[\mathbf{h}]_n\ne0$, $\forall\,n$).
For a deterministic physical rank-1 channel, the resulting effective channel will be
stochastic and will have full rank \ac{WIP}\!\!\rsp1.
The condition of $\mathbf{\Theta}$ having full rank \ac{WIP}\!\!\rsp1 may sound stringent.
It turns out, however, that a full-rank phase noise covariance matrix $\cov\{\mathbf{\Phi}\}$
is sufficient for $\mathbf{\Theta}$ to have full rank \ac{WIP}\!\!\rsp1. This statement can be
formalized as follows.

\begin{lemma}
A real Gaussian random matrix $\mathbf{\Phi}\in\numID{R}^{\MR\times\MT}$
where $\vect(\mathbf{\Phi})\equidist\realGauss(\mathbf{0},\cov\{\mathbf{\Phi}\})$
with $\det(\cov\{\mathbf{\Phi}\})>0$,
has full rank \ac{WIP}\!\!\rsp1.
The matrix $\mathbf{\Theta} = \exp\cwa(\jmath\mathbf{\Phi})$ 
has full rank \ac{WIP}\!\!\rsp1 as well.

\begin{IEEEproof}
We follow the direct proof of \cite[Th.\rsp2.3, p.\rsp712]{EaPe73j}, 
where it is shown that for an $\MR\times\MT$ random matrix
$\mathbf{X}$ to be full rank \ac{WIP}\!\!\rsp1,
it is sufficient to have the 
multivariate distribution of $\mathbf{X}$ be 
absolutely continuous \ac{WRT} $\MR\MT$-dimensional Lebesgue measure.
This condition is trivially satisfied by $\mathbf{\Phi}$ with 
$\vect(\mathbf{\Phi})\equidist\realGauss(\mathbf{0},\cov\{\mathbf{\Phi}\})$
and $\det(\cov\{\mathbf{\Phi}\})>0$ (see, e.g., \cite[Sec.\rsp4.7.2]{CaKo99b}).

The second part of the statement can be proved by using 
\cite[Lemma\rsp3]{McAL96j}, which states that $\mathbf{\Theta}$ has full rank 
if either $\real(\mathbf{\Theta})$, $\imag(\mathbf{\Theta})$, 
or $\big[\, \transp[big]{\real}{(\mathbf{\Theta})} \,\;\,
            \transp[big]{\imag}{(\mathbf{\Theta})} \, \big]^T$ has full rank.
Direct computation reveals that the multivariate \ac{PDF} of 
$\transp[big]{\real}{(\mathbf{\Theta})} = \cos\cwa(\mathbf{\Theta})$
is continuous and integrable 
(in the interval $[-1,1]^{\MT\MR}$)
so that its multivariate
\ac{CDF} is absolutely continuous \ac{WRT} $\MR\MT$-dimensional 
Lebesgue measure.
Hence, by the direct proof of \cite[Th.\rsp2.3]{EaPe73j},
$\real(\mathbf{\Theta})$ is full rank \ac{WIP}\!\!\rsp1,
and from what was said before it follows 
that $\mathbf{\Theta}$ has full rank \ac{WIP}\!\!\rsp1.
\end{IEEEproof}
\end{lemma}

Besides what was stated in \propref{prop:sep} above, relating properties
of $\cov\{\mathbf{\Phi}\}$ to properties of $\det( \mathbf{\Theta}\mathbf{\Theta}^H )$
seems difficult.
For $\MT=\MR$, we can refine the result in \propref{prop:sep} as follows.
\begin{prop}
For a rank-1 physical channel $\mathbf{H} = \mathbf{g} \mathbf{h}^T$ with
$\MT=\MR=M$ subject to phase errors with $\mathbf{\Theta}$ having
full rank \ac{WIP}\!\!\rsp1, we have
\begin{align}
\det( \widehat{\mathbf{H}}\widehat{\mathbf{H}}^H ) 
& = 
\left( \prod_{i=1}^{M} \big| [\mathbf{h}]_i \big|^2 \right)
\left( \prod_{i=1}^{M} \big| [\mathbf{g}]_i \big|^2 \right)
\det( \mathbf{\Theta}\mathbf{\Theta}^H ).
\label{eq:prop_sep2} 
\end{align}

\begin{IEEEproof}
The proof follows trivially using \lemmaref{lemma:prod} and 
noting that 
\begin{align*}
  \det(\widehat{\mathbf{H}}\widehat{\mathbf{H}}^H) 
&= 
  \det\Big( \diag(\mathbf{g}) \, \mathbf{\Theta} \, \diag(\mathbf{h}) \, 
  \diagH[big]{(\mathbf{h})} \, \mathbf{\Theta}^H \, \diagH[big]{(\mathbf{g})} \Big) \\
&= 
  \det\Big( \diagH[big]{(\mathbf{g})} \, \diag(\mathbf{g}) \Big) \,
  \det\Big( \mathbf{\Theta} \, \diag(\mathbf{h}) \, 
  \diagH[big]{(\mathbf{h})} \, \mathbf{\Theta}^H \Big) \\
&= 
  \det\Big( \diagH[big]{(\mathbf{g})} \, \diag(\mathbf{g}) \Big) \, 
  \det\Big( \diag(\mathbf{h}) \, \diagH[big]{(\mathbf{h})} \Big) \, 
  \det\big( \mathbf{\Theta} \mathbf{\Theta}^H \big)
\end{align*}
which yields \eqref{eq:prop_sep2}.
\end{IEEEproof}
\label{prop:sep2}
\end{prop}

The following Theorem allows a more specific conclusion since it shows
that for rank-1 channels $\mathbf{H}=\mathbf{g}\mathbf{h}^T$ where $\mathbf{g}$
and $\mathbf{h}$ consist of unit-modulus entries
(representative of \ac{LOS} propagation \cite{GBGP02}) 
the rank of the effective channel matrix 
is equal to the rank of $\mathbf{\Theta}$. Moreover, 
the eigenvalues of the effective channel matrix (more specifically of 
$\widehat{\mathbf{H}}\widehat{\mathbf{H}}^H$) are equal to the eigenvalues 
of $\mathbf{\Theta}\mathbf{\Theta}^H$.

\begin{prop}
\label{prop:svd}
For a rank-1 physical channel $\mathbf{H}=\mathbf{g}\mathbf{h}^T$,
where $\mathbf{g}$ and $\mathbf{h}$ are such that 
$\big| [\mathbf{g}]_{i} \big| = 1$ ($i=1,2,\ldots,\MR$) and 
$\big| [\mathbf{h}]_{i} \big| = 1$ ($i=1,2,\ldots,\MT$), we have
\begin{align*}
\lambda_i(\widehat{\mathbf{H}}\widehat{\mathbf{H}}^H) 
&= 
\lambda_i(\mathbf{\Theta}\mathbf{\Theta}^H), \quad i=1,2,\ldots,\MR, \quad \MR \le \MT \\
\lambda_i(\widehat{\mathbf{H}}^H\widehat{\mathbf{H}})
&= 
\lambda_i(\mathbf{\Theta}^H\mathbf{\Theta}), \quad i=1,2,\ldots,\MT, \quad \MR > \MT.
\end{align*}

\begin{IEEEproof}
The proof for both cases is trivially obtained using \lemmaref{lemma:prod}
and noting that the assumptions of the Theorem imply 
$\diag(\mathbf{g}) \diagH{(\mathbf{g})} = \mathbf{I}_{\MR}$ and
$\diag(\mathbf{h}) \diagH{(\mathbf{h})} = \mathbf{I}_{\MT}$. For $\MR\le
\MT$, simply note that 
\begin{align*}
  \lambda_i\mo( \widehat{\mathbf{H}}\widehat{\mathbf{H}}^H ) 
&= 
  \lambda_i\mo\Big( \diag(\mathbf{g}) \, \mathbf{\Theta} \, \diag(\mathbf{h}) \, 
  \diagH[big]{(\mathbf{h})} \, \mathbf{\Theta}^H \, \diagH[big]{(\mathbf{g})} \Big) \\
&= 
  \lambda_i\mo\Big( \diagH[big]{(\mathbf{g})} \, \diag(\mathbf{g}) \,
  \mathbf{\Theta} \mathbf{\Theta}^H \Big) \\
&= 
  \lambda_i\mo( \mathbf{\Theta} \mathbf{\Theta}^H ), 
\quad
  i=1,2,\ldots,\MR.
\end{align*}
The case $\MR>\MT$ follows exactly the same line of reasoning.
\end{IEEEproof}
\end{prop}

Since the high-\ac{SNR} \ac{MI} of $\widehat{\mathbf{H}}$ 
(for $\MR\le\MT$) is given by 
$\hat{I}\approx\log\,\det\big( (\rho/\MT)\widehat{\mathbf{H}}\widehat{\mathbf{H}}^H \big)$,
\propsref{prop:sep} and \ref{prop:sep2}
immediately yield expressions\footnote{More 
   specifically an approximation in the case of 
   \propref{prop:sep} due to the presence of the quantities 
   $\varepsilon_i$ and $\nu_i$.} 
for the high-\ac{SNR} \ac{MI} of $\widehat{\mathbf{H}}$.
However, the \ac{PDF} of the quantity
$\log\,\det(\mathbf{\Theta}\mathbf{\Theta}^H)$ is, 
in general, difficult to obtain.
Insightful analytic results are, however, possible 
by invoking the assumptions of a
separable timing matrix (as discussed in \secref{sec:freq_offs}) and of small phase noise, i.e.,
$\exp\cwa(\jmath\mathbf{\Phi})\approx\mathbf{1}+\jmath\mathbf{\Phi}$.
As demonstrated previously, a separable timing matrix is obtained by choosing 
a regular sounding pattern as in \eqref{eq:reg_pat} and the small phase noise approximation
is very well satisfied in practice as the worst-case value of 7\degreel\ \ac{RMS}
phase noise amounts to $\sigma_{\PP}^2\approx0.0149$.
The assumption of a separable timing matrix implies that 
frequency offset has no impact on \ac{MI} (see \secref{sec:freq_offs}). 
Therefore, as a consequence of the two 
simplifying assumptions, it suffices to analyze the quantity 
$\det(\widetilde{\mathbf{\Theta}}\widetilde{\mathbf{\Theta}}^H)$ with
$\widetilde{\mathbf{\Theta}}=\mathbf{1}+\jmath\mathbf{\Phi}$ instead of
$\det(\mathbf{\Theta}\mathbf{\Theta}^H)$. Interestingly, 
$\det(\widetilde{\mathbf{\Theta}}\widetilde{\mathbf{\Theta}}^H)$ can be
characterized in terms of 
chi-square \acp{RV} and a beta-distributed \ac{RV}, 
which provides the basis for tight bounds on 
$\widehat{C} = \expect\{ \hat{I} \}$ and for accurate approximations of
$\var\{ \hat{I} \}$. Before stating the corresponding results
using the exact expression for 
$\det(\widetilde{\mathbf{\Theta}}\widetilde{\mathbf{\Theta}}^H)$,
we shall, however, provide an approximation for 
$\det(\widetilde{\mathbf{\Theta}}\widetilde{\mathbf{\Theta}}^H)$ 
(in the sense of distributional equivalence), 
which turns out
to be particularly useful to derive a simple analytic lower bound on 
$\widehat{C}$ (see \propref{prop:Cbound}).
This approximation is based on the following result.

\begin{prop}
\label{prop:dist_approx}
For a separable timing matrix $\mathbf{T}$ and under 
the small phase noise approximation $\sigma^2_{\PP}\ll1$ so that
$\widetilde{\mathbf{\Theta}} = \mathbf{1}+\jmath\mathbf{\Phi}$,
assuming fully uncorrelated phase noise, i.e., 
$\vect(\mathbf{\Phi}) \equidist \realGauss(\mathbf{0}, 
\sigma_{\PP}^2 \mathbf{I}_{\MT\MR})$,
we have
\begin{align}
\begin{split}
   \det(\widetilde{\mathbf{\Theta}}\widetilde{\mathbf{\Theta}}^H) 
&\equidist
   \big( \chi^2_{\MT,\sigma_{\PP}^2} + \MT\MR \big) 
   \prod_{i=2}^{\MR} \big(   \chi^2_{\MT-i,\sigma_{\PP}^2} 
                           + Z(\lrv^{(i)}) \big), 
\quad 
   \MR\le \MT 
\\
   \det(\widetilde{\mathbf{\Theta}}^H \widetilde{\mathbf{\Theta}}) 
&\equidist
   \big( \chi^2_{\MR,\sigma_{\PP}^2} + \MT\MR \big) 
   \prod_{i=2}^{\MT} \big(   \chi^2_{\MR-i,\sigma_{\PP}^2}
                           + Z(\lrv^{(i)}) \big), 
\quad 
   \MR>\MT
\end{split}
\label{eq:chi2approx} 
\end{align}
where the
$\chi^2_{n,\sigma^2}$ are statistically independent\footnote{Note 
   that the product over $i$ on the \ac{RHS} of 
   \eqref{eq:chi2approx} is equal to 1 if $\MR=1$ (in the case $\MR\le\MT$)
   and $\MT=1$ (in the case $\MR>\MT$).}
and $Z(\lrv^{(i)}) 
= \sigma_{\PP}^2(\lrv^{(i)}X^{(i)}_{1}+(1-\lrv^{(i)})X^{(i)}_{2})$
with $X^{(i)}_{1},X^{(i)}_{2}$ i.i.d. as $\chi^{2}_{1,1}$ and the 
$\lrv^{(i)}$ being \acp{RV} with \ac{PDF} supported 
in the interval [0,1] $\forall\:i$.

\begin{IEEEproof}
We provide the proof for $\MR\le\MT$ only. 
The case $\MR>\MT$ follows exactly the same line of reasoning. 
Let us start by noting that the 
singular value decomposition of $\mathbf{1}_{\MR,\MT}$ is
given by $\mathbf{1}_{\MR,\MT} = \mathbf{V} \mathbf{\Sigma} \mathbf{W}^T$, 
where $\mathbf{V}$ is of dimension $\MR\times\MR$,
$\mathbf{W}$ is $\MT\times\MT$,
and the $\MR\times\MT$ matrix $\mathbf{\Sigma}$ is given by
\begin{align}
[\mathbf{\Sigma}]_{\vrow,\vcol} 
= \left\{ \begin{array}{ll}
\mysqrt{\MT\MR}, & \vrow = \vcol = 1 \\
0              , & \text{else.}
\end{array}\right.
\label{eq:Sigma1}
\end{align}
Defining the $\MR\times\MT$ matrix
$\mathbf{S} = -\jmath\mathbf{\Sigma} + \widetilde{\mathbf{\Phi}}$ with
$\widetilde{\mathbf{\Phi}} = \mathbf{V}^T\mathbf{\Phi}\mathbf{W}$
(and hence $\widetilde{\mathbf{\Phi}}\equidist\mathbf{\Phi}$), it follows
that $\det(\widetilde{\mathbf{\Theta}}\widetilde{\mathbf{\Theta}}^H) =
\det(\mathbf{S}\mathbf{S}^H)$.
With $\mathbf{S} = [ \, \mathbf{s}_1 \,\;\, \mathbf{s}_2 \,\;\,
\cdots \,\;\, \mathbf{s}_{\MR} \, ]^T$ being a square ($\MR = \MT$) or 
a wide matrix ($\MR < \MT$), 
a basic result in geometry
(e.g., \cite[Th.\rsq7.5.1]{Ande03b}, 
\cite[Sec.\rsp3.2.2]{Math99b},
which can be shown to hold in the complex case upon replacing transposition 
by conjugate transposition) yields
\begin{align}
  \mysqrt{\det(\mathbf{S}\mathbf{S}^H)}
= 
  \vol(P_\mathbf{S})
= 
  \|\mathbf{s}_1^\bot\| \, \|\mathbf{s}_2^\bot\| \cdots \|\mathbf{s}_{\MR}^\bot\|
\label{eq:eqF}
\end{align}
where $\vol(P_\mathbf{S})$ stands for the volume or $\MR$-content
of the parallelotope spanned by the $\MR$
row vectors of $\mathbf{S}$, 
$\mathbf{s}^\bot_1 = \mathbf{s}_1$, and
$\mathbf{s}_i^\bot$ ($i>1$) denotes the component of $\mathbf{s}_i$ orthogonal 
to the span of the vectors
$\mathbf{s}_{1}^\bot, \mathbf{s}_{2}^\bot, \ldots, \mathbf{s}_{i-1}^\bot$. 
The orthogonal vectors $\mathbf{s}_i^\bot$ ($i=2,3,\ldots,\MR$)
are obtained using Gram-Schmidt orthogonalization and are given by
\begin{align}
\mathbf{s}_i^\bot 
&= 
\left( 
    \mathbf{I}_{\MT} - \sum_{n=1}^{i-1} \frac{ \mathbf{s}^\bot_n\mathbf{s}_n^{\bot H} }{
    \|\mathbf{s}_n^\bot\|^2 } \right) \mathbf{s}_i 
 =  \mathbf{A}_i \mathbf{s}_i. 
\label{eq:GSstep}
\end{align}
It is well known that applying the decomposition \eqref{eq:eqF} to 
an \ac{IID} complex Gaussian random matrix 
$\mathbf{S}$ with $\cplxGauss(0,1)$ elements
results in independent chi-square distributed factors 
$\|\mathbf{s}^\bot_i\|^2$ ($i=1,2,\ldots,\MR$) 
\cite[Th.\rsp3.4 ff.]{Edel89t}.
The problem at hand differs, however, from the \ac{IID} complex Gaussian case in two aspects, namely the fact that
the elements in $\mathbf{\Phi}$ and hence $\widetilde{\mathbf{\Phi}}$ are real-valued Gaussian 
and the presence of the deterministic component $-\jmath\mathbf{\Sigma}$.

It follows
trivially from the definition of $\mathbf{S}$ that 
$\|\mathbf{s}_1^\bot\|^2 \equidist  \chi^2_{\MT,\sigma_{\PP}^2}+\MT\MR$.
From \eqref{eq:GSstep} we can see that, conditioned on 
$\mathbf{s}_1^\bot, \mathbf{s}_2^\bot, \ldots, \mathbf{s}_{i-1}^\bot$, 
the vectors $\mathbf{s}_i^\bot$ ($i=2,3,\ldots,\MR$)
are \ac{JG} and hence the $\|\mathbf{s}^\bot_i\|^2$ ($i=2,3,\ldots,\MR$) are chi-square distributed. 
Using the fact that $\mathbf{s}_i\in\numID{R}^{\MT}$ ($i=2,3,\ldots,\MR$) 
and $\mathbf{A}_i^H\mathbf{A}_i=\mathbf{A}_i$,
it follows immediately that
\begin{align*}
  \| \mathbf{s}_i^\bot \|^2 = \mathbf{s}_i^T \mathbf{A}_i \mathbf{s}_i, 
\quad
  i=2,3,\ldots,\MR.
\end{align*}
Next, noting that
\begin{align*}
\|\mathbf{s}_i^\bot\|^2 &= \mathbf{s}_i^T \, \big( \real(\mathbf{A}_i) 
                                          + \jmath\imag(\mathbf{A}_i) \big) \, \mathbf{s}_i \\
                        &= \mathbf{s}_i^T \, \real(\mathbf{A}_i) \, \mathbf{s}_i 
                         + \jmath \mathbf{s}_i^T \, \imag(\mathbf{A}_i) \, \mathbf{s}_i,
\quad i=2,3,\ldots,\MR
\end{align*}
has to be real-valued for all $\mathbf{s}_i$, it follows that
\begin{align}
\|\mathbf{s}_i^\bot\|^2  = \mathbf{s}_i^T \, \real(\mathbf{A}_i) \, \mathbf{s}_i 
                         = \mathbf{s}_i^T \left( \mathbf{I}_{\MT} 
                         - \sum_{n=1}^{i-1} \frac{ \real(\mathbf{s}^\bot_n\mathbf{s}_n^{\bot H}) }
                                                 { \|\mathbf{s}_n^\bot\|^2 } \right) 
                           \mathbf{s}_i.
\label{eq:eqU1}
\end{align}
Based on \eqref{eq:eqU1}, we can now invoke \lemmaref{lemma:lemma2} in the Appendix
to conclude that the eigenvalues of $\real(\mathbf{A}_i)$ are given by 
\begin{align*}
\{ \sigma_{k}^{(i)} \} = \big\{ \underbrace{1,\ldots,1}_{\MT-i} \, , \, \underbrace{0,\ldots,0}_{i-2} \, , \, 
                              \lrv^{(i)}, 1-\lrv^{(i)} \big\}, \quad k = 1,2,\ldots,\MT
\end{align*}
where $\lrv^{(i)} = \lrv^{(i)}(\mathbf{s}_1^\bot, \mathbf{s}_2^\bot, \ldots, \mathbf{s}_{i-1}^\bot)$
is a \ac{RV} with \ac{PDF} supported in the interval $[0,1]$. 
Consequently, using \cite[Eq.\rsp(4.1.1)]{MaPr92b}, we obtain
\begin{align*}
\mathbf{s}_i^T \, \real(\mathbf{A}_i) \, \mathbf{s}_i
&\equidist 
\sigma^2_{\varphi} \sum_{k=1}^{\MT} \sigma_{k}^{(i)} \, X_i =  \chi^2_{\MT-i,\sigma_{\PP}^2} + \underbrace{\sigma_{\PP}^2(\lrv^{(i)}X^{(i)}_{\MT-1}+(1-\lrv^{(i)})X^{(i)}_{\MT})}_{\equidist \, Z(\lrv^{(i)})}
\end{align*}
where the $X_i \equidist \chi^2_{1,1}$ are independent.
\end{IEEEproof}
\end{prop}

We shall next show that for $\sigma^2_{\PP}\ll1$, 
$Z(\lrv^{(i)})\apprdist\chi^2_{1,\sigma_{\PP}^2}$,
which then implies that
\begin{align}
  \|\mathbf{s}_i^\bot\|^2
\apprdist 
  \chi^2_{\MT-i,\sigma_{\PP}^2} + \chi^2_{1,\sigma_{\PP}^2} = \chi^2_{\MT-i+1,\sigma_{\PP}^2} 
\label{eq:sibot}
\end{align}
thereby allowing an approximation 
of \eqref{eq:chi2approx} as\footnote{We would like to use this chance to 
   point out that the distributional equivalence in \cite[Prop.\rsp4]{BaBo04c} 
   should be an approximate equivalence (as in \eqref{eq:chi2approxtrue}).
   Furthermore, $\cplxGauss(\mathbf{0}, \sigma_{\Phi}^2 \mathbf{I}_{\MT\MR})$
   in \cite[Prop.\rsp4 and Prop.\rsp5]{BaBo04c} should be replaced by
   $\realGauss(\mathbf{0}, \sigma_{\Phi}^2 \mathbf{I}_{\MT\MR})$.}
\begin{align}
\begin{split}
   \det(\widetilde{\mathbf{\Theta}}\widetilde{\mathbf{\Theta}}^H) 
& \apprdist
   \big( \chi^2_{\MT,\sigma_{\PP}^2} + \MT\MR \big) 
   \prod_{i=1}^{\MR-1} \chi^2_{\MT-i,\sigma_{\PP}^2},
\qquad
   \MR\le \MT 
\\
   \det(\widetilde{\mathbf{\Theta}}^H \widetilde{\mathbf{\Theta}}) 
& \apprdist
   \big( \chi^2_{\MR,\sigma_{\PP}^2} + \MT\MR \big) 
   \prod_{i=1}^{\MT-1} \chi^2_{\MR-i,\sigma_{\PP}^2},
\qquad 
   \MR>\MT.
\end{split}
\label{eq:chi2approxtrue}
\end{align}
In order to see that $Z(\lrv^{(i)})\apprdist\chi^2_{1,\sigma_{\PP}^2}$,
we start by noting that the \ac{PDF} of $Z(\lrv)$ conditional on $\lrv$ 
is given by \cite[Eq.\rsp(5.7)]{Simo02b}
\begin{align}
  \pdf_{Z|\lrv}(x)
&=
  \frac{1}{2\sigma_{\PP}^2\sqrt{\lrv(1-\lrv)}} \,
  e^{ -\frac{x}{4\sigma_{\PP}^2\lrv(1-\lrv)}} \,
  I_0\mo\left( \frac{1-2\lrv}{4\sigma_{\PP}^2\lrv(1-\lrv)} \, x \right)
\label{eq:sumchis}
\end{align}
where $I_0(z)$ is the modified Bessel function of the first kind \cite[Sec.\rsp9.6]{AbSt72b}.
For $\sigma_{\PP}^2$ small, we can invoke the large-$|z|$ expansion of 
$I_0\mo(z)$ \cite[Eq.\rsp9.7.1]{AbSt72b} according to
\begin{align*}
  I_0\mo(z)
&=
  \frac{1}{\sqrt{2\pi z}} \, e^{z} \left( 1 + \frac{1}{8z} + \frac{3^2}{2!(8z)^2} + \frac{3^2 5^2}{3!(8z)^3} + \cdots \right)
 \approx
  \frac{1}{\sqrt{2\pi z}} \, e^{z} 
\end{align*}
which, when used in \eqref{eq:sumchis},
upon renormalizing so that 
$\int_{x=0}^{\infty} \pdf_{Z|\lrv}(x) \, \ud x = 1$, yields
\begin{align*}
  \pdf_{Z|\lrv}(x)
&\approx
  \frac{1}{\sqrt{2\pi\sigma_{\PP}^2(1-\lrv) x}} \, 
  e^{ -\frac{x}{2\sigma_{\PP}^2(1-\lrv)}}
=
  \pdf_{\chi^2_{1,\sigma_{\PP}^2(1-\lrv)}}(x).
\end{align*}
This means that 
$Z|\lrv \equidist \chi^2_{1,\sigma_{\PP}^2(1-\lrv)}$ for $0<\lrv<1$ if
$\sigma_{\PP}^2$ is small. 
We shall next see that $\lrv^{(i)},\forall i,$ is small, in general, 
which then directly results in the (unconditional) pdf of $Z(\lrv^{(i)})$ satisfying 
$Z(\lrv^{(i)})\apprdist\chi^2_{1,\sigma_{\PP}^2},\forall i$.
Recall that $\{\lrv^{(i)},1-\lrv^{(i)}\}$ 
are the nonzero, nonunity eigenvalues 
of $\real(\mathbf{A}_i)$ in \eqref{eq:eqU1}.
The first pair of such eigenvalues is obtained for $i=2$. 
Due to the symmetry of the eigenvalues, 
we may investigate $\mathbf{I} - \real(\mathbf{A}_2)$ 
instead of $\real(\mathbf{A}_2)$, which, using $\mathbf{s}_1^\bot = \mathbf{s}_1$, can be written as
\begin{align*}
   \mathbf{I} - \real(\mathbf{A}_2)
&= \frac{ \real(\mathbf{s}_1\mathbf{s}_1^H) }{ \|\mathbf{s}_1\|^2 }
 = \frac{ \real(\mathbf{s}_1)\realT{(\mathbf{s}_1)} }{ \|\mathbf{s}_1\|^2 } 
 + \frac{ \imag(\mathbf{s}_1)\imagT{(\mathbf{s}_1)} }{ \|\mathbf{s}_1\|^2 }
\\
&= \frac{ \bm{\varphi} \bm{\varphi}^T }{ \|\bm{\varphi}\|^2+\|\bm{\sigma}\|^2 } 
 + \frac{ \bm{\sigma}  \bm{\sigma}^T  }{ \|\bm{\varphi}\|^2+\|\bm{\sigma}\|^2 }
\end{align*}
where $\bm{\varphi} = \real(\mathbf{s}_1)$ 
and $\bm{\sigma} = -\imag(\mathbf{s}_1) 
= [\, \mysqrt{\MT\MR} \,\;\, 0 \,\;\, 0 \,\;\, \cdots \,\;\, 0 \,]^T$.
In the following, we denote $\pr_i = [\bm{\pr}]_i$.
The nonzero eigenvalues of $\mathbf{I} - \real(\mathbf{A}_2)$ 
are equal to the eigenvalues of
\begin{align*}
   \frac{ \big[\, \bm{\varphi} \,\;\, \bm{\sigma} \,\big]^T \big[\, \bm{\varphi} \,\;\, \bm{\sigma} \,\big] }
        { \|\bm{\varphi}\|^2 + \|\bm{\sigma}\|^2 }
=
   \frac{ 1 }
        { \|\bm{\varphi}\|^2 + \|\bm{\sigma}\|^2 }
   \left[
   \begin{array}{cc}
   \|  \bm{\varphi} \|^2 & \mysqrt{\MT\MR} {\pr}_1 \\
   \mysqrt{\MT\MR} {\pr}_1 & \MT\MR
   \end{array}
   \right]
\end{align*}
given by
\begin{align}
  \{ \lrv^{(2)}, 1-\lrv^{(2)} \} 
&= 
  \frac{1}{2} \pm \frac{1}{2}\frac{ \mysqrt{ 4 \MT\MR \pr_1^2 + (\MT\MR - \|\bm{\varphi}\|^2)^2 }}
                                  { \MT\MR + \|\bm{\varphi}\|^2 }.
\label{eq:lrvEVs}
\end{align}
For $\MT$ sufficiently large, with
$\|\bm{\varphi}\|^2 = \varphi_1^2 + \varphi_\mathrm{s}^2$,
where
$\varphi_\mathrm{s}^2 = \sum_{i=2}^{\MT} \varphi_i^2$,
we can replace \eqref{eq:lrvEVs} by 
\begin{align*}
  \{ \lrv^{(2)}, 1-\lrv^{(2)} \} 
&\approx 
  \frac{1}{2} \pm \frac{1}{2}\frac{ \MT\MR - \varphi_\mathrm{s}^2 }{ \MT\MR + \varphi_\mathrm{s}^2 }.
\end{align*}
Next, since $\varphi_\mathrm{s}^2$ is small compared to $\MT\MR$, 
we obtain the first-order Taylor series expansion
\begin{align*}
  \{ \lrv^{(2)}, 1-\lrv^{(2)} \} 
&\approx 
  \frac{1}{2} \pm \frac{1}{2}\left(1 - 2\frac{ \varphi_\mathrm{s}^2 }{ \MT\MR }\right)
\equidist
  \big\{ \chi^2_{\MT-1,\sigma^2_\varphi/(\MT\MR)}, 1 - \chi^2_{\MT-1,\sigma^2_\varphi/(\MT\MR)} \big\}.
\end{align*}
Hence, we have
\begin{align*}
  \expect\{ \lrv^{(2)} \}
&=
  \sigma^2_\varphi \frac{\MT-1}{\MT\MR}
&
  \var\{ \lrv^{(2)} \}
&=
  2 \sigma^4_\varphi \frac{\MT-1}{\MT^2 \MR^2}
\end{align*}
which shows that, for sufficiently large $\MT,\MR$, $\lrv^{(2)}$ is indeed small.
For $\MR\times1$ and $1\times\MT$ systems, i.e., 
for \ac{SIMO} and \ac{MISO} systems, respectively,
we can therefore immediately conclude that the 
approximation \eqref{eq:chi2approxtrue} is very accurate. 
In the case of general $\MT$ and $\MR$, it seems difficult 
to prove that $\lrv^{(i)} \approx 0$ for $i\ge3$.
We do, however, have strong numerical evidence that
this is, indeed, the case.

\begin{figure}[htbp]
\centering
\psfrag{16  x  16, 2G RMS }{\scriptsize{$16\times16$, 2\degreel\ \acs{RMS}}}
\psfrag{16  x  16, 4G RMS }{\scriptsize{$16\times16$, 4\degreel\ \acs{RMS}}}
\psfrag{16  x  16, 7G RMS }{\scriptsize{$16\times16$, 7\degreel\ \acs{RMS}}}
\psfrag{4  x  4, 2G RMS }{\scriptsize{$4\times4$, 2\degreel\ \acs{RMS}}}
\psfrag{4  x  4, 4G RMS }{\scriptsize{$4\times4$, 4\degreel\ \acs{RMS}}}
\psfrag{4  x  4, 7G RMS }{\scriptsize{$4\times4$, 7\degreel\ \acs{RMS}}}
\psfrag{16  x  16, 3.5G RMS }{\scriptsize{$16\times16$, 3.5\degreel\ \acs{RMS}}}
\psfrag{4  x  8, 3.5G RMS }{\scriptsize{$4\times8$, 3.5\degreel\ \acs{RMS}}}
\psfrag{4  x  8, 7G RMS }{\scriptsize{$4\times8$, 7\degreel\ \acs{RMS}}}
\psfrag{4  x  4, 3.5G RMS }{\scriptsize{$4\times8$, 3.5\degreel\ \acs{RMS}}}
\psfrag{Proposition xyz}{\scriptsize{Approx.\ \eqref{eq:chi2approxtrue}}}
\psfrag{MC, linear approx}{\scriptsize{$\log\,\det(\tilde{\mathbf{\Theta}}\tilde{\mathbf{\Theta}}^H)$}}
\psfrag{MC}{\scriptsize{$\log\,\det(\mathbf{\Theta}\mathbf{\Theta}^H)$}}
\psfrag{log2 det (TTH)}{\scriptsize{$\log\,\det(\tilde{\mathbf{\Theta}}\tilde{\mathbf{\Theta}}^H)$}}
\psfrag{CDF}{\scriptsize{\acs{CDF}}}
\icg{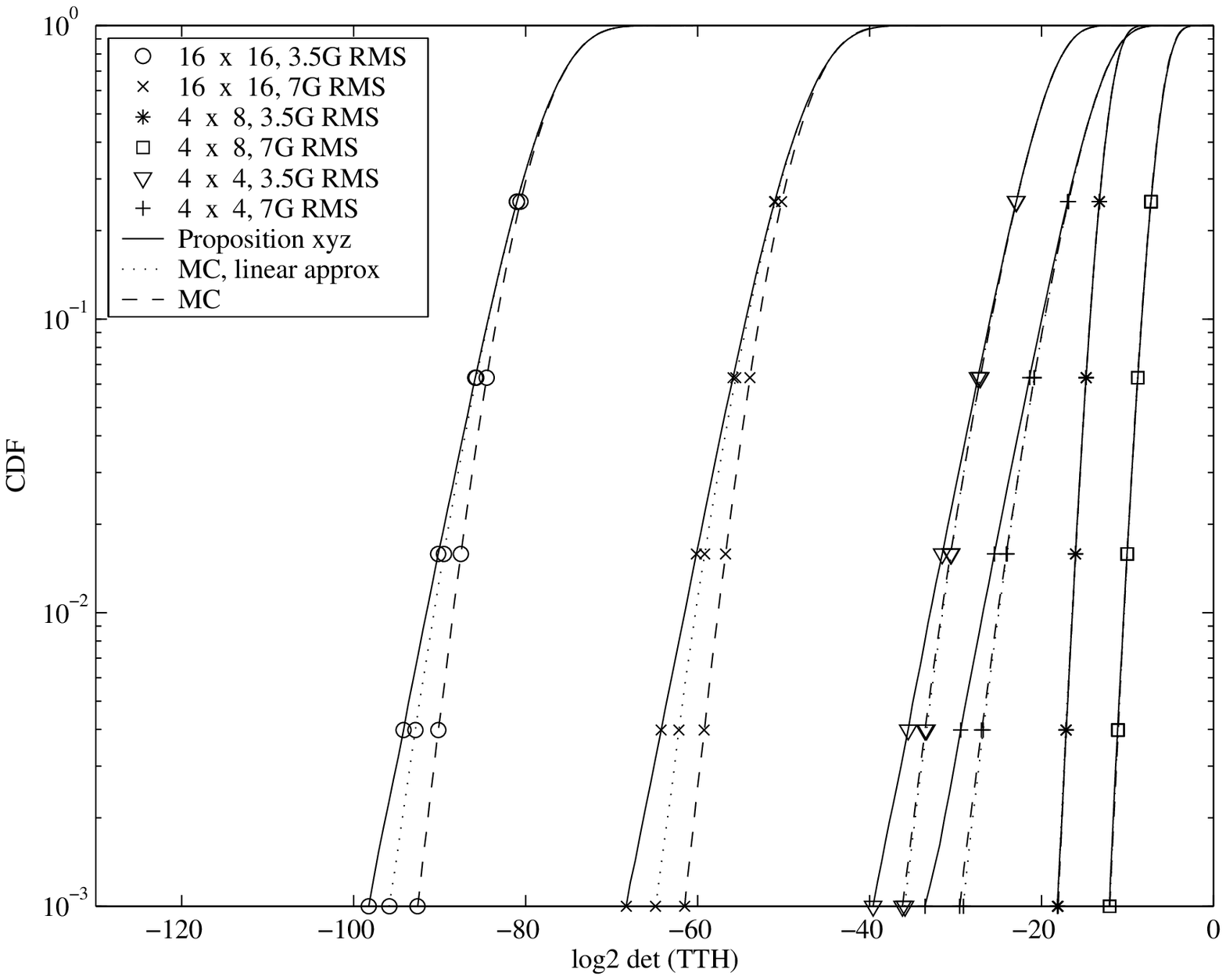}
\vspace{-2mm}
\caption{\label{fig:total}
\acs{CDF} of $\log\,\det(\tilde{\mathbf{\Theta}}\tilde{\mathbf{\Theta}}^H)$
and analytic approximation \eqref{eq:chi2approxtrue},
for various $\MT,\MR$ and different phase noise levels.
For comparison, $\log\,\det(\mathbf{\Theta}\mathbf{\Theta}^H)$ is also shown.
A regular sounding pattern according to \eqref{eq:reg_pat} was assumed.
All results are based on 300\,000 Monte Carlo runs.}
\end{figure}

Recalling that $\sigma_{\PP}^2\approx0.0149$ for the worst-case 
phase noise value of 7\degreel\ \ac{RMS}, 
we can conclude that the assumption $\sigma_{\PP}^2\ll1$ 
made in \propref{prop:dist_approx} and in \eqref{eq:chi2approxtrue}
is very well satisfied in practice.
\figref{fig:total} shows the \ac{CDF} of 
$\log\,\det(\widetilde{\mathbf{\Theta}}\widetilde{\mathbf{\Theta}}^H)$
corresponding to the approximation \eqref{eq:chi2approxtrue} 
along with the exact \ac{CDF}\footnote{Note that ``exact \ac{CDF}''
   means exact under the linear phase noise approximation.}
(in both cases obtained through Monte Carlo methods). 
We observe that the approximation is excellent in general and, 
indeed, becomes better for smaller 
$\sigma^2_{\PP}$ and/or for less symmetric (in terms of the number
of transmit and receive antennas) configurations. 
We finally note that comparing \eqref{eq:chi2approxtrue} 
to \cite[Eq.\rsp(3)]{HaBo04c} suggests that 
for fully uncorrelated phase noise with $\sigma_{\PP}^2\ll1$,
the effective \ac{MIMO} channel behaves like 
a physical \ac{MIMO} channel consisting of
a rank-1 Ricean component plus an \ac{IID} Rayleigh fading component with the
difference that in our case the chi-square \acp{RV} have 
half the order of those in \cite{HaBo04c}
(reflecting the fact that here we are dealing with real-valued Gaussian \acp{RV}).

We proceed by stating the result on the exact distribution of 
$\det(\widetilde{\mathbf{\Theta}}\widetilde{\mathbf{\Theta}}^H)$. 

\begin{prop}
\label{prop:dist_exact}
For a separable timing matrix $\mathbf{T}$ 
and under the small phase noise approximation 
$\sigma_\PP^2\ll1$ so that
$\widetilde{\mathbf{\Theta}}=\mathbf{1}+\jmath\mathbf{\Phi}$,
assuming fully uncorrelated phase noise, i.e., 
$\vect(\mathbf{\Phi}) \equidist \realGauss(\mathbf{0}, 
\sigma_{\PP}^2 \mathbf{I}_{\MT\MR})$, we have
\begin{align}
\begin{split}
  \det(\widetilde{\mathbf{\Theta}}\widetilde{\mathbf{\Theta}}^H) 
&\equidist
  \Bigg( \chi^2_{M_\DX+1,\sigma_{\PP}^2} 
         + \MT\MR \, \beta\mo\bigg(\frac{M_\DX+1}{2},\frac{\MR-1}{2}\bigg) \Bigg)
  \prod_{i=1}^{\MR-1} \chi^2_{\MT-i+1,\sigma_{\PP}^2},
\quad 
  \MR \le \MT 
\\
  \det(\widetilde{\mathbf{\Theta}}^H \widetilde{\mathbf{\Theta}}) 
&\equidist
  \Bigg( \chi^2_{M_\DX+1,\sigma_{\PP}^2} 
         + \MT\MR \, \beta\mo\bigg(\frac{M_\DX+1}{2},\frac{\MT-1}{2}\bigg) \Bigg)
  \prod_{i=1}^{\MT-1} \chi^2_{\MR-i+1,\sigma_{\PP}^2}, 
\quad 
  \MR > \MT
\end{split}
\label{eq:chi2exact} 
\end{align}
where $M_\DX = |\MT-\MR|$,
and the $\chi^2_{n,\sigma^2}$ are statistically independent.

\begin{IEEEproof}
Again, we provide the proof for $\MR\le \MT$ only. The
case $\MR>\MT$ follows exactly the same line of reasoning. 
We start by noting that the matrix $\mathbf{S}$ 
defined in the proof of \propref{prop:dist_approx}
is unitarily equivalent to the matrix 
$\mathbf{S}'=-\jmath\mathbf{\Sigma}'+\widetilde{\mathbf{\Phi}}'$, 
where the $\MR\times\MT$ matrix $\mathbf{\Sigma}'$ is given by
\begin{align*}
[\mathbf{\Sigma}']_{\vrow,\vcol} 
= \left\{ \begin{array}{ll}
\mysqrt{\MT\MR}, & \vrow = \vcol = \MR \\
0              , & \text{else}
\end{array}\right.
\end{align*}
and 
$\widetilde{\mathbf{\Phi}}' \equidist \widetilde{\mathbf{\Phi}} \equidist \mathbf{\Phi}$.
In what follows, 
we shall work with $\mathbf{S}'$ and, by slight abuse of notation, denote it as $\mathbf{S}$.
The \ac{PDF} of $\|\mathbf{s}_1^\bot\|^2$ follows
trivially from the definition of $\mathbf{S}$ and is given by $\chi^2_{\MT,\sigma_{\PP}^2}$.
Applying Gram-Schmidt orthogonalization, due to the nonzero entry in $\mathbf{\Sigma}'$
being at position $\vrow = \vcol = \MR$, we can conclude that for $i = 2,3,\ldots,\MR-1$, 
the matrix $\real(\mathbf{A}_i)$ has only two distinct eigenvalues,
namely 0 with multiplicity $i-1$ and 1 with multiplicity $\MT-i+1$. 
Consequently, 
\eqref{eq:eqU1} implies that $\|\mathbf{s}_i^\bot\|^2$ $(i=1,2,\ldots,\MR-1)$ conditioned on
$\mathbf{s}_1^\bot,\mathbf{s}_2^\bot,\ldots,\mathbf{s}_{i-1}^\bot$ 
is distributed as $\chi^2_{\MT-i+1,\sigma_{\PP}^2}$. 
Since the eigenvalues of $\real(\mathbf{A}_i)$ 
do not depend on $\mathbf{s}_1^\bot,\mathbf{s}_2^\bot,\ldots,\mathbf{s}_{i-1}^\bot$
and the statistics of $\| \mathbf{s}_i^\bot \|^2$ depend on $\mathbf{A}_i$ only
through the eigenvalues of $\mathbf{A}_i$, we can conclude that 
the unconditional distribution of $\|\mathbf{s}_i^\bot\|^2$ satisfies
$\|\mathbf{s}_i^\bot\|^2 
\equidist \chi^2_{\MT-i+1,\sigma_{\PP}^2}$ ($i=1,2,\ldots,\MR-1$).
For $i=\MR$, noting that $\mathbf{A}_{\MR}$ is
a real-valued matrix, we have
\begin{align}
\| \mathbf{s}_{\MR}^\bot \|^2
= \| \mathbf{A}_{\MR} \real(\mathbf{s}_{\MR}) \|^2 + \| \mathbf{A}_{\MR} \imag(\mathbf{s}_{\MR}) \|^2.
\label{eq:2term}
\end{align}
The distribution of the first term on the \ac{RHS} of \eqref{eq:2term} can be shown, using the same line of
reasoning as for $i=1,2,\ldots,\MR-1$, to satisfy 
$\| \mathbf{A}_{\MR} \real(\mathbf{s}_{\MR}) \|^2 \equidist \chi^2_{\MT-\MR+1,\sigma_{\PP}^2}$.
The second term on the \ac{RHS} of \eqref{eq:2term} can be expanded as
\begin{align*}
  \| \mathbf{A}_{\MR} \imag(\mathbf{s}_{\MR}) \|^2
&= 
  \imagT[big]{(\mathbf{s}_{\MR})} \, \mathbf{A}_{\MR}^T \mathbf{A}_{\MR} \, \imag(\mathbf{s}_{\MR}) \\
&= 
  \imagT[big]{(\mathbf{s}_{\MR})} \, \mathbf{A}_{\MR} \, \imag(\mathbf{s}_{\MR})
\end{align*}
where we made use of the fact that $\mathbf{A}_{\MR}$ is real-valued and hence
$\mathbf{A}_{\MR}^H \mathbf{A}_{\MR}=\mathbf{A}_{\MR}$ reduces to  
$\mathbf{A}_{\MR}^T \mathbf{A}_{\MR}=\mathbf{A}_{\MR}$. Next, we note that
\begin{align*}
  \mathbf{A}_{\MR} 
&= 
  \mathbf{I}_{\MT} - \mathbf{G} \mathbf{G}^T \\
&= 
  [ \, \mathbf{G} \,\;\, \mathbf{K} \, ] 
  \left[ 
  \begin{array}{cc}\mathbf{0}&\mathbf{0}\\\mathbf{0}&\mathbf{I}_{\MT-\MR+1}\end{array}
  \right] [ \, \mathbf{G} \,\;\, \mathbf{K} \, ]^T \\
&= 
  \sum_{n=\MR}^{\MT} \mathbf{u}_n \mathbf{u}_n^T
\end{align*}
with 
\begin{align*}
\mathbf{G}  = \left[ \, \frac{\mathbf{s}_1^\bot}{\|\mathbf{s}_1^\bot\|} \,\;\,
                        \frac{\mathbf{s}_2^\bot}{\|\mathbf{s}_2^\bot\|} \,\;\, \cdots \,\;\,
                        \frac{\mathbf{s}_{\MR-1}^\bot}{\|\mathbf{s}_{\MR-1}^\bot\|} \, \right] 
\quad \text{and} \quad
\mathbf{K}  = \left[ \, \mathbf{u}_{\MR} \,\;\, 
                        \mathbf{u}_{\MR+1} \,\;\, 
                        \cdots \,\;\,
                        \mathbf{u}_{\MT} \, \right]
\end{align*}
and the vectors $\mathbf{u}_n$ ($n=\MR,\MR+1,\ldots,\MT$) have to be chosen such that the matrix 
$\mathbf{U} = [ \, \mathbf{G} \,\;\, \mathbf{K} \, ]$ satisfies 
$\mathbf{U}\mathbf{U}^T = \mathbf{G}\mathbf{G}^T + \mathbf{K}\mathbf{K}^T = \mathbf{I}$.
Recognizing that the vectors $\mathbf{s}_i^\bot/\|\mathbf{s}_i^\bot\|$ $(i = 1,2,\ldots,\MR-1)$ 
are obtained by applying the Gram-Schmidt procedure 
to the real-valued $(\MR-1)\times\MT$ 
\ac{IID} Gaussian matrix 
$[ \, \mathbf{s}_1 \,\;\, \mathbf{s}_2 \,\;\, \cdots \,\;\, \mathbf{s}_{\MR-1} \, ]^T$
with zero-mean entries, we can take the stacked matrix 
$\mathbf{U} = [ \, \mathbf{G} \,\;\, \mathbf{K} \, ]$ 
to be given by the Q-matrix
obtained by applying the QR-decomposition 
to an $\MT\times\MT$ \ac{IID} real-valued Gaussian matrix with zero-mean entries. 
Note that using the Gram-Schmidt procedure for QR-decomposition
yields the unique factorization characterized by
positive entries on the main diagonal of the R-matrix
\cite[Th.\rsp2.6.1]{HoJo85b}.
Next, realizing that
\begin{align*}
  \| \mathbf{A}_{\MR} \imag(\mathbf{s}_{\MR}) \|^2
&= 
  \imagT[big]{(\mathbf{s}_{\MR})} 
  \left( \sum_{n=\MR}^{\MT} \mathbf{u}_n \mathbf{u}_n^T \right) \imag(\mathbf{s}_{\MR}) \\
&= 
  \MT\MR \sum_{n=\MR}^{\MT} [\mathbf{U}]^2_{\MR,n}
\end{align*}
the proof is complete upon deriving the \ac{PDF} of 
$\sum_{n=\MR}^{\MT} [\mathbf{U}]^2_{\MR,n}$.
It is well known that, applying any procedure for QR-decomposition 
leading to the unique factorization
where the elements on the main diagonal of the 
R-matrix are positive, the resulting Q-matrix $\mathbf{Q}$
is distributed such that 
$\mathbf{A}\mathbf{Q}\mathbf{B}\equidist\mathbf{Q}$ 
for any orthonormal\footnote{The matrix $\mathbf{A}$ is said to be orthonormal 
if $\mathbf{A}\mathbf{A}^T=\mathbf{I}$.}
$\mathbf{A}$ and $\mathbf{B}$ 
\cite[Th.\rsq3.2]{Stew80j}. 
Choosing $\mathbf{A}$ and $\mathbf{B}$ to be permutation matrices, we can conclude that
the rows and columns of $\mathbf{Q}$, and hence $\mathbf{U}$ in our case,
are all equally distributed. Now, the quantity we are interested in is the sum of
squares of the elements $\{ \MR,\MR+1,\ldots,\MT \}$ in any such row or column. 
Specifically, if the
Gram-Schmidt procedure is used to obtain the QR-decomposition, the first
column of $\mathbf{U}$ is given explicitly as $\mathbf{s}_1/\|\mathbf{s}_1\|$. 
From \cite[Def.\rsp1.4]{FaKN90b}
we know that the quantities $[\mathbf{s}_1]_n^2/\|\mathbf{s}_1\|^2$
are jointly Dirichlet distributed, i.e., 
\begin{align*}
\mathbf{s} =
\left[ \, 
       \frac{[\mathbf{s}_1]_1^2}{\|\mathbf{s}_1\|^2} \,\;\,
       \frac{[\mathbf{s}_1]_2^2}{\|\mathbf{s}_1\|^2} \,\;\,
       \cdots                                        \,\;\,
       \frac{[\mathbf{s}_1]_{\MT}^2}{\|\mathbf{s}_1\|^2} \, \right]
\equidist D_{\MT}\mo\left( \frac{1}{2}, \frac{1}{2}, \ldots, \frac{1}{2} \right).
\end{align*}
Partitioning $\mathbf{s}$ into subvectors of length $\MT-\MR+1$ and $\MR-1$, respectively,
and employing \cite[Th.\rsq1.4 and Th.\rsq1.5]{FaKN90b}
(reproduced as \thref{theorem:dirichlet1} and \thref{theorem:dirichlet2},
respectively, in the Appendix for convenience), it follows that
\begin{align*}
\MT\MR\sum_{n=\MR}^{\MT} [\mathbf{U}]_{\MR,n}^2
&\equidist
\MT\MR \, \beta\mo\left(\frac{\MT-\MR+1}{2},\frac{\MR-1}{2}\right)
\end{align*}
where $\beta(a,b)$ is a beta-distributed \ac{RV} with parameters $a$ and $b$
as defined in the Notations section.
\end{IEEEproof}
\end{prop}

Note that even though the results 
in \eqref{eq:chi2approxtrue} and \propref{prop:dist_exact} 
have a striking similarity and \eqref{eq:chi2approxtrue} 
provides an approximation for the exact
result in \propref{prop:dist_exact},
it seems difficult to derive \eqref{eq:chi2approxtrue}
directly from \propref{prop:dist_exact}.

We are now ready 
to state an analytic lower bound on the ergodic capacity of an effective channel
resulting from a rank-1 physical channel with unit-modulus entries
subject to fully uncorrelated phase noise.

\begin{prop}
\label{prop:CuZF04}
For a separable timing matrix $\mathbf{T}$ 
and under the small phase noise approximation $\sigma_\PP^2\ll1$ so that
$\widetilde{\mathbf{\Theta}}=\mathbf{1}+\jmath\mathbf{\Phi}$ with
$\vect(\mathbf{\Phi}) \equidist \realGauss(\mathbf{0}, \sigma_{\PP}^2 \mathbf{I}_{\MT \MR})$,
assuming that $\mathbf{H} = \mathbf{g} \mathbf{h}^T$ with 
$|[\mathbf{g}]_i|=1$ $(i=1,2,\ldots,\MR)$ and 
$|[\mathbf{h}]_i|=1$ $(i=1,2,\ldots,\MT)$, 
the ergodic capacity of the effective channel $\widehat{\mathbf{H}}$ satisfies
\begin{align}
\begin{split}
  \widehat{C} 
&\geq 
  \log \left( 1 + \sum_{n=1}^{\MR} \left(\frac{\rho}{\MT}\right)^n \binom{\MR}{n} \, 
  \prod_{i=0}^{n-1} \left( \delta_i n \MT  + 2 \sigma^2_{\varphi} 
                         e^{\digamma\left( \frac{\MT-i}{2} \right)} \right)\right),
\quad \MR\le\MT
\\
  \widehat{C} 
&\geq 
  \log \left( 1 + \sum_{n=1}^{\MT} \left(\frac{\rho}{\MT}\right)^n \binom{\MT}{n} \,
  \prod_{i=0}^{n-1} \left( \delta_i n \MR  
              + 2 \sigma^2_{\varphi} e^{\digamma\left( \frac{\MR-i}{2} \right)} \right)\right),
\quad \MR>\MT.
\end{split}
\label{eq:coolbound}
\end{align}

\begin{IEEEproof}
We provide the proof for $\MR\le \MT$ only. 
The case $\MR>\MT$ follows exactly the same line of reasoning. 
We start by using \lemmaref{lemma:prod} and noting that our assumptions imply that
\begin{align}
  \hat{I}
&= 
  \log \, \det \left( \mathbf{I} + \frac{\rho}{\MT} \widehat{\mathbf{H}} \widehat{\mathbf{H}}^H \right)
= \log \, \det \left( \mathbf{I} + \frac{\rho}{\MT} \widetilde{\mathbf{\Theta}} \widetilde{\mathbf{\Theta}}^H \right) 
\label{eq:logdetTT}
\\
&= 
  \log \, \det \left( \mathbf{I} + \frac{\rho}{\MT} \mathbf{S} \mathbf{S}^H \right)
\nonumber
\end{align}
where $\mathbf{S} = -\jmath\mathbf{\Sigma}+\widetilde{\mathbf{\Phi}}$ 
was defined in the proof of \propref{prop:dist_approx}.
Next, using \cite[Eq.\rsp(25)]{ZhCL05j}, it follows that
\begin{align}
  \widehat{C} 
&\ge 
  \log\left( 1 + \sum_{i=1}^{\MR} \left(\frac{\rho}{\MT}\right)^i 
  \sum_{l_1<l_2<\ldots<l_i} e^{
    \expect\big\{ \ln\,\det\big( (\mathbf{S}\mathbf{S}^H)_{l_1<l_2<\ldots<l_i} \big) \big\}
  } \right)
\label{eq:Zhang}
\end{align}
where $(\mathbf{S}\mathbf{S}^H)_{l_1<l_2<\ldots<l_i}$ denotes the submatrix of
$\mathbf{S}\mathbf{S}^H$ obtained by retaining the rows $l_1<l_2<\ldots<l_i$ and 
the columns $l_1<l_2<\ldots<l_i$. The summation in \eqref{eq:Zhang} is over all
ordered tuples $(l_1,l_2,\ldots,l_i)$
chosen from the set $\{1,2,\ldots,\MR\}$.
Next, we note that the \ac{PDF} of $\det\big( (\mathbf{S}\mathbf{S}^H)_{l_1<l_2<\ldots<l_i} \big)$
follows in a straightforward fashion from the results 
developed in the proof of \propref{prop:dist_approx}.
In particular, $\det\big( (\mathbf{S}\mathbf{S}^H)_{l_1<l_2<\ldots<l_i} \big)$ is the determinant
of the matrix $\widetilde{\mathbf{S}}\widetilde{\mathbf{S}}^H$ where the $i\times\MT$
matrix $\widetilde{\mathbf{S}}$ is obtained from $\mathbf{S}$ by retaining the rows
$\{l_1,l_2,\ldots,l_i\}$. Distinguishing between the terms, in the summation over
$l_1<l_2<\ldots<l_i$ on the \ac{RHS} of \eqref{eq:Zhang}, that have $l_1=1$ and those
where $l_1>1$, we obtain
\begin{align}
  \expect\big\{ \ln\,\det\big( (\mathbf{S}\mathbf{S}^H)_{l_1<l_2<\ldots<l_i} \big) \big\}
&\stackrel{\text{(a)}}{\ge}
  \expect\left\{ \ln\left( \big( \chi^2_{\MT,\sigma_{\PP}^2} + \MT\MR \big) 
                 \prod_{l=1}^{i-1} \chi^2_{\MT-l,\sigma_{\PP}^2} \right) \right\}
\label{eq:Zd1}
\end{align}
in the former case and
\begin{align}
  \expect\big\{ \ln\,\det\big( (\mathbf{S}\mathbf{S}^H)_{l_1<l_2<\ldots<l_i} \big) \big\}
&=
  \expect\left\{ \ln\left(\prod_{l=0}^{i-1}\chi^2_{\MT-l,\sigma_{\PP}^2}\right) \right\}
\label{eq:Zd2}
\end{align}
in the latter case, where (a) is obtained as follows.
Recognizing that 
$\det((\mathbf{S}\mathbf{S}^H)_{l_1<l_2<\ldots<l_i})
=\det((\widetilde{\mathbf{\Theta}}\widetilde{\mathbf{\Theta}}^H)_{l_1<l_2<\ldots<l_i})$
and applying \eqref{eq:chi2approxtrue} 
properly modified to account for
the fact that we are interested in the submatrix of
$\widetilde{\mathbf{\Theta}}\widetilde{\mathbf{\Theta}}^H$
obtained by retaining the rows
$\{l_1<l_2<\ldots<l_i\}$ in $\widetilde{\mathbf{\Theta}}$ 
would yield an approximate expression for the \ac{LHS}
in \eqref{eq:Zd1}.
However, using \propref{prop:dist_approx} 
and invoking \propref{prop:E_Z} in the Appendix,
we can show that the lower bound in \eqref{eq:Zd1} holds firmly. 
Specifically, starting from \eqref{eq:chi2approx} 
and setting
$\Elchi_{\MT,\sigma_{\PP}^2}' 
= \expect\{ \ln( \MT\MR + \chi^2_{\MT,\sigma^2_{\PP}}) \}$,
for brevity, we can rewrite the \ac{LHS} of \eqref{eq:Zd1} as
\begin{align}
  \expect\big\{ \ln\,\det\big( (\mathbf{S}\mathbf{S}^H)_{l_1<l_2<\ldots<l_i} \big) \big\}
&= 
  \Elchi_{\MT,\sigma_{\PP}^2}'
+ \sum_{l=2}^{i} \expect_{X_l, X_{1},X_{2},\lrv^{(l)}}\big\{ \ln\big( 
    \underbrace{\chi^2_{\MT-l,\sigma_{\PP}^2}}_{X_l} + Z( \lrv^{(l)} ) \big) \big\} 
\label{eq:Zd1exact}
\\
\begin{split}
&= 
  \Elchi_{\MT,\sigma_{\PP}^2}'
+ \sum_{l=2}^{i} \expect_{X_l}\expect_{\lrv^{(l)}}\expect_{X_1,X_2|X_l,\lrv^{(l)}}
                 \big\{ \ln \big( X_l + Z( \lrv^{(l)} ) \big) \big\}
\\
&\stackrel{\text{(a)}}{\ge} 
  \Elchi_{\MT,\sigma_{\PP}^2}'
+ \sum_{l=2}^{i} \expect_{X_l}\expect_{Y|X_l}
  \big\{ \ln \big( X_l + \underbrace{\chi^2_{1,\sigma_{\PP}^2}}_{Y} \big) \big\}
\\
&= 
  \Elchi_{\MT,\sigma_{\PP}^2}'
+ \sum_{l=1}^{i-1} \expect\big\{ \ln\big( \chi^2_{\MT-l,\sigma_{\PP}^2} \big) \big\}
\end{split}
\nonumber
\end{align}
where (a) follows from \propref{prop:E_Z} in the Appendix.
The relation in \eqref{eq:Zd2} is obtained
in exactly the same fashion
upon noting that the term $-\jmath\mysqrt{\MT\MR}$ 
is absent in the sets $(l_1,l_2,\ldots,l_i)$ where $l_1>1$.
The number of terms in the first group (where $l_1=1$) is given 
by $\binom{\MR-1}{i-1}$ whereas the number of terms in the second group
is $\binom{\MR}{i}-\binom{\MR-1}{i-1}$.
It remains to find analytic expressions for the \ac{RHS} of \eqref{eq:Zd1} and
of \eqref{eq:Zd2}. 
It is well known \cite{Lee97b}
that $\expect\{ \ln( \chi^2_{n,\sigma^2_{\PP}}) \} 
= \ln (2 \sigma^2_{\PP}) + \digamma( n/2 ) \defas \Elchi_{n,\sigma_{\PP}^2}$. 
The term $\Elchi_{\MT,\sigma_{\PP}^2}' = 
\expect\{ \ln( \chi^2_{\MT,\sigma_{\PP}^2} + \MT\MR) \}$ has a closed-form
analytic expression in terms of the generalized exponential integral
$E_\nu(z) = \int_1^\infty t^{-\nu} e^{-z t} \, \ud t$, $\real(z)>0$.
For our purposes, we shall, however, be content with 
a simple lower bound obtained by applying 
Jensen's inequality
to the function $f(x) = \ln(e^x + a)$,
which results in
\begin{align}
\begin{split}
  \Elchi_{\MT,\sigma_{\PP}^2}'
&\ge
  \ln\Big( e^{\expect\big\{ \ln\big( \chi^2_{\MT,\sigma_{\PP}^2} \big) \big\}} + \MT\MR \Big) \\
&= 
  \ln\Big( e^{\Elchi_{\MT,\sigma_{\PP}^2}} + \MT\MR \Big).
\end{split}
\label{eq:JBterm1}
\end{align}
Putting the pieces together, we get
\begin{align*}
  e^{\expect\big\{ \ln\,\det\big( (\mathbf{S}\mathbf{S}^H)_{l_1<l_2<\ldots<l_i} \big) \big\}}
&\ge
  \Big( \MT\MR + e^{\Elchi_{\MT,\sigma_{\PP}^2}} \Big)
  e^{ \sum_{l=1}^{i-1} \Elchi_{\MT-l,\sigma_{\PP}^2}  } \\
&=
  \MT\MR e^{\sum_{l=1}^{i-1} \Elchi_{\MT-l,\sigma_{\PP}^2} } 
+ e^{\sum_{l=0}^{i-1} \Elchi_{\MT-l,\sigma_{\PP}^2} } 
\end{align*}
for \eqref{eq:Zd1} and
\begin{align*}
  e^{\expect\big\{ \ln\,\det\big( (\mathbf{S}\mathbf{S}^H)_{l_1<l_2<\ldots<l_i} \big) \big\}}
&=
  e^{ \sum_{l=0}^{i-1} \Elchi_{\MT-l,\sigma_{\PP}^2} }
\end{align*}
for \eqref{eq:Zd2}. 
Combining our results and noting
that the term 
$\exp\big( \sum_{l=0}^{i-1} \Elchi_{\MT-l,\sigma_{\PP}^2} \big)$
occurs in both cases so that its
total number of occurences is $\binom{\MR}{i}$,
finally yields
\begin{align}
\begin{split}
&\sum_{l_1<l_2<\ldots<l_i} e^{\expect\big\{\ln\,\det\big( (\mathbf{S}\mathbf{S}^H)_{l_1<l_2<\ldots<l_i} \big)\big\}} 
\\
&\qquad\qquad
\ge 
  \binom{\MR-1}{i-1}\MT\MR e^{ \sum_{n=1}^{i-1} \Elchi_{\MT-n,\sigma_{\PP}^2} } 
+ \binom{\MR}{i}           e^{ \sum_{n=0}^{i-1} \Elchi_{\MT-n,\sigma_{\PP}^2} }
\end{split}
\label{eq:multiplicity}
\end{align}
which, upon inserting into \eqref{eq:Zhang} and reorganizing terms, 
concludes the proof.
\end{IEEEproof}
\end{prop}

The result in \eqref{eq:coolbound} can be made more explicit 
by using the simplifications for the digamma function 
at positive integer multiples of $1/2$ given by \eqref{eq:dig_simpl}.
Furthermore, we note that \propref{prop:CuZF04} 
can be generalized to the cases where 
i)  $|[\mathbf{h}]_i|=1$, $\forall\,i$, $\mathbf{g}$ is general and $\MR\le\MT$ and
ii) $|[\mathbf{g}]_i|=1$, $\forall\,i$, $\mathbf{h}$ is general and $\MR>\MT$.
The corresponding results are stated, without proof, as
\begin{align*}
\begin{split}
  \text{i)}\quad
  \widehat{C} 
&\geq 
  \log \left( 1 + \sum_{n=1}^{\MR} \left(\frac{\rho}{\MT}\right)^n K_n(\mathbf{g}) \, 
  \prod_{i=0}^{n-1} \left( \delta_i n \MT  
              + 2 \sigma^2_{\PP} e^{\digamma\left( \frac{\MT-i}{2} \right)} \right)\right),
\quad \MR\le\MT
\\
  \text{ii)}\quad
  \widehat{C} 
&\geq 
  \log \left( 1 + \sum_{n=1}^{\MT} \left(\frac{\rho}{\MT}\right)^n K_n(\mathbf{h}) \,
  \prod_{i=0}^{n-1} \left( \delta_i n \MR  
              + 2 \sigma^2_{\PP} e^{\digamma\left( \frac{\MR-i}{2} \right)} \right)\right),
\quad \MR>\MT
\end{split}
\end{align*}
where $K_n(\mathbf{x}) 
= \sum_{\mathbf{s}\in\set{S}_{n,l\mo(\mathbf{x})}} \prod_{i=1}^{n} |[\mathbf{x}]_{s_i}|^2$,
$\set{S}_{k,m}$ is the set of all possible ordered $k$-tuples 
$\mathbf{s} = (s_1,s_2,\ldots,s_k)$ with $1\le s_1<s_2<\ldots<s_k\le m$,
and $l\mo(\mathbf{x})$ is the number of elements in the vector $\mathbf{x}$.

Again assuming 
$|[\mathbf{g}]_i|=1$ $(i=1,2,\ldots,\MR)$ and
$|[\mathbf{h}]_i|=1$ $(i=1,2,\ldots,\MT)$,
further lower-bounding \eqref{eq:coolbound} by ignoring the first term inside
the ``log'' and retaining only the highest-order (in $\rho$) term yields
\begin{align*}
\begin{split}
  \widehat{C} 
&\ge
  \MR \, \log \left(\frac{\rho}{\MT}\right) 
+ \log \left( \prod_{i=0}^{\MR-1} \left( \delta_i \MT \MR 
                + 2 \sigma^2_{\PP} e^{\digamma\left( \frac{\MT-i}{2} \right)} \right)\right),
\quad \MR\le\MT
\\
  \widehat{C} 
&\ge
  \MT \, \log \left(\frac{\rho}{\MT}\right) + \log \left( 
  \prod_{i=0}^{\MT-1} \left( \delta_i \MT \MR  
                + 2 \sigma^2_{\PP} e^{\digamma\left( \frac{\MR-i}{2} \right)} \right)\right),
\quad \MR>\MT
\end{split}
\end{align*}
which clearly shows that the effective channel has full rank 
and hence its multiplexing gain is given by $\min(\MT,\MR)$. 
Put differently, phase noise can cause a rank-1 physical channel
to appear like a full-rank channel.
In Section VII, we shall show, based on measurement results, 
that significant rank increase does, indeed, occur in practice.

We shall next provide 
a slightly looser (than \eqref{eq:coolbound})
lower bound on $\widehat{C}$ 
with a simpler structure.

\begin{prop}
\label{prop:Cbound}
For a separable timing matrix $\mathbf{T}$ and 
under the small phase noise approximation $\sigma_\PP^2\ll1$
so that
$\widetilde{\mathbf{\Theta}}=\mathbf{1}+\jmath\mathbf{\Phi}$
with $\vect(\mathbf{\Phi}) \equidist \realGauss(\mathbf{0}, 
\sigma_{\PP}^2 \mathbf{I}_{\MT\MR})$, 
assuming that $\mathbf{H}=\mathbf{g}\mathbf{h}^T$ with 
$|[\mathbf{g}]_i|=1$ ($i=1,2,\ldots,\MR$) 
and $|[\mathbf{h}]_i|=1$ ($i=1,2,\ldots,\MT$),
the ergodic capacity of the effective channel $\widehat{\mathbf{H}}$ satisfies
\begin{align}
\begin{split}
  \widehat{C} 
& \geq 
  \sum_{i=0}^{\MR-1} \log\Bigg( 1 + \frac{\rho}{\MT} 
    \bigg( \MT\MR \delta_i + 2 \sigma^2_{\PP} 
    e^{ \digamma\left( \frac{\MT-i}{2} \right) } \bigg) \Bigg), \quad \MR\le\MT
\\
  \widehat{C} 
& \geq
  \sum_{i=0}^{\MT-1} \log\Bigg( 1 + \frac{\rho}{\MT} 
    \bigg( \MT\MR \delta_i + 2 \sigma^2_{\PP} 
    e^{ \digamma\left( \frac{\MR-i}{2} \right) } \bigg) \Bigg), \quad \MR>\MT.
\end{split}
\label{eq:Cbound} 
\end{align}

\begin{IEEEproof}
We provide the proof for $\MR\le \MT$ only. The
case $\MR>\MT$ follows exactly the same line of reasoning.
We start by noting that \eqref{eq:Zd1} and \eqref{eq:Zd2} can be combined as
\begin{align*}
  \expect\big\{ \ln\,\det\big( (\mathbf{S}\mathbf{S}^H)_{l_1<l_2<\ldots<l_i} \big) \big\}
\ge
  \sum_{l=0}^{i-1} \expect\big\{ \ln \big( \MT\MR \delta_l \delta_{l_1-1} 
                                  + \chi^2_{\MT-l,\sigma_{\PP}^2} \big) \big\}
\end{align*}
which, upon inserting into \eqref{eq:Zhang}, yields
\begin{align*}
\widehat{C} \ge 
\log\left(
1 + \sum_{i=1}^{\MR} \left(\frac{\rho}{\MT}\right)^i \sum_{l_1<l_2<\ldots<l_i}
\prod_{l=0}^{i-1} e^{ \expect\big\{ \ln\big( \MT\MR \delta_l \delta_{l_1-1} 
                                              + \chi^2_{\MT-l,\sigma_{\PP}^2} \big) \big\} } 
\right) = C_2.
\end{align*}
The proof will be completed by showing that $2^{C_2} \ge 2^{C_1}$ with
\begin{align}
C_1 = \log \prod_{i=0}^{\MR-1} \left( 1 + \frac{\rho}{\MT}e^{
  \expect\big\{\ln\big( \MT\MR\delta_i + \chi^2_{\MT-i,\sigma_{\PP}^2} \big) \big\} 
} \right)
\label{eq:C1}
\end{align}
and noting that the \ac{RHS} of \eqref{eq:Cbound} 
is obtained by lower-bounding the term corresponding to $i=0$ in \eqref{eq:C1}
according to \eqref{eq:JBterm1}.
Setting $X_n = \Elchi_{\MT-n,\sigma_{\PP}^2}$ $(n=0,1,\ldots,\MR-1)$ 
and $X_0' = \expect \{ \ln (\MT\MR+\chi^2_{\MT,\sigma_{\PP}^2}) \}$,
and expanding 
\begin{align*}
  2^{C_1} 
&= 
                      \left( 1 + \frac{\rho}{\MT} e^{ X_0' } \right)
  \prod_{i=1}^{\MR-1} \left( 1 + \frac{\rho}{\MT} e^{ X_i }  \right)
\end{align*}
we get
\begin{align}
\begin{aligned}
&& 2^{C_1} 
=
1 &+ \left(\frac{\rho}{\MT}\right)       \Big( e^{X_0'} + e^{X_1} + \cdots + e^{X_{\MR-1}} \Big)
\\
\text{2a)} \quad
&&&+ \left(\frac{\rho}{\MT}\right)^2     \Big( e^{X_0'}e^{X_1} + e^{X_0'}e^{X_2} + \cdots + e^{X_0'}e^{X_{\MR-1}}
\\
\text{2b)} \quad
&&& \qquad\qquad\qquad\qquad                 + e^{X_1}e^{X_2} + e^{X_1}e^{X_3} + \cdots + \cdots + e^{X_{\MR-2}}e^{X_{\MR-1}} \Big)
\\
\text{3a)} \quad
&&&+ \left(\frac{\rho}{\MT}\right)^3     \Big( e^{X_0'}e^{X_1}e^{X_2} + e^{X_0'}e^{X_1}e^{X_3} + \cdots + e^{X_0'}e^{X_{\MR-2}}e^{X_{\MR-1}}
\\
\text{3b)} \quad
&&& \qquad\qquad\qquad\qquad                 + e^{X_{1}}e^{X_{2}}e^{X_{3}} + \cdots + \ldots + e^{X_{\MR-3}}e^{X_{\MR-2}}e^{X_{\MR-1}} \Big)
\\
&&&+ \;\:\quad \vdots \qquad \qquad \vdots
\\
&&&+ \left(\frac{\rho}{\MT}\right)^{\MR} \Big( e^{X_0'} e^{X_1} \cdots e^{X_{\MR-1}} \Big)
\end{aligned}
\label{eq:LHSC}
\end{align}
where lines 2a) and 2b) contain $\binom{\MR-1}{1}$ and $\binom{\MR-1}{2}$ terms, respectively,
and   lines 3a) and 3b) contain $\binom{\MR-1}{2}$ and $\binom{\MR-1}{3}$ terms, respectively.
Expanding 
\begin{align*}
  2^{C_2}
=
  1 + \sum_{i=1}^{\MR} \left( \frac{\rho}{\MT} \right)^i 
  \left( \sum_{1=l_1<l_2<\ldots<l_i} \left( X_0' \prod_{l=1}^{i-1} e^{ X_l } \right)
        +\sum_{1<l_1<l_2<\ldots<l_i} \left( X_0  \prod_{l=1}^{i-1} e^{ X_l } \right) \right)
\end{align*}
we obtain
\begin{align}
\begin{aligned}
&& 2^{C_2} = 
1 &+ \left(\frac{\rho}{\MT}\right) \Big( e^{X_0'} + e^{X_0} + \cdots + e^{X_0} \Big) \\
\text{2a)} &&&+ \left(\frac{\rho}{\MT}\right)^2 \Big( e^{X_0'}e^{X_1} + e^{X_0'}e^{X_1} + \cdots + e^{X_0'}e^{X_1} \\
\text{2b)} &&& \qquad\qquad\qquad\qquad             + e^{X_0}e^{X_1}  + e^{X_0}e^{X_1}  + \cdots + \cdots + e^{X_0}e^{X_1} \Big) \\
\text{3a)} &&&+ \left(\frac{\rho}{\MT}\right)^3 \Big( e^{X_0'}e^{X_1}e^{X_2} + e^{X_0'}e^{X_1}e^{X_2} + \cdots + e^{X_0'}e^{X_1}e^{X_2} \\
\text{3b)} &&& \qquad\qquad\qquad\qquad             + e^{X_0}e^{X_1}e^{X_2}  + e^{X_0}e^{X_1}e^{X_2}  + \cdots + \cdots + e^{X_0}e^{X_1}e^{X_2} \Big) \\
&&&+ \;\:\quad \vdots \qquad \qquad \vdots\\
&&&+ \left(\frac{\rho}{\MT}\right)^{\MR} \Big( e^{X_0'} e^{X_1} \cdots e^{X_{\MR-1}} \Big)
\end{aligned}
\label{eq:RHSC}
\end{align}
where the number of terms in lines 2a), 2b), 3a), and 3b) is the same as 
in the corresponding lines in the expansion of $2^{C_1}$.
In both cases, the number of terms associated with the factor
$(\rho/\MT)^i$ is given by $\binom{\MR}{i}$. 
The proof is completed by comparing \eqref{eq:LHSC} and \eqref{eq:RHSC} 
term by term and noting that the monotonicity of the digamma function implies
$\Elchi_{\MT-n,\sigma_{\PP}^2} \ge \Elchi_{\MT-n-k,\sigma_{\PP}^2}$ 
and hence $X_n\ge X_{n+k}$ for $k\ge 1$.
\end{IEEEproof}
\end{prop}

So far we derived lower bounds on $\widehat{C}$. We shall next show that the
result in \propref{prop:dist_approx} together with a technique first proposed in 
\cite{ZhCL05j} (and used to derive the lower bound in \propref{prop:CuZF04}) can be
employed to derive a tight analytic upper bound on $\widehat{C}$. 

\begin{prop}
\label{prop:Ubound}
Under the assumptions in \propref{prop:Cbound}, the ergodic capacity of the
effective \ac{MIMO} channel can be upper-bounded as 
\begin{align}
\widehat{C} 
&\le \log \left( 1 + \sum_{n=1}^{\MR} \left(\frac{\rho}{\MT}\right)^n (\sigma^2_{\PP})^n
\binom{\MR}{n} \binom{\MT}{n} n! \left(1+\frac{n}{\sigma^2_{\PP}}\right) \right).
\label{eq:eqZ4}
\end{align}

\begin{IEEEproof}
We provide the proof for $\MR\le \MT$ only. The
case $\MR>\MT$ follows exactly the same line of reasoning. 
The proof starts from \cite[Eq.\rsp(19)]{ZhCL05j} which,
specialized to our case, reads
\begin{align}
  \widehat{C} 
\le 
  \log \left( 1 + \sum_{n=1}^{\MR} \left(\frac{\rho}{\MT}\right)^n \sum_{l_1<l_2<\ldots<l_n}
  \expect\big\{ \det\big( (\mathbf{S}\mathbf{S}^H)_{l_1<l_2<\ldots<l_n} \big) \big\} \right)
\label{eq:UBZhang}
\end{align}
where $\mathbf{S}$ was defined in the proof of \propref{prop:dist_approx}.
The main point of the proof is to recognize that we can obtain analytic expressions for
the terms $\expect\{ \det((\mathbf{S}\mathbf{S}^H)_{l_1<l_2<\ldots<l_n}) \}$
using \eqref{eq:eqF}. 
As already shown in the proof of \propref{prop:CuZF04}, 
the terms in $\sum_{l_1<l_2<\ldots<l_n} \expect\{ \cdot \}$
on the \ac{RHS} of \eqref{eq:UBZhang}
fall into two groups depending on whether 
$l_1=1$ or $l_1>1$. 
Specifically, for $l_1=1$ we have (cf.\ \eqref{eq:Zd1exact})
\begin{align*}
  \expect\{ \det((\mathbf{S}\mathbf{S}^H)_{l_1<l_2<\ldots<l_n}) \} 
&=
  \MT\MR \prod_{i=2}^{n} \expect\big\{ \chi^2_{\MT-i,\sigma_{\PP}^2} + Z(\eta^{(i)}) \big\}
\\
&\qquad\qquad
+ \expect\big\{ \chi^2_{\MT,\sigma_{\PP}^2} \big\} 
   \prod_{i=2}^{n} \expect\big\{ \chi^2_{\MT-i,\sigma_{\PP}^2} + Z(\eta^{(i)}) \big\}
\end{align*}
and for $l_1>1$ (cf.\ \eqref{eq:Zd2})
\begin{align*}
  \expect\big\{ \det\big( (\mathbf{S}\mathbf{S}^H)_{l_1<l_2<\ldots<l_n} \big) \big\} 
&=
  \expect\big\{ \chi^2_{\MT,\sigma_{\PP}^2} \big\} 
  \prod_{i=2}^{n} \expect\big\{ \chi^2_{\MT-i,\sigma_{\PP}^2} + Z(\eta^{(i)}) \big\}.
\end{align*}
Using $\expect\{ \chi^2_{\MT-i,\sigma^2_{\PP}} \} 
=
  (\MT-i) \sigma^2_{\PP}
$, 
noting that
\begin{align*}
  \expect\big\{ Z(\eta^{(i)}) \big\}
&=
  \expect_{\lrv^{(i)}}\Big\{
  \expect\big\{
  \chi^2_{1,\sigma_{\PP}^2\lrv^{(i)}} + \chi^2_{1,\sigma_{\PP}^2(1-\lrv^{(i)})} \big| \lrv^{(i)}
  \big\} \Big\}
\\
&=
  \expect_{\lrv^{(i)}}\big\{ 
  \sigma_{\PP}^2\lrv^{(i)} + \sigma_{\PP}^2(1-\lrv^{(i)}) 
\big\}
=
  \expect_{\lrv^{(i)}}\big\{ \sigma_{\PP}^2 \big\}
=
  \sigma_{\PP}^2
\end{align*}
and counting the multiplicity of the terms as in \eqref{eq:multiplicity},
we obtain
\begin{multline*}
  \sum_{l_1<l_2<\ldots<l_n} 
  \expect\big\{ \det\big( (\mathbf{S}\mathbf{S}^H)_{l_1<l_2<\ldots<l_n} \big) \big\} 
\\
= 
  \binom{\MR-1}{n-1} \MT\MR \prod_{i=2}^{n} \big( (\MT-i+1) \sigma^2_{\PP} \big)
+ \binom{\MR}{n} \MT \sigma^2_{\PP}
  \prod_{i=2}^{n} \big( (\MT-i+1) \sigma^2_{\PP} \big)
\\
= 
  \binom{\MR-1}{n-1} \MT\MR \frac{(\MT-1)!}{(\MT-n)!} (\sigma^2_{\PP})^{n-1} 
+ \binom{\MR}{n} \MT \frac{(\MT-1)!}{(\MT-n)!} (\sigma^2_{\PP})^{n}
\\
= 
  \binom{\MR}{n} \frac{n \MT!}{(\MT-n)!} (\sigma^2_{\PP})^{n-1}
+ \binom{\MR}{n} \frac{\MT!}{(\MT-n)!} (\sigma^2_{\PP})^n.
\end{multline*}
Putting the pieces together, we get \eqref{eq:eqZ4}, which concludes the proof.
\end{IEEEproof}
\end{prop}
We note that the proof of \propref{prop:Ubound} can alternatively 
be carried out by obtaining analytic expressions for the terms 
$\expect\{ \det((\mathbf{S}\mathbf{S}^H)_{l_1<l_2<\ldots<l_n}) \}$
using properly modified versions of \eqref{eq:chi2exact}.

\subsubsection*{Numerical results}

\begin{figure}[htb]
\centering
\psfrag{exact, no phase noise}{\scriptsize{exact, no phase noise }}
\psfrag{exact, with phase noise}{\scriptsize{exact, with phase noise}}
\psfrag{lower bound, Theorem xyz}{\scriptsize{lower bound, \propref{prop:CuZF04}}}
\psfrag{lower bound, Theorem zyx}{\scriptsize{lower bound, \propref{prop:Cbound}}}
\psfrag{upper bound, Theorem xyz}{\scriptsize{upper bound, \propref{prop:Ubound}}}
\psfrag{Ergodic capacity C [bit/s/Hz]}{\scriptsize{Ergodic capacity $\widehat{C}$ [bit/s/Hz]}}
\psfrag{SNR w [dB]}{\scriptsize{\acs{SNR} $\rho$ [dB]}}
\icg{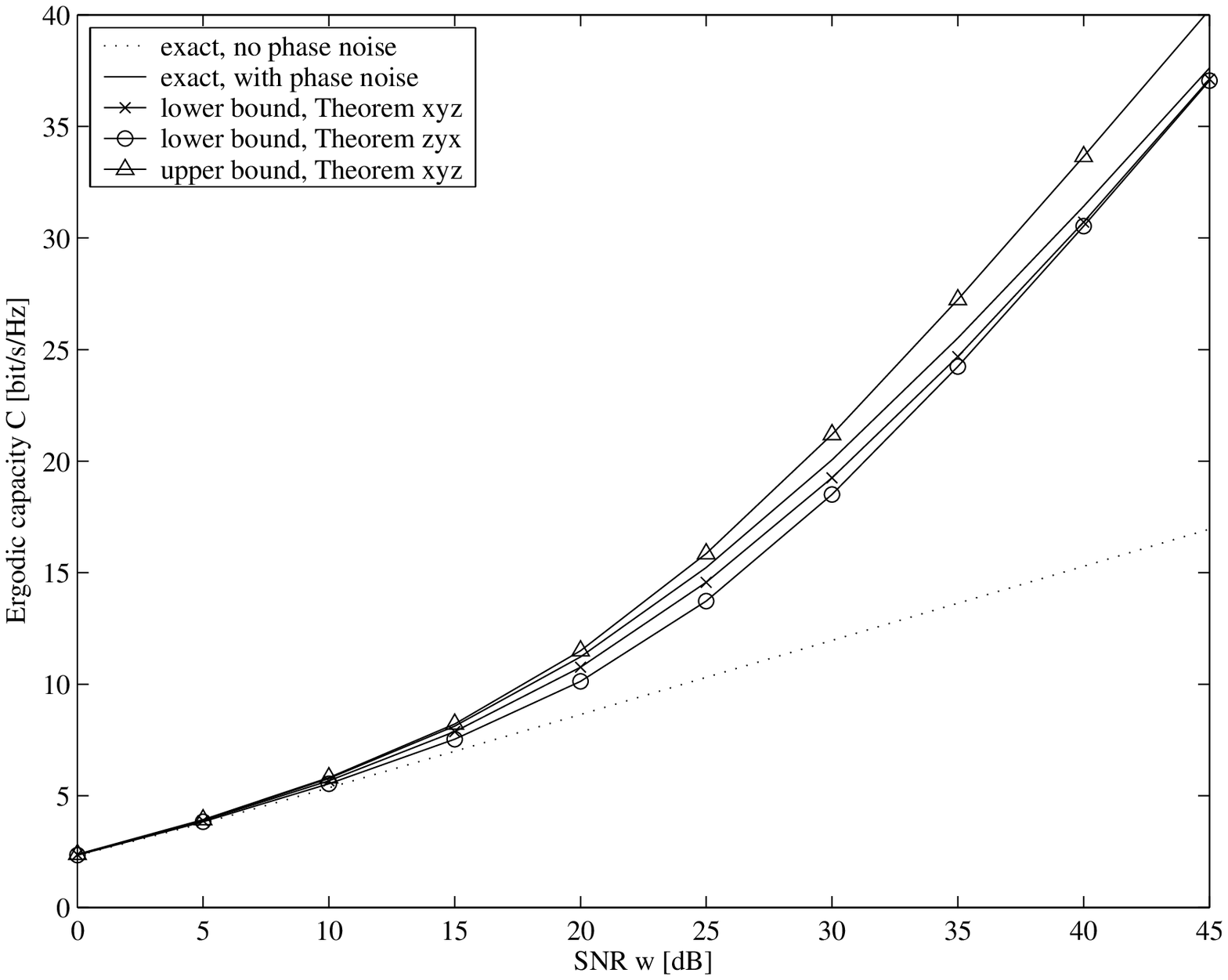}
\vspace{-2mm}
\caption{\label{fig:lbounds_left}
Ergodic capacity of a $4\times4$ rank-1 physical channel with unit-modulus entries
subject to 7\degreel\ \acs{RMS} fully uncorrelated phase noise. 
Exact results are obtained through Monte Carlo simulation.}
\end{figure}

We shall next provide a numerical example that serves to quantify the quality of 
the lower bounds in \propsref{prop:CuZF04} and \ref{prop:Cbound}, 
and the upper bound in \propref{prop:Ubound}.
For a $4\times 4$ deterministic physical channel 
$\mathbf{H}=\mathbf{g}\mathbf{h}^T$ with
$|[\mathbf{g}]_i|=1$, $\forall\,i$, and 
$|[\mathbf{h}]_i|=1$, $\forall\,i$,
subject to 7\degreel\ \ac{RMS}
fully uncorrelated phase noise,
\figref{fig:lbounds_left} shows that the (ergodic) capacity of the effective channel
starts deviating from the capacity of the rank-1 physical \ac{MIMO} channel at $\rho\approx15$\unit{dB},
and that significant capacity estimation errors (up to around 100\%) 
occur in the high-\ac{SNR} regime. 
This behavior is consistent with our observation that the low-\ac{SNR} capacity is not
influenced by phase noise. 
Moreover, we observe that the lower and upper bounds \eqref{eq:coolbound} and \eqref{eq:eqZ4},
respectively, 
very accurately predict the capacity behavior of the effective channel.

\subsubsection*{High-\acs{SNR} Variance of \acs{MI}}

Considering an ergodic block-fading \ac{MIMO} channel, it was shown in \cite{ONBP03j} that
the (high-SNR) variance of \ac{MI} can be interpreted as quantifying
the amount of ``spatial averaging'' that occurs \emph{on a per-stream basis} 
in each fading block. The smaller the variance of \ac{MI} the more spatial 
averaging occurs. As shown in \cite{ONBP03j,HoMT04j},
$\sigma^2_I = \var\{I\}$ for fixed $\MR$, as a function of $\MT$, 
has its maximum at $\MT=\MR$.
For more details on the 
interpretation of $\sigma^2_I$ as a measure of the amount of spatial diversity,
the interested reader is referred to \cite{ONBP03j}.
We have seen that phase noise (and frequency offset) can have a significant impact
on the rank of the \ac{MIMO} channel and hence its spatial multiplexing gain.
In the following, we shall characterize the increase in spatial diversity due to
phase noise by analyzing the variance of the high-\ac{SNR} \ac{MI} 
of the effective \ac{MIMO} channel. 
Finding exact expressions for $\sigma^2_{\hat{I}}=\var\{\hat{I}\}$ seems difficult. 
Under the assumptions in \propref{prop:Cbound}, 
we can, however, provide accurate and analytically
tractable approximations, which are obtained as follows. 
Considering, for simplicity, the case $\MR\le\MT$, 
we can infer from \eqref{eq:logdetTT} and \eqref{eq:chi2approxtrue} 
that in the high-\ac{SNR} regime 
\begin{align}
\hat{I} 
&\approx \log \left( \frac{\rho}{\MT} \Big( \chi^2_{\MT,\sigma_{\PP}^2} + \MT\MR \Big) 
              \prod_{i=1}^{\MR-1} \chi^2_{\MT-i,\sigma_{\PP}^2} \right) \nonumber \\
&= \log \left(\frac{\rho}{\MT}\right) 
 + \log \left( \MT\MR + \chi^2_{\MT,\sigma_{\PP}^2} \right) 
 + \sum_{i=1}^{\MR-1} \log \left( \chi^2_{\MT-i,\sigma_{\PP}^2} \right).
\label{eq:hSNRMI}
\end{align}
Writing the second term in \eqref{eq:hSNRMI} as
\begin{align*}
  \log \left( \MT\MR + \chi^2_{\MT,\sigma_{\PP}^2} \right) 
&=
  \log ( \MT\MR ) + \log \left( 1 + \frac{\chi^2_{\MT,\sigma_{\PP}^2}}{\MT\MR} \right)
\end{align*}
and noting that for $\MT\MR$ large and $\sigma_{\PP}^2$ small, we have
\begin{align*}
  \log \left( 1 + \frac{\chi^2_{\MT,\sigma_{\PP}^2}}{\MT\MR} \right) 
&\approx
  \log(e) \frac{\chi^2_{\MT,\sigma_{\PP}^2}}{\MT\MR}
\end{align*}
it follows that
\begin{align*}
  \var\{\hat{I}\} 
&\approx  
  \var \left\{   \log(e) \frac{\chi^2_{\MT,\sigma_{\PP}^2}}{\MT\MR}
               + \sum_{i=1}^{\MR-1} \log \left( \chi^2_{\MT-i,\sigma_{\PP}^2} \right) \right\}.
\end{align*}
Using 
\begin{align*}
  \var \left\{ \ln \left( \chi^2_{\MT-i,\sigma^2_\PP} \right) \right\} 
&=
  \var \left\{ \ln(\sigma^2_\PP) + \ln \left( \chi^2_{\MT-i,1} \right) \right\}
\\ 
&=
  \var \left\{ \ln \left( \chi^2_{\MT-i,1} \right) \right\} 
 =
  \digammader\left( \frac{\MT-i}{2} \right)
\end{align*}
and \cite[App.\rsp A.7]{Lee97b},
we finally get
\begin{align}
  \var\{\hat{I}\} 
\approx 
  \fsquare[big]{\log(e)} \left(
  \frac{2 \sigma^4_{\PP}}{\MT \MR^2} 
+ \sum_{i=1}^{\MR-1} \sum_{p=1}^\infty \frac{1}{\big( p+\frac{\MT-i}{2}-1 \big)^2} \right).
\label{eq:varMI}
\end{align}
Comparing \eqref{eq:varMI} to the expression \cite[Eq.\rsp(31)]{ONBP03j} for 
the variance of the high-\ac{SNR} \ac{MI} of 
an \ac{IID} $\MR\times\MT$ complex Gaussian channel, 
we can show that, for the same number of transmit and receive antennas,
the variance of the high-\ac{SNR} \ac{MI} 
of a rank-1 physical channel as in \propref{prop:Cbound} subject to
fully uncorrelated phase noise 
is higher than that in the \ac{IID} complex Gaussian case, i.e., 
\begin{align}
  \frac{2 \sigma^4_{\PP}}{\MT \MR^2} 
+ 
  \sum_{i=1}^{\MR-1} \digammader\left( \frac{\MT-i}{2} \right)
&\ge
  \sum_{i=1}^{\MR} \digammader\left( \MT-i+1 \right).
\label{eq:compdiga}
\end{align}
To prove \eqref{eq:compdiga}, we omit the first term on the \ac{LHS}
and use \lemmref{lemma:dig_der} in the Appendix which leaves us with
having to show that 
\begin{align}
  2 \sum_{i=1}^{\MR-1} \digammader\left( \MT-i \right)
&\ge
  \sum_{i=0}^{\MR-1} \digammader\left( \MT-i \right).
\label{eq:compdiga2}
\end{align}
Subtracting the common terms on both sides, 
it follows that \eqref{eq:compdiga2} is equivalent to 
\begin{align*}
  \sum_{i=1}^{\MR-1} \digammader\left( \MT-i \right)
&\ge
  \digammader\left( \MT \right).
\end{align*}
The final result follows from the monotonicity property 
$\digammader\left( \MT-i \right) \ge \digammader\left( \MT \right)$, $i\ge1$.
We note that \eqref{eq:compdiga2} suggests that the variance of \ac{MI}
in the phase noise case is essentially twice that obtained for an \ac{IID}
Gaussian channel with the same number of transmit and receive antennas.
The underlying reason lies in the fact that the individual chi-square
terms in the phase noise case (cf.\ \eqref{eq:Zd1} and \eqref{eq:Zd2})
have half the number of degrees of freedom when compared to the 
Gaussian channel case.
The following numerical example corroborates the factor-2 statement motivated
by the inequality \eqref{eq:compdiga2}.

\begin{figure}[htbp]
\centering
\psfrag{exact, rank-1 with phase noise}{\scriptsize{exact, rank-1 with phase noise}}
\psfrag{approximation (xy)}{\scriptsize{approximation \eqref{eq:varMI}}}
\psfrag{exact, i.i.d. Rayleigh fading}{\scriptsize{exact, \acs{IID} Rayleigh fading}}
\psfrag{approximation [Oyman,(xy)]}{\scriptsize{approximation \cite[Eq.\rsp(31)]{ONBP03j}}}
\psfrag{MT}{\scriptsize{$\MT$}}
\psfrag{wI2 [(bit/s/Hz)2]}{\scriptsize{$\sigma^2_{\hat{I}}$ [(bit/s/Hz)$^2$]}}
\icg{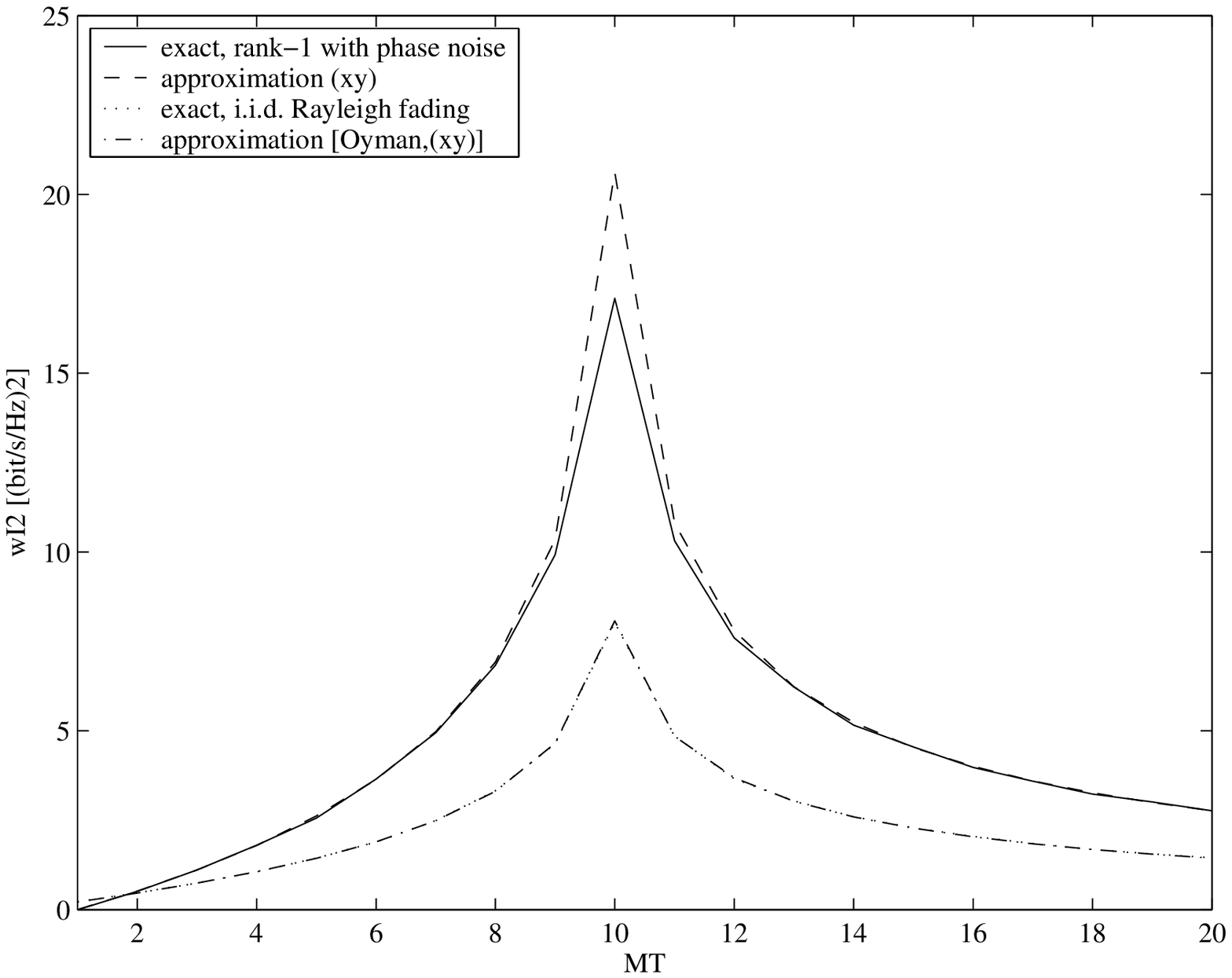}
\vspace{-2mm}
\caption{\label{fig:lbounds_right}
Exact variance of \acl{MI} and analytic approximation \eqref{eq:varMI} at high \acs{SNR} 
for $\MR=10$ as a function of $\MT$ 
for a rank-1 physical channel with unit-modulus entries
subject to 3.5\degreel\ \ac{RMS} fully uncorrelated phase noise, 
and for an \ac{IID} Rayleigh fading channel. 
Exact results are obtained through Monte Carlo simulation.}
\end{figure}

\emph{Numerical result: }
For a deterministic rank-1 physical channel 
$\mathbf{H} = \mathbf{g} \mathbf{h}^T$ with 
$|[\mathbf{g}]_i|=1$, $\forall\,i$, and 
$|[\mathbf{h}]_i|=1$, $\forall\,i$, 
subject to 3.5\degreel\ \ac{RMS} fully uncorrelated phase noise,
\figref{fig:lbounds_right} shows $\sigma_{\hat{I}}^2$ according to the approximation
\eqref{eq:varMI} along with the exact result (obtained from Monte Carlo simulation)
for $\MR=10$ as a function of $\MT$.
We can see that the approximation is very tight for $\MT>\MR$ and $\MT<\MR$
and predicts the location of the
maximum of $\sigma_{\hat{I}}^2$ accurately.
For comparison, we show $\sigma_I^2$ 
according to the approximation \cite[Eq.\rsp(31)]{ONBP03j} along with the exact result 
(obtained from Monte Carlo simulation) for an \ac{IID} Rayleigh fading channel
with the same number of transmit and receive antennas. 
It is clearly seen that both types of channels exhibit similar \ac{MI} variance 
behavior as a function of $\MT$,
and that $\sigma_{\hat{I}}^2 \approx 2 \sigma_{I}^2$.

\section{Measurements}
\label{sec:meas}
In this section, we provide results from measurements taken with 
a commercially employed \ac{TDMS}-based \ac{MIMO} channel sounder
by applying the rank-1 ``calibration procedure'' 
discussed in the beginning of \secref{sec:rank1}.
Before elaborating on the measurement results and comparing them
with our analytic results,
we shall verify assumptions on the system model and on 
the phase error characteristics that were made 
throughout the paper.

\subsection{Description of the Measurement Setup}
\label{sec:meas_setup}

In the following, a \ac{MIMO} channel snapshot indexed by the superscript $^{(l)}$
with $l = 1,2,\ldots,L$ consists of one snapshot of each of the $\MT\MR$ scalar
subchannels. 
Since the physical channel is static (i.e., the cable transfer function is static),
time-alignment,
as discussed in \secref{sec:MIMOSigMod}, is not needed.
The scalar subchannel snapshots in the $l$th \ac{MIMO} snapshot 
are taken at times $t_k^{(l)}$ $(k = 1,2,\ldots,\MT\MR)$.
All measurement results in this section are based on $L = 1100$ \ac{MIMO} snapshots.

Parameters specifying the channel sounder employed and the measurement setup 
are provided in \tabref{tab:meas}.
The sounder is based on circular convolution and uses a regular sounding pattern according to \eqref{eq:reg_pat},
implying that frequency offset has no impact on \ac{MI}.
As a consequence of \eqref{eq:reg_pat}, the \ac{SISO} snapshot 
distance is given by $T_\RX$. 
The sounder inserts a ``dummy'' receive antenna (i.e., one
unused \ac{SISO} snapshot) to accomodate switching between transmit antennas,
so that $T_\TX = (\MR+1)T_\RX$. Note that \eqref{eq:reg_pat} implies that the SISO snapshot corresponding
to the dummy receive antenna is simply omitted.
The sounding sequence results from a periodically extended m-sequence 
of length $N$ (chips) with period $N T$, where $1/T$ is the chip rate.

The phase noise process, in general, has a distinctive low-pass characteristic.
Increasing the \ac{SISO} snapshot time distances, 
and thereby reducing the rate at which
the continuous-time phase noise process is effectively sampled,
results in stronger decorrelation of the time-discrete phase noise
process underlying the effective MIMO channel.
This, in turn, leads to increased error in estimating the \ac{MI}. 
On the other hand, the minimum
\ac{SISO} snapshot time distance 
is determined by the antenna switching speed, the duty cycle (as 
explained below) and, in particular, the sounding sequence length.
Specifically, reducing the sounding sequence length
leads to a degradation of the sequence-correlation properties
and hence a reduction in sequence \ac{SNR}, which in turn implies lower
measurement \ac{SNR}. 
In summary, there is a tradeoff between the
sequence \ac{SNR}, determined by the 
time-discrete sounding sequence, in particular by its length, and the \ac{MI}
estimation error due to decorrelation of the phase noise
process underlying the effective MIMO channel.
To further understand this tradeoff,
we performed measurements based on two sounder setups,
as defined in \tabref{tab:meas}, with different sequence lengths.

\begin{table}[htbp]
\caption{\label{tab:meas}Channel Sounder and (Two Different) Measurement Setup Parameters}
\vspace{-0.2cm}
\centering
\begin{tabular}{|l|c|c|}
\hline
Carrier frequency & \multicolumn{2}{c|}{5.25~GHz} \\
Frequency generation & \multicolumn{2}{c|}{Rubidium reference and \ac{PLL}-VCO} \\
Sounding sequence & \multicolumn{2}{c|}{bipolar m-sequence, rectangular chip pulses} \\
Sounding pattern & \multicolumn{2}{c|}{regular (see \secref{sec:freq_offs})} \\
Chip rate & \multicolumn{2}{c|}{100\unit{MHz}} \\ \hline\hline
Sequence length (chips) & \phantom{abcdefghi} 511 \phantom{abcdefghi}  & 31 \\
\ac{MIMO} configuration ($\MR \times \MT$) & 16$\,\times\,$16 & 23$\,\times\,$23 \\
\ac{MIMO} snapshot distance $T_\text{MIMO}$  & 19.46\unit{ms} & 10.27\unit{ms} \\
Receive-side \ac{SISO} snapshot distance $T_{\RX}$ & 10.22\unit{$\upmu$s} & 0.93\unit{$\upmu$s} \\
Ratio $T_{\TX}/T_{\RX}$ & 17 & 24 \\
\ac{SISO} snapshot duration $\ilen{\INTVL_{\mathrm{P}}}$ & 5.11\unit{$\upmu$s} & 0.31\unit{$\upmu$s} \\
\hline
\end{tabular}
\end{table}

We would like to point out that the
various sounder settings (including sounding sequence length)
can usually not be chosen independently due to 
hardware limitations of the sounder.
One such typical limitation is the overall duty cycle, 
i.e., the ratio $\eta = \MT\MR N T / T_\text{MIMO}$
of sounding time (where the received signal is recorded and processed) 
to total measurement time,
where $T_\text{MIMO}$
is the \ac{MIMO} snapshot distance.
Channel sounders typically employ a small overall duty-cycle 
to limit the real-time signal processing and data-storage requirements.
The overall duty-cycle can be separated into
one within and one between \ac{MIMO} snapshot periods given by  
$\eta_\mathrm{intra} = \MT\MR NT
/ (t_{\MT\MR}^{(l)}-t_1^{(l)}+NT)$
and 
$\eta_\mathrm{inter} 
= (t_{\MT\MR}^{(l)}-t_1^{(l)} + NT) / T_\text{MIMO}$,
respectively, with $\eta = \eta_\mathrm{intra} \eta_\mathrm{inter}$.
In general, $\eta_\mathrm{inter}$ has no influence on the estimated \ac{MI}
as the correlation properties of the channel and/or phase noise across \ac{MIMO}
snapshot periods do not play a role in our considerations. 
If $\eta$ is fixed, it is 
therefore preferable to have $\eta_\mathrm{inter}$ small and 
$\eta_\mathrm{intra}$ large. 
In the two setups considered here, this is indeed the case, with
$\eta_\mathrm{inter} = 0.1431$ and $\eta_\mathrm{intra} = 0.4697$ for
the length-511 sequence, 
and $\eta_\mathrm{inter} = 0.05$ and $\eta_\mathrm{intra} = 0.3193$ for the
length-31 sequence. 

The transfer function of the cable used to connect transmitter and receiver
is flat (recall that we are performing the rank-1 calibration procedure 
decribed in \secref{sec:rank1})
over the frequency range of interest. 
In the sounder under consideration,
the overall (i.e., effective) channel
induced by $s(t)$ and $r(t)$
along with the cable exhibits, however, 
some delay spread. 
This is mainly due to 
oversampling of the signal at the output of the receive frontend filter by a factor of 2 
(relative to the chip rate).
We estimated the \ac{PDP} of the resulting effective channel
and identified the position of its peak. 
For further processing we used only the signal corresponding to the peak 
of the effective channel's \ac{PDP}.

\subsection{Verifying Assumptions on Phase Error Characteristics and System Model}
\label{sec:ph_err_char}

The purpose of this section is to investigate the general phase error characteristics
of the sounder under consideration, to verify our assumptions on the phase error
statistics stated in \secref{sec:instabilities}, and to verify the system model
assumptions stated in \secref{sec:narrowband}.
Correspondingly, the following discussion is organized into three parts.

\subsubsection{General phase error characteristics}

\begin{figure}[htbp]
\centering
\subfigure[Phase characteristics]{
\psfrag{time [s]}{\scriptsize{time [s]}}
\psfrag{phase [degrees]}{\scriptsize{phase [degrees]}}
\icg{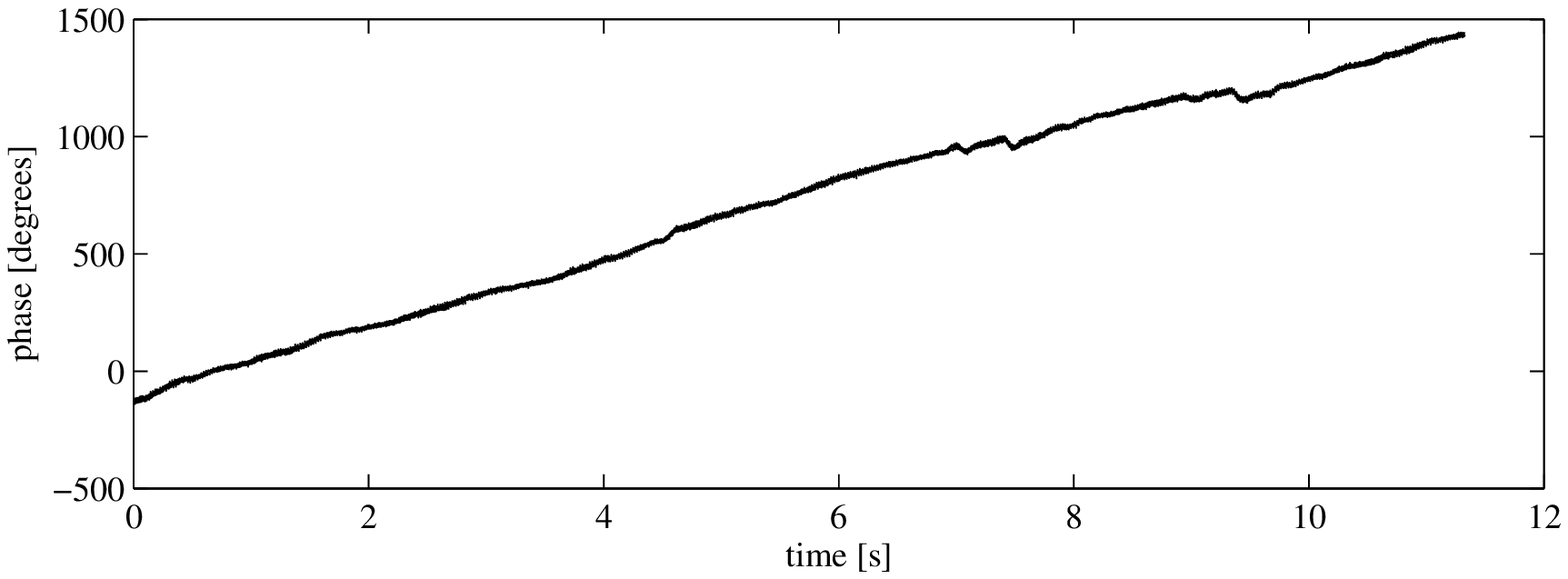}
\label{fig:long_term_pna}
} \\ 
\subfigure[Phase characteristics after removal of estimated linear component]{
\psfrag{time [s]}{\scriptsize{time [s]}}
\psfrag{phase [degrees]}{\scriptsize{phase [degrees]}}
\icg{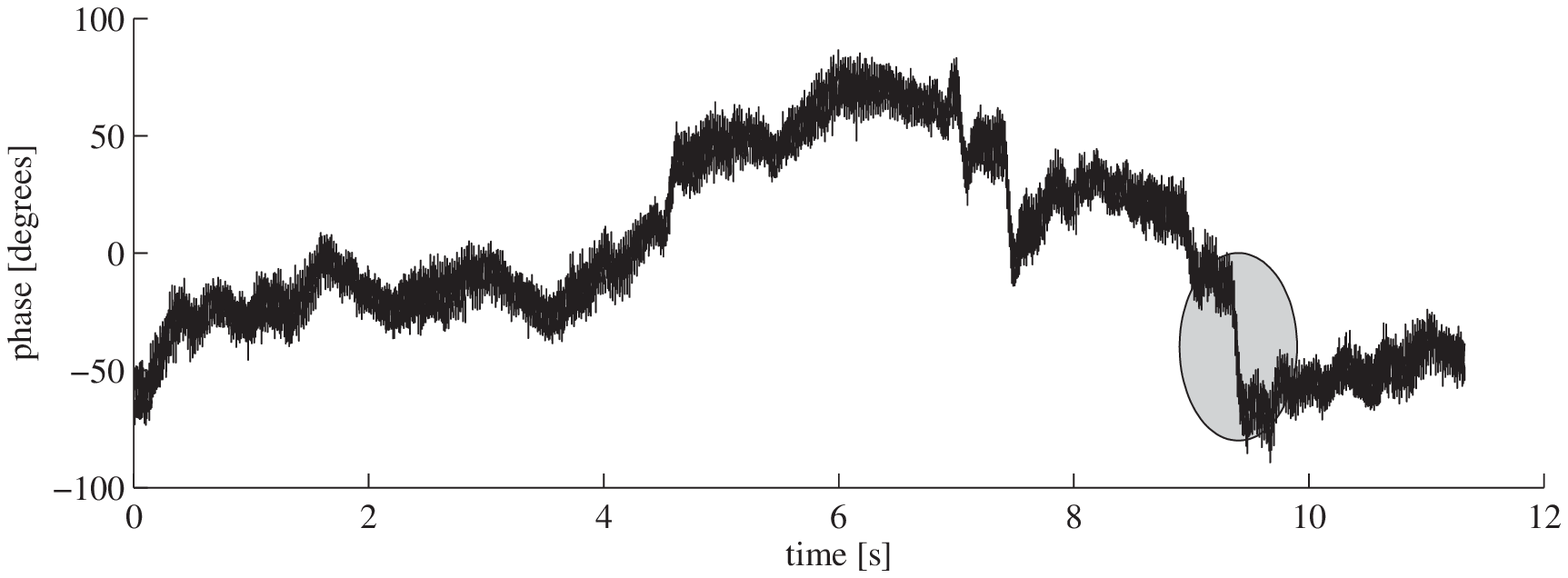}
\label{fig:long_term_pnb}
}
\vspace{-2mm}
\caption{\label{fig:long_term_pn}
Long-term phase noise behavior of a commercially employed
\acs{TDMS}-based \acs{MIMO} channel sounder (length-31 sounding sequence).}
\end{figure}

\begin{figure}[htbp]
\centering
\subfigure[Phase-noise trace of a single \acs{MIMO} snapshot (length-31 sounding sequence)]{
\psfrag{time [s]}{\scriptsize{time [s]}}
\psfrag{phase [degrees]}{\scriptsize{phase [degrees]}}
\icg{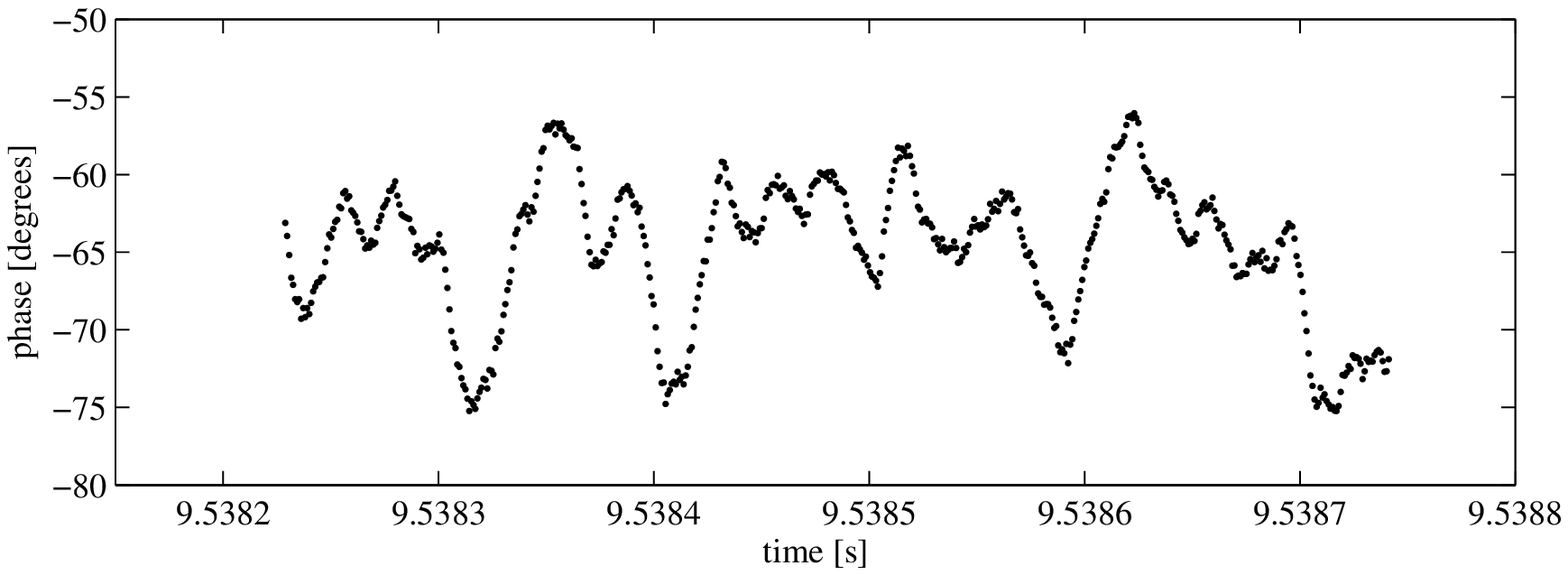}
\label{fig:short_term_pna}
} \\
\subfigure[Phase-noise trace of a single \acs{MIMO} snapshot (length-511 sounding sequence)]{
\psfrag{time [s]}{\scriptsize{time [s]}}
\psfrag{phase [degrees]}{\scriptsize{phase [degrees]}}
\icg{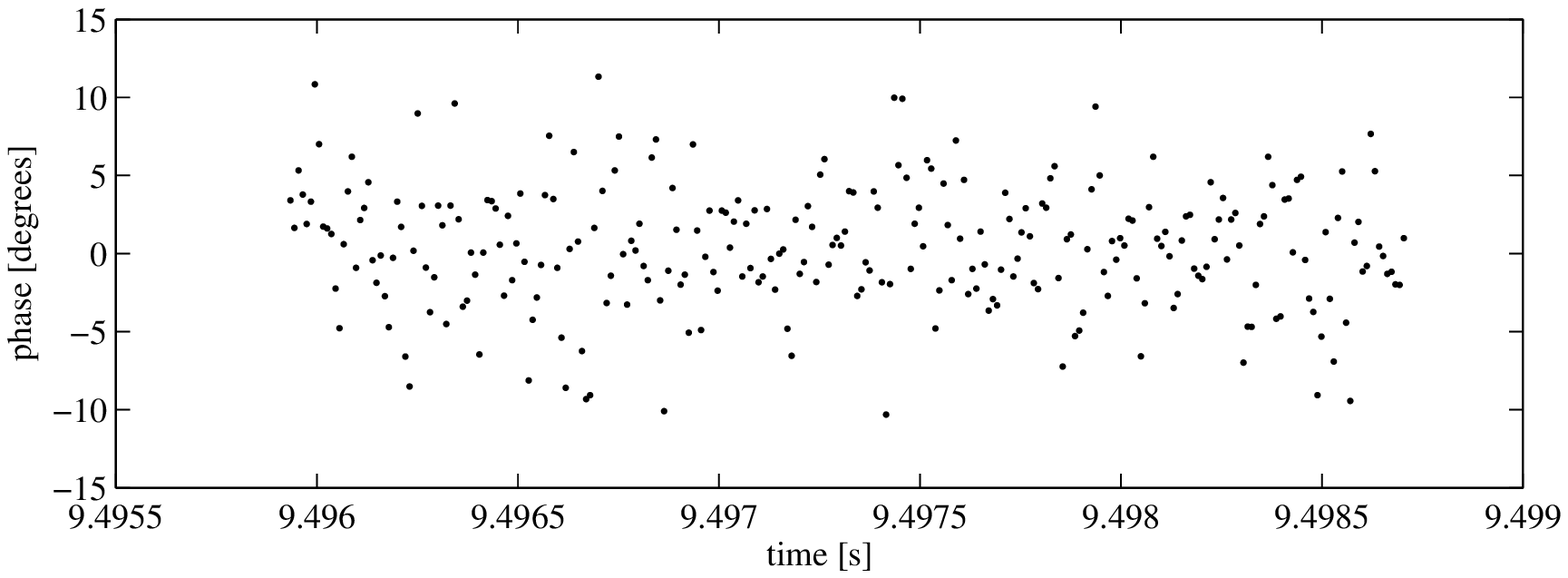}
\label{fig:short_term_pnb}
}
\vspace{-2mm}
\caption{\label{fig:short_term_pn}
Short-term phase noise behavior of a commercially employed
\ac{TDMS}-based \ac{MIMO} channel sounder.}
\end{figure}

We start by giving an impression of the phase variation 
characteristics over different time horizons. 
\sfigref{fig:long_term_pn}{fig:long_term_pna} 
shows the long-term behavior (multiple seconds)
of the phase-unwrapped raw channel estimates.\footnote{In 
   \figsref{fig:long_term_pn} and \ref{fig:short_term_pn}, the time axis
   corresponds to the sequence of scalar subchannel measurements, with a total of
   $\MT\MR L$ (with $L=1100$), taken according to the regular sounding pattern
   \eqref{eq:reg_pat}.}
Each physical \ac{SISO} channel measured corresponds to the cable
connecting transmitter and receiver. Since the cable
exhibits a frequency-flat and constant (over time) transfer function, 
any variation (over time) in the measured channel 
must necessarily come from channel sounder nonidealities,
or more specifically from phase noise and frequency offset in the transmit 
and receive \acp{LO}.
It is clearly seen that the unwrapped phase contains a linear
component resulting from the carrier frequency offset between transmitter and receiver.
\sfigref{fig:long_term_pn}{fig:long_term_pnb} 
shows the trace of the residual phase obtained by removing 
the linear component from the overall observed phase. 
This linear component was estimated\footnote{Note that 
   the low duty-cycle between \ac{MIMO} snapshots
   increases the possibility of the phase containing jumps larger than $2\pi$
   between consecutive \ac{MIMO} snapshots, thereby resulting in erroneous 
   unwrapping and hence carrier frequency offset estimation errors.
   Estimation was performed through least-squares fitting.}
to correspond to 
a relative (\ac{WRT} the carrier frequency of 5.25\unit{GHz}) frequency offset
of $7.2\cdot10^{-11}$, indicating excellent performance in terms of carrier frequency accuracy.
Moreover, as already pointed out, since we are employing a regular sounding pattern according to \eqref{eq:reg_pat},
a carrier frequency offset
does not have impact on \ac{MI}. 
Taking a closer look at the phase residual 
in \sfigref{fig:long_term_pn}{fig:long_term_pnb},
we can see that it exhibits two different 
constituents: a constant ``thickness'' 
corresponding to essentially uncorrelated (on the time scale used)
random fluctuations and a component of comparatively
larger amplitude and higher temporal correlation
containing at times abrupt changes
(see, e.g., the area marked in \sfigref{fig:long_term_pn}{fig:long_term_pnb}). 
These abrupt phase changes may be caused by spontaneous phase jumps of 
the reference oscillators, 
smoothed out by the \ac{PLL}. 
If the \ac{PLL} bandwidth is small compared to the \ac{MIMO} measurement rate 
(i.e., the rate at which \ac{MIMO} snapshots are taken),
we can, however, neglect the impact of the abrupt phase changes.

\figref{fig:short_term_pn} shows the phase noise trace 
(i.e., the overall phase after removal of the estimated linear component)
during one \ac{MIMO} snapshot period 
taken with\footnote{Note that the length of the sounding sequences is in fact $31T$ and $511T$, respectively.} length-31 and length-511 sounding sequences, respectively. 
Each dot in the two figures represents the phase of 
one scalar subchannel measurement (note the different absolute
time scales in \sfigsref{fig:short_term_pn}{fig:short_term_pna} and 
\sref{fig:short_term_pn}{fig:short_term_pnb}).
In accordance with what was said earlier,
one can immediately see that the phase trace corresponding 
to the length-31 sequence 
shows significantly higher correlation between successive \ac{SISO} snapshots
than that for the length-511 sequence 
(which can essentially be considered \ac{IID}).
While the shorter (length-31) sequence thus is clearly preferable regarding
the phase noise properties, it may fail to yield sufficiently high sequence 
and consequently measurement \ac{SNR}
(see the discussion in \secref{sec:meas_setup}).
The corresponding estimated \ac{RMS} standard deviation $\hat{\sigma}_{\PP}$ 
(estimated by averaging the standard deviation per \ac{MIMO} snapshot
over all $L=1100$ \ac{MIMO} snapshots) 
was found to be 3.9\degreel\ for the length-31 sequence
and 3.8\degreel\ for the length-511 sequence.
These values agree very well with the 3.5\degreel\ \ac{RMS} value
used as ``typical'' case throughout the paper.

\subsubsection{Verification of assumptions on phase error statistics}
\label{sec:vapes}

\begin{figure}[htbp]
\centering
\psfrag{  1e-6}{\scriptsize{10$^\text{-6}$}}
\psfrag{  1e-5}{\scriptsize{10$^\text{-5}$}}
\psfrag{  1e-4}{\scriptsize{10$^\text{-4}$}}
\psfrag{  1e-3}{\scriptsize{10$^\text{-3}$}}
\psfrag{  1e-2}{\scriptsize{10$^\text{-2}$}}
\psfrag{  1e-1}{\scriptsize{10$^\text{-1}$}}
\psfrag{   0.5}{\scriptsize{\!\!0.5}}
\psfrag{1-1e-1}{\scriptsize{1-10$^\text{-1}$}}
\psfrag{1-1e-2}{\scriptsize{1-10$^\text{-2}$}}
\psfrag{1-1e-3}{\scriptsize{1-10$^\text{-3}$}}
\psfrag{1-1e-4}{\scriptsize{1-10$^\text{-4}$}}
\psfrag{1-1e-5}{\scriptsize{1-10$^\text{-5}$}}
\psfrag{1-1e-6}{\scriptsize{1-10$^\text{-6}$}}
\psfrag{data1}{\scriptsize{length-31 sequence}}
\psfrag{data2}{\scriptsize{length-511 sequence}}
\psfrag{reference distribution N( 0, 3.9 )}{\scriptsize{reference $\realGauss(0,(3.9\degree)^2)$ distr.}}
\psfrag{w 1  std.  dev.  to  ref.  distr.}{\scriptsize{$\pm$1 std.\ dev.\ to ref.\ distr.}}
\psfrag{phase [degrees]}{\scriptsize{phase [degrees]}}
\psfrag{CDF}{\scriptsize{\acs{CDF}}}
\icg{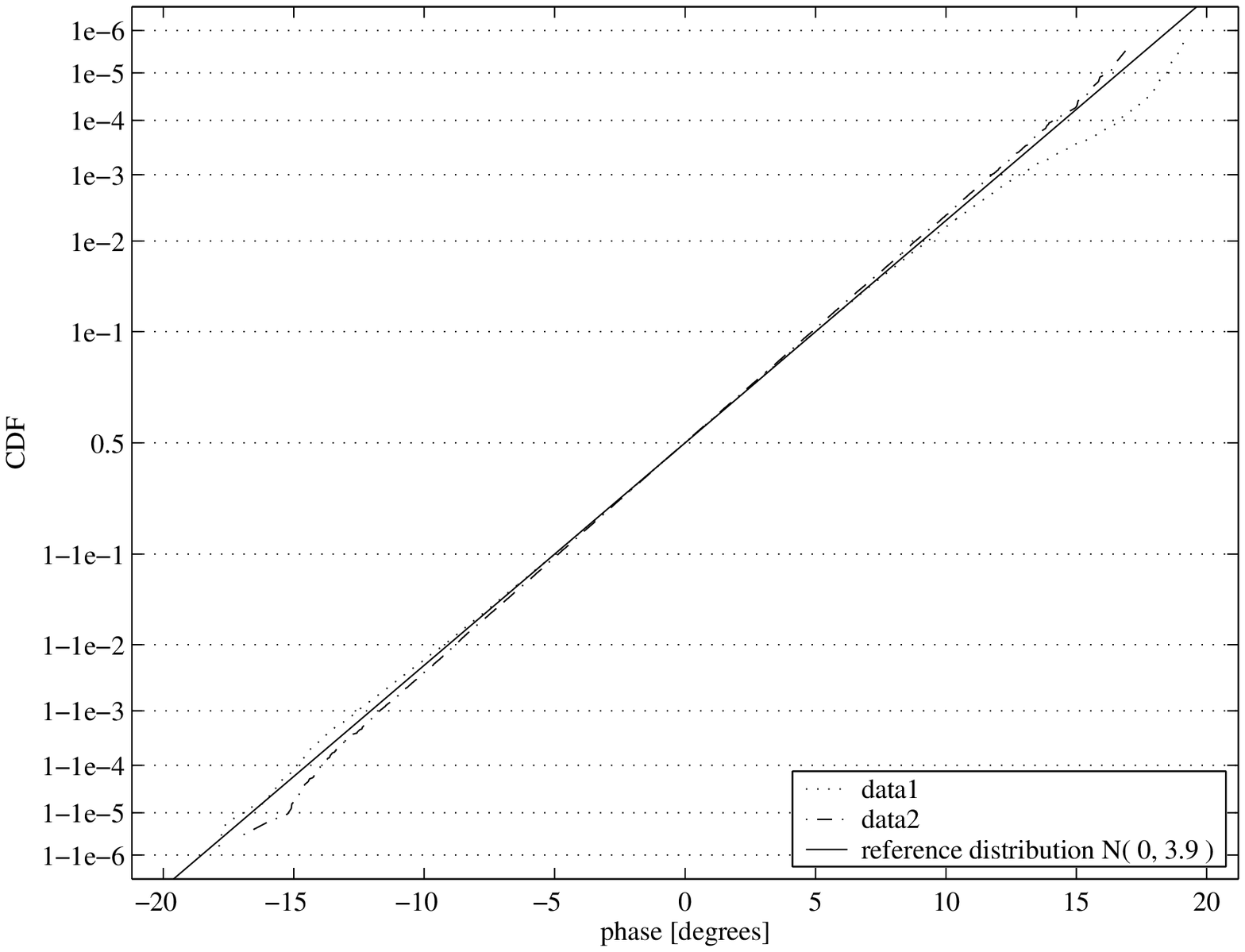}
\vspace{-2mm}
\caption{\label{fig:pn_normal}
\acs{CDF} of the measured \ac{MIMO} snapshot phase noise traces (i.e., overall phase after
removal of estimated linear component) compared to $\realGauss(0,(3.9\degree)^2)$ 
reference distribution.}
\end{figure}

We start by showing, in \figref{fig:pn_normal}, the \ac{CDF} of the 
measured \ac{MIMO} snapshot phase noise traces (i.e., the overall phase 
after removal of the estimated linear component), normalized to zero-mean, and plotted
on a scale where Gaussian distributions show a linear behavior. 
The normalization of the mean was performed by computing and subtracting the empirical mean on a
MIMO snapshot by MIMO snapshot basis.
It is clearly seen 
that both for the length-31 and the length-511 sounding sequence an excellent
match with an $\realGauss(0, (3.9\degree)^2)$ distribution is obtained. 
This allows us to conclude that 
the assumption of a Gaussian phase noise process 
(see \secref{sec:instabilities})
is well justified. 

The main differentiating factor between the phase noise models 
discussed in \secref{sec:instabilities} is stationarity. 
It therefore remains to verify the stationarity assumption (on the time-scale of one MIMO snapshot)
made throughout the paper. In particular, we need to show that the phase noise 
sequence $\pr_n$ resulting from samples taken at the time instants $nT_R$
exhibits stationary behavior.
This can be done by examining the phase
differences $\alpha(k,n) = \pr_k - \pr_{k+n}$,
or more specifically the variance 
$\sigma_{\alpha}^2(k,n) = \expect\{ \alpha^2(k,n) \}$ 
(note that $\expect\{\alpha(k,n)\}=0$).
In the case of stationary phase noise, we have 
$\sigma_{\alpha}^2(k,n)= \sigma_{\alpha}^2(n)
= 2 \sigma_{\PP}^2 \big( 1 - r_{\varphi\varphi}(n) \big)$,
where $r_{\varphi\varphi}(n)$
denotes the phase noise autocorrelation function.
Assuming that the stationary process has limited memory, i.e.,
$|r_{\varphi\varphi}(n)|\to0$ for $|n|\to\infty$, we have 
$\lim_{n\to\infty} \sigma_{\alpha}^2(n) = 2 \sigma_{\PP}^2$.
On the other hand, if the nonstationary phase noise model, 
described in \secref{sec:instabilities}, would be applicable,
we would have $\sigma_{\alpha}^2(k,n)=\sigma_{\alpha}^2(n)
\propto |n|$ \cite[Lemma\rsq8.2]{DeMR00j}.

\figref{fig:acf_pn} shows the estimate
$\mysqrt{ \hat{\sigma}_{\alpha}^2(n)/2 }$ for the 
length-31 and the length-511 sounding sequence, where
$\hat{\sigma}_{\alpha}^2(n)$ is obtained as follows.
The measured phase noise sequence $\pr_k$ corresponds to the time instants $t_k$. Consequently, $\pr_k$ is the
result of sampling at a rate of $1/T_R$, 
omitting, as already pointed out, the samples corresponding
to the dummy receive antenna.
Finally, $\expect\{\alpha^{2}(k,n)\}$ is estimated
by averaging $\alpha^{2}(k,n)$ in each MIMO snapshot, over all indices $k$ and $n$ satisfying $t_{k+n}-t_{k}=nT_R$ with 
$k\,\in\,[1,\MT \MR]$ and $k+n\,\in\,[1,\MT \MR]$,
and then averaging
the results over all snapshot periods $l\,\in\,[1,L]$.
We can clearly see that, for both sampling sequence lengths, 
$\mysqrt{ \hat{\sigma}_{\alpha}^2(n)/2}$ levels out
at a constant value, which allows us to conclude that over the time
frame of interest (i.e., one MIMO snapshot period) the assumption of stationary phase noise is well
justified. Moreover, we can observe that the large-$n$ limit of
$\mysqrt{ \hat{\sigma}_{\alpha}^2(n)/2 }$ is close to 4\degree, which is very well
in accordance with the 3.8\degreel\ and 3.9\degreel\ \ac{RMS} values
estimated earlier.

\subsubsection{Verification of the system model assumptions}

Based on our measurement results, we shall next 
quantify the impact of phase errors on sequence \ac{SNR}
and discuss the validity of the assumptions 
\eqref{eq:smallChangeCirc} and \eqref{eq:assDSCirc}. 

While we have seen in \secref{sec:narrowband}
that it is safe to assume the absence of 
peak-shifting in the sounder under consideration (recall that the sounder employs m-sequences), phase errors
will, in general, lead to a reduction in sequence \ac{SNR}
and hence also in measurement \ac{SNR}.
Defining 
\begin{align*}
   \mathsf{SNR}_{i}
= 
   \frac{|c'(0)|^2}{|c'(iT)|^2},
\qquad
   i = 1,2,\ldots,N-2
\end{align*}
we estimate (Monte Carlo) the histogram of the overall SNR, defined through its histogram to be given by
the average of the histograms corresponding to the quantities $\mathsf{SNR}_{i}$;
10\,000 phase noise realizations per lag were used.
In the absence of phase noise an m-sequence
has a (deterministic) sequence \ac{SNR} of $20\,\log_{10}(m)$, 
which, for $m=511$, equals 54.2\unit{dB}.
The phase noise in the sounder under consideration
randomizes this value to 
a mean of 53.8\unit{dB} and 
a level below 49.1\unit{dB} \ac{WIP} 0.1\%.
We will see later that 49.1\unit{dB}
lies above the sounder's \ac{SNR} caused by thermal noise. Consequently, the effect
of phase noise on sequence SNR, albeit noticeable, 
can be considered insignificant.

\begin{figure}[htbp]
\centering
\psfrag{length-511 sequence}{\scriptsize{length-511 sequence}}
\psfrag{length-31 sequence}{\scriptsize{length-31 sequence}}
\psfrag{time difference [s]}{\scriptsize{time difference [s]}}
\psfrag{S(w2(n)/2) [degrees]}{\scriptsize{$\mysqrt{\hat{\sigma}_{\alpha}^2(n)/2}$ [degrees]}}
\icg{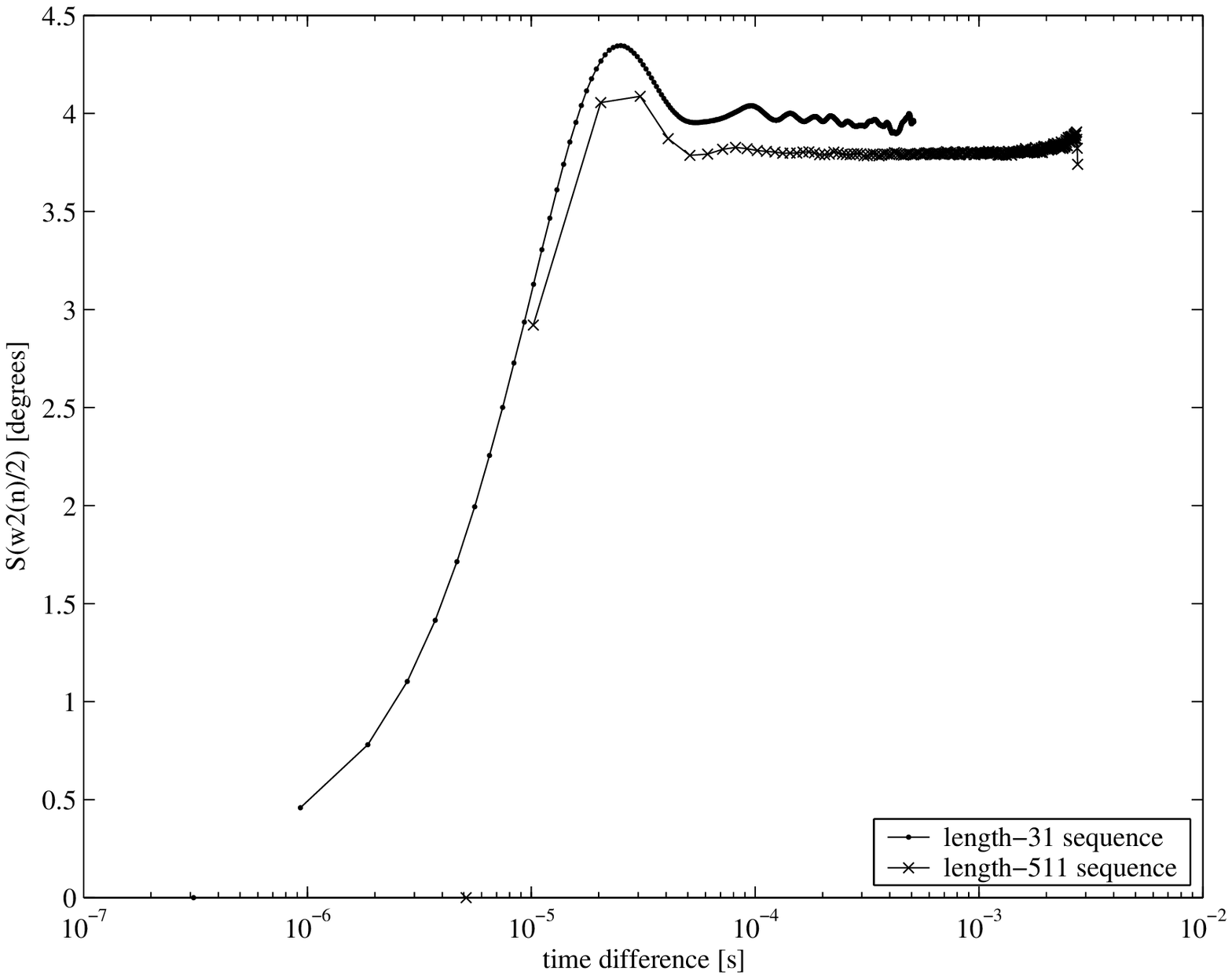}
\vspace{-2mm}
\caption{\label{fig:acf_pn}
$\mysqrt{ \hat{\sigma}_{\alpha}^2(n)/2 }$ for length-31 and length-511 sounding sequences.}
\end{figure}

Condition \eqref{eq:smallChangeCirc} is verified as follows. 
We start by noting that, 
for a periodically extended m-sequence and circular convolution
in the receiver, as employed in the sounder under consideration,
the integration interval in \eqref{eq:smallChangeCirc} 
has length $\ilen{\INTVL_{\mathrm{P}}} = NT$. 
With $\Delta\omega = 2\pi\Delta f = 2.4$\unit{rad/s},
$N T = 310$\unit{ns} 
for the length-31 sounding sequence, and 
$N T = 5110$\unit{ns} 
for the length-511 sounding sequence, we have
$|\Delta\omega\xi| \le 12.3\cdot10^{-6}$\unit{rad},
which implies that we can safely assume that 
$\exp(\jmath\Delta\omega\xi) \approx 1$.
If the inverse of the bandwidth of $\varphi_\RX(t)$ and 
$\varphi_\TX(t)$ is much larger than $NT$ 
and/or the variance of $\varphi_\RX(t)$ and $\varphi_\TX(t)$ is small, 
the term 
$\exp\big(-\jmath[ \varphi_\RX(t_k+\xi) - \varphi_\TX(t_k-\tau_{k}^{(l)}+\xi) ]\big)$
will remain essentially constant over an interval of length $NT$ thereby, together
with $\exp(\jmath\Delta\omega\xi) \approx 1$, resulting in
\begin{align}
  \left| \int_{\INTVL_{\mathrm{P}}}
  e^{ \jmath {\Delta\omega} \xi } \,
  e^{-\jmath \pr_\RX(t_k+\xi) } \,
  e^{ \jmath \pr_\TX(t_k-\tau_{k}^{(l)}+\xi) } \,
  \tilde{w}(\xi-\tau_{k}^{(l)}) \, 
  \ud\xi \right| \approx 
  \left| \int_{\INTVL_{\mathrm{P}}}
  \tilde{w}(\xi-\tau_{k}^{(l)}) \, 
  \ud\xi \right| = 1
\end{align}
which confirms \eqref{eq:smallChangeCirc}.
From \figref{fig:acf_pn}, we can see that the phase noise process' bandwidth
is small enough for phase noise to sustain a correlation coefficient of
at least $0.75$ over the duration of a length-511 sounding sequence.
In addition the phase noise variance is very small as well.

Condition
\eqref{eq:assDSCirc} 
requires that the inverse of the coherence bandwidth of the
phase noise process $\varphi_\TX(t)$ is large compared to the maximum multipath delay
difference. 
From \figref{fig:acf_pn}, we can conclude that for the sounder
under consideration this is the case only for multipath delay differences of up to 
$5$\unit{$\upmu$s}, i.e., only for
indoor or small-cell cellular outdoor propagation scenarios. 
Consequently, assumption \eqref{eq:assDSCirc}
needs to be carefully verified on a case by case basis for 
outdoor measurments.

Finally, even though only relevant in the case of linear convolution-based sounders,
we would like to comment on the condition 
$\Delta\omega\tau_k^{(l)} \approx 0, \forall k,l$.
We start by noting that typical 
values for maximum multipath delay differences 
range from 1\unit{$\upmu$s}
for indoor environments, 
over 10\unit{$\upmu$s} for outdoor small-cell cellular systems 
(with a few kilometers cell size),
to 100\unit{$\upmu$s} for a propagation range of
30\unit{km} encountered in outdoor large-cell cellular systems.
With $\Delta\omega = 2.4$\unit{rad/s}, this corresponds to 
$\Delta\omega\tau_k^{(l)}$ ranging from $2.4\cdot10^{-6}$\unit{rad} 
to $2.4\cdot10^{-4}$\unit{rad}. We can therefore conclude that
the condition $\Delta\omega\tau_k^{(l)} \approx 0$ is well satisfied in practice.

\paragraph*{Other imperfections}
Finally, before quantifying the impact of phase errors
on estimated \ac{MI} in our calibration measurement,
we need to convince ourselves that the errors observed are,
indeed, phase errors and are not caused by other measurement 
imperfections.
This can be done as follows. We start by noting that, in polar coordinates,
phase errors are visible only in the argument of the effective
channel and not in the magnitude. 
Assuming that all additional error sources can be subsumed
as additive noise $n$ which is circularly symmetric distributed, we can
measure the \ac{SNR}
in the magnitude direction which, due to 
\begin{align*}
  \big| h \, e^{\jmath \pr} + n \big|^{2} 
\equidist 
  \big| h \, e^{\jmath \pr} + n \, e^{\jmath \pr} \big|^{2} 
\equidist 
  \big| h + n \big|^{2},
\end{align*}
is simply the \ac{SNR} of $h + n$. By circular symmetry
of the additive noise term, the result of this measurement provides us with the
noise caused by other error sources in the direction 
orthogonal to the magnitude direction which, 
by the small phase noise
assumption is the phase direction.
This leads to an \ac{SNR} level (in phase direction) of 
below $37.5$\unit{dB} \ac{WIP} 1\% for the length-511 sequence.
Compared to the corresponding 
phase noise \ac{SNR} level of below 15.1\unit{dB} \ac{WIP} 1\%, we can conclude that
phase noise is, indeed, the dominating error source.

\subsection{Estimated \ac{MI} for a Rank-1 Physical Channel}
\label{sec:calibration}

Next, we illustrate the impact of phase errors on estimated \ac{MI} 
by performing the rank-1 ``calibration procedure'' described in 
the first paragraph of \secref{sec:rank1}.
For brevity, we restrict ourselves to the length-511 
sounding sequence (where phase noise is essentially fully uncorrelated). 

\begin{figure}[htbp]
\centering
\subfigure[
Mutual information 
for measured channel and for reference rank-1 physical channel]{
\psfrag{MIMO snapshot index l}{\scriptsize{\ac{MIMO} snapshot index $l$}}
\psfrag{Mutual Information I [bit/s/Hz]}{\scriptsize{\acl{MI} $\hat{I}$ [bit/s/Hz]}}
\psfrag{measured}{\scriptsize{measured}}
\psfrag{rank-1}{\scriptsize{rank-1}}
\psfrag{rank-1 AWGN}{\scriptsize{rank-1 \acs{AWGN}}}
\icg{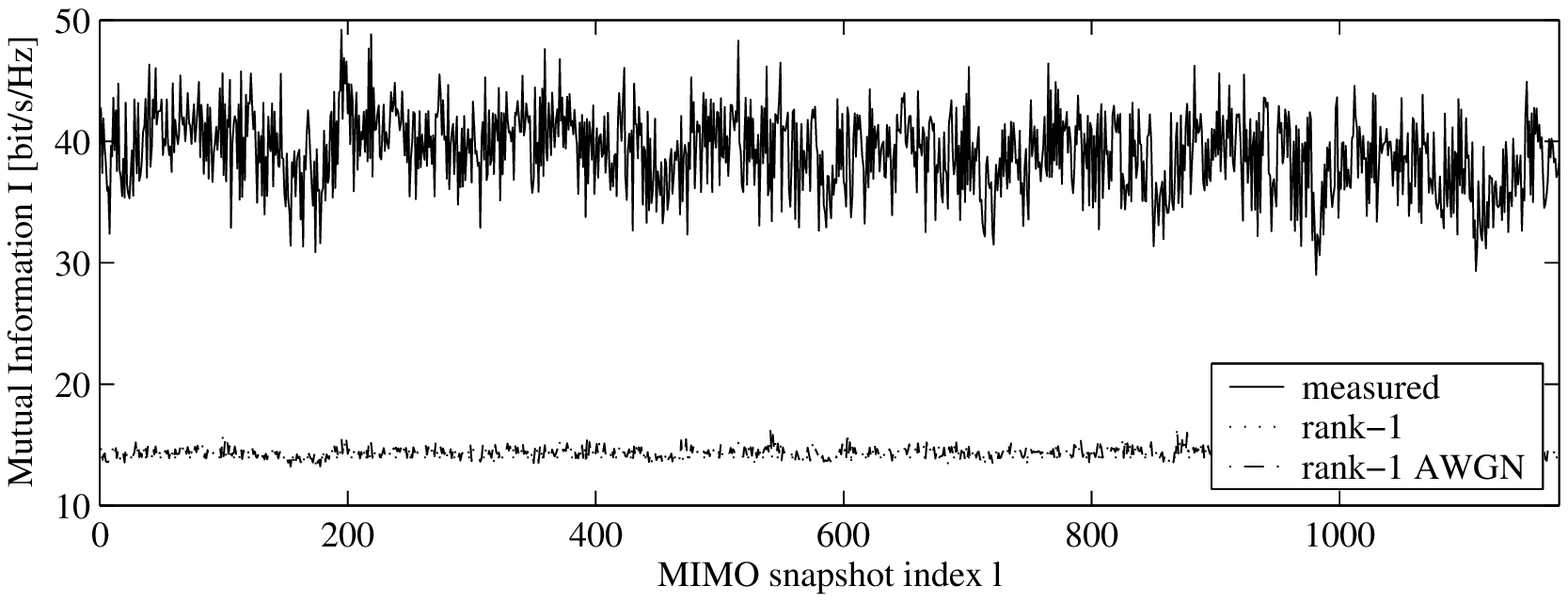}
\label{fig:corr_mi_pna}
} \\
\subfigure[Number of degrees of freedom $m_\FREE$ corresponding to 
results in (a)]{
\psfrag{MIMO snapshot index l}{\scriptsize{\acs{MIMO} snapshot index $l$}}
\psfrag{degrees of freedom DF free}{\scriptsize{degrees of freedom $m_\FREE$}}
\psfrag{measured}{\scriptsize{measured}}
\psfrag{rank-1}{\scriptsize{rank-1}}
\psfrag{rank-1 AWGN}{\scriptsize{rank-1 \acs{AWGN}}}
\icg{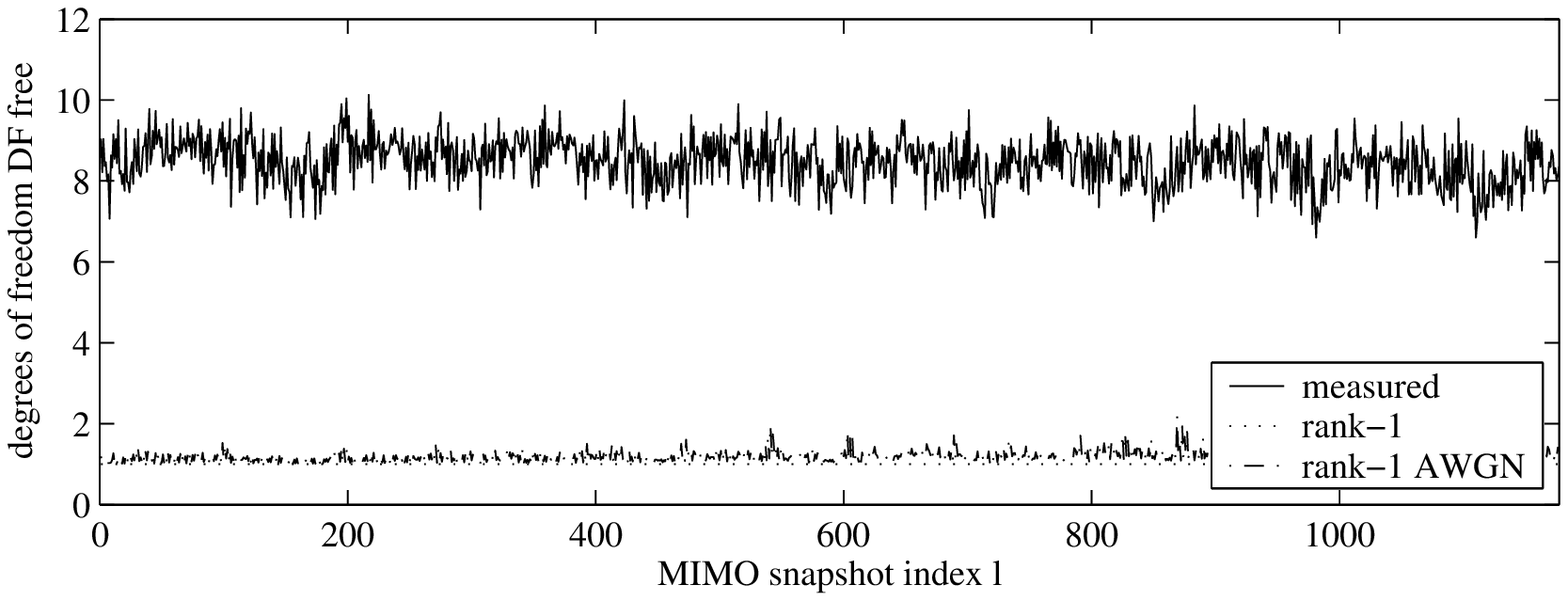}
\label{fig:corr_mi_pnb}
}
\vspace{-2mm}
\caption{\label{fig:corr_mi_pn}
Mutual information and corresponding number of degrees of freedom $m_\FREE$ 
at \acs{SNR} $\rho=30$\unit{dB} for $\MT=\MR=16$ configuration of \tabref{tab:meas}
and a rank-1 ``calibration measurement'' conducted on a commercially employed
\acs{TDMS}-based \acs{MIMO} channel sounder.
For reference, we also show the \acl{MI} corresponding to the rank-1 physical channel
$\mathbf{H}=\mathbf{1}$ and the mutual information corresponding to a 
rank-1 physical channel subject to \ac{AWGN} on the channel coefficients (see \secref{sec:calibration}).}
\end{figure}

\begin{figure}[htbp]
\centering
\psfrag{measured}{\scriptsize{measured}}
\psfrag{lower bound Theorem ww}{\scriptsize{lower bound, \propref{prop:CuZF04}}}
\psfrag{upper bound Theorem ww}{\scriptsize{upper bound, \propref{prop:Ubound}}}
\psfrag{SNR w [dB]}{\scriptsize{\ac{SNR} $\rho$ [dB]}}
\psfrag{Ergodic capacity C [bit/s/Hz]}{\scriptsize{Ergodic capacity $\widehat{C}$ [bit/s/Hz]}}
\icg{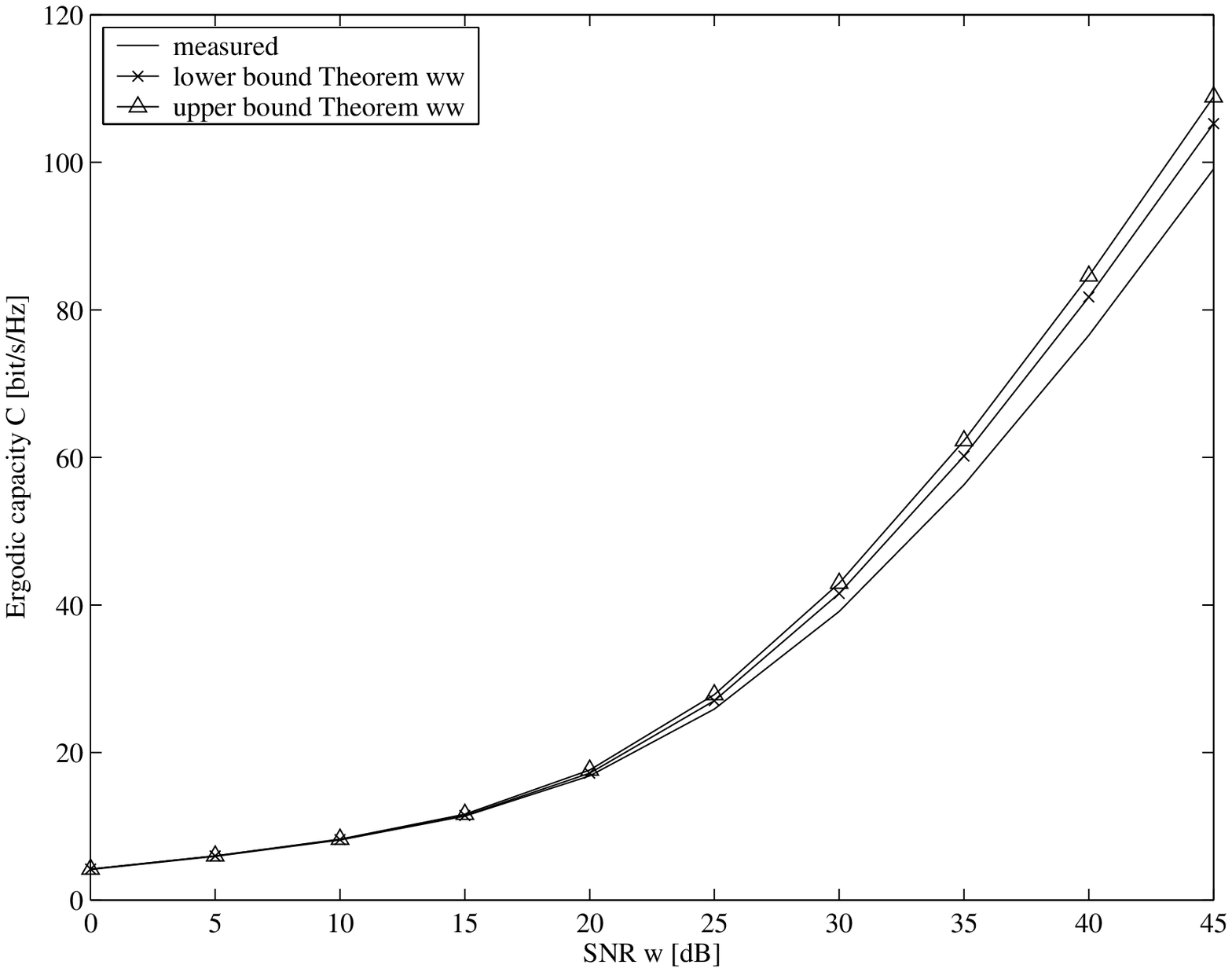}
\vspace{-2mm}
\caption{\label{fig:CDFandBounds}
Measured ergodic capacity along with corresponding analytic lower and upper bound
in \propsref{prop:CuZF04} and \ref{prop:Ubound}, respectively.}
\end{figure}

For an \ac{SNR} of $\rho = 30$\unit{dB}, 
\sfigref{fig:corr_mi_pn}{fig:corr_mi_pna} shows 
the \ac{MI} obtained from the measurements along with 
the corresponding \ac{MI} of the rank-1 physical channel $\mathbf{H}=\mathbf{1}$.
To illustrate the impact of additive noise on the channel coefficients, we also show
the capacity 
of a rank-1 physical channel subject to \ac{AWGN} on the channel coefficients, i.e., 
$\vect(\widehat{\mathbf{H}}) \equidist 
        \cplxGauss(\alpha \mathbf{1}, \sigma^2_N\mathbf{I})$ where $\alpha$, real-valued and positive, and $\sigma^2_N$ were
estimated (using the moment-matching method \cite{MiEG99c}) from the measurement results\footnote{The entries of the effective MIMO 
channel are, in general, not unit-modulus (due to, e.g., additive thermal
noise).} on a MIMO snapshot by MIMO snapshot basis.
The corresponding per-subchannel SNR is hence given by $\mbox{SNR}=\alpha^2/\sigma^2_N$.
While additive errors on the channel coefficients do not impact the capacity significantly, we can see that
the \ac{MI} of the effective channel is more than 200\% higher than
the \ac{MI} of the underlying physical channel. 
This significant measurement error can alternatively be quantified
in terms of the \ac{MIMO} channel's number of spatial degrees of freedom, 
using the definition provided in \cite{Gans02} as
\begin{align*}
m_\FREE = I_{\rho} - I_{\rho/2}.
\end{align*}
\sfigref{fig:corr_mi_pn}{fig:corr_mi_pnb} shows 
$m_\FREE$ corresponding to 
\sfigref{fig:corr_mi_pn}{fig:corr_mi_pna}.
We can see that the presence of phase noise increases $m_\FREE$
from 1 for the rank-1 physical channel
to 8-9 for the measured channel. 
An effective full-rank channel, i.e., 
a channel where all eigenvalues are significant at the given \ac{SNR},
would correspond to $m_\FREE=16$. 
Moreover, in \figref{fig:CDFandBounds} we show the 
ergodic capacity corresponding to the measurement results in 
\figref{fig:corr_mi_pn}, i.e., averaged over $L=1100$ \ac{MIMO} snapshots at each SNR,
along with the
ergodic capacity predicted by our analytic results in \propsref{prop:CuZF04}
and \ref{prop:Ubound} for 3.8\degreel\ \ac{RMS} fully uncorrelated phase noise. 
We can see that the analytic lower and upper bounds are slightly higher
than the measured ergodic capacity. This is probably due to 
the residual (very low since we are using length-511 sounding sequences)
correlation in the phase noise process, neglected in the analytic
results. Apart from this effect, the measurement results exhibit an
excellent match with the analytic results. 

\section{Conclusions}
\label{sec:conc}

We showed that phase errors (caused by phase noise and carrier frequency offset)
in \acf{TDMS}-based \acf{MIMOs} radio channel sounders 
can alter the channel statistics significantly
and thereby lead to severe \acf{MI} and capacity estimation errors.
The impact of phase errors is most pronounced for
low-rank physical channels where overestimation by several hundred percent can occur. 
A detailed analysis of the rank-1 physical \ac{MIMO} channel revealed
that realistic phase noise properties lead to a
decorrelation of the channel matrix and result in a full-rank effective channel. 
Our analytic results are supported by measurements conducted with 
a commercially employed
\ac{TDMS}-based \ac{MIMO} channel sounder.
In the particular measurement setup considered, 
phase noise turned a physical rank-1 channel into an effective rank-8 channel.
In general, for a given sounder,
the impact of phase errors will be more
pronounced for a larger number of antennas and for larger 
\ac{SISO} snapshot time distances.
Our analysis furthermore demonstrated that the 
\ac{MI} measurement error
induced by phase errors can depend significantly 
on the order in which the individual subchannels of
the overall \ac{MIMO} channel matrix are measured. 

Based on matrix differential calculus and matrix-variate 
Wirtinger calculus, we characterized the sensitivity of MI w.r.t. phase errors.

The presence of a carrier frequency offset between transmitter and receiver
was found to be less problematic as it does not impact the
\ac{MI} of the physical channel 
under quite general conditions on the sounding pattern.
Moreover, the impact of a carrier frequency offset is of deterministic nature 
and can be compensated by estimating the 
offset and subsequently removing it.

In the light of the main findings of this paper, 
the results obtained through
\ac{MIMO} channel measurement campaigns using 
\ac{TDMS}-based MIMO channel sounders
should be interpreted with great care. 
To the best of our knowledge, the large majority of 
commercially available \ac{MIMO} channel sounders 
is \ac{TDMS}-based.
In particular, measurement results reporting the absence of 
pin-hole or key-hole channels \cite{GBGP02,CFGV02}, seem questionable
unless a channel sounder with separate \acf{RF} chains for the individual transmit
and/or receive antenna elements is used. 

Simple averaging (\ac{WRT} phase noise realizations)
of \ac{MI} is not a viable way to mitigate the effect of phase errors as 
\ac{MI} is a nonlinear function of the phase noise samples and of 
the physical channel realization.
Averaging of entries of the effective channel
matrix over different \ac{MIMO} snapshots 
is not a viable option either as it leads to a 
significant reduction in measurement \ac{SNR}; 
this is due to the phase offsets between \ac{MIMO}
snapshots essentially being arbitrary. 
While this problem could, in principle, 
be mitigated through an increased 
\ac{MIMO} measurement rate, the latter is often limited by requirements
on the minimum sequence length, maximum duty cycle, and 
minimum switching times. 

A mitigation method based on taking multiple measurements per scalar subchannel at a higher rate with the goal of improving the phase
stability across antennas and, ideally, emulating one-sided
switching, is described, along with the corresponding tradeoffs, in \cite{diss_baum}. This method yields excellent results
in the case of small $\mbox{min}(\MT,\MR)$.

The safest, but most expensive, solution to avoid measurement errors 
due to phase errors
is to employ a channel sounder with separate \ac{RF} chains
for the individual transmit and/or receive antenna elements.
We hasten to add, however,
that such an architecture has other problems associated with it,
such as gain/phase imbalances and crosstalk between the different \ac{RF} chains.
Nevertheless, at least in theory,
as shown in \secref{sec:freq_offs}, if either the transmit antenna
array or the receive antenna array employs a separate \ac{RF} chain 
for each antenna element and \ac{TDMS} is used on the other side of the link, 
\ac{MI} is not affected by phase errors. 
If such a sounder architecture cannot be realized,
the \acfp{LO} should be selected very carefully or even custom-designed
which can be very expensive. 
As a last resort, we recommend to synchronize the \acp{LO} at transmitter and receiver 
through a cable. This solution, however, makes outdoor measurements difficult
and limits the flexibility of the sounding device, in particular \ac{WRT} the distance 
between transmitter and receiver.

\startappendix

\begin{theorem}[{Fang, Kotz, and Ng \cite[Th.\rsp1.4]{FaKN90b}}]
\label{theorem:dirichlet1}
Let $\mathbf{x}\equidist D_n(\mathbf{a})$ be partitioned into $k$ subvectors
$\mathbf{x}^{(1)},\mathbf{x}^{(2)},\ldots,\mathbf{x}^{(k)}$ and $\mathbf{a}$
into the corresponding subvectors 
$\mathbf{a}^{(1)},\mathbf{a}^{(2)},\ldots,\mathbf{a}^{(k)}$. 
Let $y_i$ and $b_i$ be, respectively, the sums of the components of 
$\mathbf{x}^{(i)}$ and $\mathbf{a}^{(i)}$, and set
$\mathbf{z}^{(i)} = \mathbf{x}^{(i)}/y_i$. The following statements
hold:
\begin{enumerate}
\item[i)]   The vectors $\mathbf{z}^{(i)},\mathbf{z}^{(i+1)},\ldots,\mathbf{z}^{(k)}$,
            $\mathbf{y} = [\,y_1\,\;\,y_2\,\;\,\cdots\,\;\,y_k\,]^T$ 
            are statistically independent.
\item[ii)]  $\mathbf{y}$ is distributed as $D_k(b_1,b_2,\ldots,b_k)$.
\item[iii)] $\mathbf{z}^{(i)}$ is distributed as $D_{n_i}(\mathbf{a}^{(i)})$, 
            $i=1,2,\ldots,k$, where $n_i$ is the number of elements in the
            subvectors $\mathbf{a}^{(i)}$ and $\mathbf{z}^{(i)}$.
\end{enumerate}
\end{theorem}
From \propref{theorem:dirichlet1} it follows immediately 
that the vectors $\mathbf{x}^{(i)}$ 
are Dirichlet (or beta) distributed. 
This result can be stated formally as follows.
\begin{theorem}[{Fang, Kotz, and Ng \cite[Th.\rsp1.5]{FaKN90b}}]
\label{theorem:dirichlet2}
If 
$[\, x_1 \,\;\, x_2 \,\;\, \cdots \,\;\, x_{n-1} \,]
\equidist D_{n-1}(a_1, a_2, \ldots,$\linebreak
$a_{n-1}; a_n)$, 
then for any $k<n-1$, we have 
$[\, x_1 \,\;\, x_2 \,\;\, \cdots \,\;\, x_{k} \,] 
\equidist D_{k}(a_1, a_2, \ldots, a_{k}; b)$,
where $b = a_{k+1}+a_{k+2}+\cdots+a_{n}$.
In particular, for any $i$ ($i=1,2,\ldots,n-1$), we have 
$x_i \equidist \beta(a_i, a - a_i)$ with $a = \sum_{i=1}^n a_i$.
\end{theorem}

\begin{lemma}
\label{lemma:dig_der}
The first derivative $\digammader(z) = \ud\digamma(z)/\ud z$
of the digamma function $\digamma(z)$ defined in \eqref{eq:dig_def} 
satisfies
\begin{align*}
  \digammader\left( z \right) 
\le 
  \frac{1}{2} \, \digammader\left( \frac{z}{2} \right).
\end{align*}

\begin{IEEEproof}
Using \eqref{eq:dig_der}, we have
\begin{align*}
\digammader\left( \frac{x}{2} \right) 
&= 
  \sum_{p=0}^\infty \frac{1}{\left(p+\frac{x}{2}\right)^2}
= \sum_{p=0}^\infty \frac{4}{(2p+x)^2} 
= 4\left(\frac{1}{x^2} + \frac{1}{(2+x)^2} + \frac{1}{(4+x)^2} + \cdots \right)
\\
&= 
  4 \left( \sum_{p=0}^\infty \frac{1}{(p+x)^2} 
         - \sum_{p=0}^\infty \frac{1}{\big( (2p+1)+x \big)^2} \right)
= 
  4 \, \digammader\left( x \right) 
- \digammader\left( \frac{x+1}{2} \right)
\end{align*}
which upon noting that 
\begin{align*}
\digammader\left( \frac{x}{2} \right) 
\ge
\digammader\left( \frac{x+1}{2} \right)
\end{align*}
completes the proof. 
\end{IEEEproof}
\end{lemma}

\begin{lemma}
\label{lemma:lemma2}
The eigenvalues of the matrix 
\begin{align*}
\real( \mathbf{A}_i )
=
\mathbf{I}_{\MT} 
- \sum_{n=1}^{i-1} \frac{ \real( \mathbf{s}^\bot_n\mathbf{s}_n^{\bot H} ) }{ \|\mathbf{s}_n^\bot\|^2 },
\qquad 
i=2,3,\ldots,\MR
\end{align*}
with $\mathbf{A}_i$ and $\mathbf{s}^\bot_n$ defined in \eqref{eq:GSstep}
and $\mathbf{s}_n$ defined through 
$\mathbf{S} 
= [\, \mathbf{s}_1 \,\;\, \mathbf{s}_2 \,\;\, \ldots \,\;\, \mathbf{s}_{\MR} \,]^T 
= -\jmath\mathbf{\Sigma} + \widetilde{\mathbf{\Phi}}$ 
with $\mathbf{\Sigma}$ as in \eqref{eq:Sigma1},
are given by ($\vk = 1,2,\ldots,\MT$)
\begin{align*}
  \{ \sigma_{\vk}^{(i)} \}
= 
  \big\{ \underbrace{1,\ldots,1}_{\MT-i} \, , \, 
  \underbrace{0,\ldots,0}_{i-2}   \, , \,
  \lrv^{(i)}, 1-\lrv^{(i)} \big\}
\end{align*}
where $\lrv^{(i)} = \lrv^{(i)}\mo(\mathbf{s}^\bot_1, \mathbf{s}^\bot_2, \ldots, \mathbf{s}^\bot_{i-1}) 
\in [0,1]$.

\begin{IEEEproof}
We start by writing 
\begin{align*}
  \real( \mathbf{A}_i )
&=
\mathbf{I}_{\MT} 
- \sum_{n=1}^{i-1} 
\frac{ \real(\mathbf{s}^\bot_n) \realT{(\mathbf{s}_n^{\bot})} }
     { \|\mathbf{s}_n^\bot\|^2 }
- \sum_{n=1}^{i-1} 
\frac{ \imag(\mathbf{s}^\bot_n) \imagT{(\mathbf{s}_n^{\bot})} }
     { \|\mathbf{s}_n^\bot\|^2 } \\
&= 
\mathbf{I}_{\MT} - \mathbf{G}_i \mathbf{G}_i^T
\end{align*}
where 
\begin{align*}
\mathbf{G}_i = \big[ \, \real(\mathbf{N}_i) \,\;\, \imag(\mathbf{N}_i) \, \big] 
               \qquad \text{with} \qquad
\mathbf{N}_i = 
\left[ \, \frac{ \mathbf{s}^\bot_1 }{ \| \mathbf{s}^\bot_1 \| } \,\;\,
          \frac{ \mathbf{s}^\bot_2 }{ \| \mathbf{s}^\bot_2 \| } \,\;\,
          \cdots \,\;\,          \frac{ \mathbf{s}^\bot_{i-1} }{ \| \mathbf{s}^\bot_{i-1} \| } \, \right].
\end{align*}
By slight abuse of notation, in the remainder of this proof, 
we let $\lambda_k(\mathbf{X})$ denote the unordered eigenvalues 
of the matrix $\mathbf{X}$.
It follows that 
$\sigma_k^{(i)} = 1 - \lambda_k(\mathbf{G}_i \mathbf{G}_i^T)$,
$k=1,2,\ldots,\MT$.
Invoking \lemmaref{lemma:lemma3}, we can conclude that 
the $2(i-1)$ eigenvalues $\lambda_k(\mathbf{G}_i^T \mathbf{G}_i)$ 
are given by 
\begin{align}
  \left\{ \frac{1}{2} + \frac{1}{2}\mysqrt{\mu_1^{(i)}}, \frac{1}{2} - \frac{1}{2}\mysqrt{\mu_1^{(i)}}, 
  \ldots, \frac{1}{2} + \frac{1}{2}\mysqrt{\mu_{i-1}^{(i)}}, 
          \frac{1}{2} - \frac{1}{2}\mysqrt{\mu_{i-1}^{(i)}} \right\}
\label{eq:2kev}
\end{align}
with $\mu_k^{(i)}\in\numID{R}$ $(k = 1,2,\ldots,i-1)$.
Therefore, when paired properly,
the $\lambda_k(\mathbf{G}_i^T \mathbf{G}_i)$
pairwise add up to 1. This property
will next allow us to show that $\mathbf{G}_i^T \mathbf{G}_i$ 
has $i-2$ eigenvalues equal to 1,
$i-2$ eigenvalues equal to 0,
and one pair of eigenvalues given by 
$\{ \lrv^{(i)}, 1 - \lrv^{(i)} \}$.
We start by noting that, using \eqref{eq:GSstep}, it follows for $i=2,3,\ldots,\MR$
that (recall that $\mathbf{s}_i\in\numID{R}^{\MT}$)
\begin{align}
  \imag\left( \frac{\mathbf{s}_i^\bot}{\|\mathbf{s}_i^\bot\|} \right) \nonumber
&=
- \sum_{n=1}^{i-1} \frac{\imag(\mathbf{s}_n^\bot \mathbf{s}_n^{\bot H})}
                        {\|\mathbf{s}_n^\bot\|^2} \, 
                   \frac{\mathbf{s}_i}{\|\mathbf{s}_i^\bot\|} \nonumber \\
&=
- \sum_{n=1}^{i-1} \frac{ \imag(\mathbf{s}^\bot_n) \realT{(\mathbf{s}_n^{\bot})} 
                        - \real(\mathbf{s}^\bot_n) \imagT{(\mathbf{s}_n^{\bot})} }
                          { \|\mathbf{s}_n^\bot\|^2 } \, 
                   \frac{\mathbf{s}_i}{\|\mathbf{s}_i^\bot\|} \nonumber \\
&= \phantom{-}
\sum_{n=1}^{i-1} \left(
\imag(\mathbf{s}^\bot_n) \, \xi_n  +  \real(\mathbf{s}^\bot_n) \, \zeta_n \right)
\label{eq:lincomb}
\\
  \real\left( \frac{\mathbf{s}_i^\bot}{\|\mathbf{s}_i^\bot\|} \right) \nonumber
&=
  \left( \mathbf{I}_{\MT} 
- \sum_{n=1}^{i-1} 
  \frac{ \real(\mathbf{s}^\bot_n) \realT{(\mathbf{s}_n^{\bot})} }
       { \|\mathbf{s}_n^\bot\|^2 }
- \sum_{n=1}^{i-1} 
  \frac{ \imag(\mathbf{s}^\bot_n) \imagT{(\mathbf{s}_n^{\bot})} }
       { \|\mathbf{s}_n^\bot\|^2 } \right)
  \frac{\mathbf{s}_i}{\|\mathbf{s}_i^\bot\|} \nonumber 
\\
&=
  \frac{\mathbf{s}_i}{\|\mathbf{s}_i^\bot\|} 
+ \sum_{n=1}^{i-1} \left(
  \real(\mathbf{s}^\bot_n) \, \xi_n
+ \imag(\mathbf{s}^\bot_n) \, \zeta_n \right)
\label{eq:indcomb}
\end{align}
with $\xi_n, \zeta_n \in \numID{R}$. 
The significance of \eqref{eq:lincomb} 
and \eqref{eq:indcomb} is that it can be used to show
that $\rank(\mathbf{G}_{\vi})=\vi$. 
More specifically, starting with $i=2$, we note that
$\mathbf{G}_2 = [\, \real(\mathbf{s}_1/\|\mathbf{s}_1\|) \,\;\, 
                    \imag(\mathbf{s}_1/\|\mathbf{s}_1\|) \,]$ 
has rank 2 \ac{WIP}\!\!\rsp1 as will be shown first. 
Noting that $\real(\mathbf{s}_1)
\equidist \realGauss(\mathbf{0},\sigma^2_\varphi\mathbf{I})$
and
$\imag(\mathbf{s}_1)
= [\,-\MT\MR \,\;\, \mathbf{0}_{1,\MT-1}\,]^T$, 
it follows that $\|\mathbf{s}_1\|>0$ \ac{WIP}\!\!\rsp1
and hence $r(\mathbf{G}_2) 
= r([\, \real(\mathbf{s}_1) \,\;\, \imag(\mathbf{s}_1) \,])$.
By definition, 
$\real(\mathbf{s}_1)$ and $\imag(\mathbf{s}_1)$
are linearly independent \ac{WIP}\!\!\rsp1
and hence $r(\mathbf{G}_2) = 2$ \ac{WIP}\!\!\rsp1.
Now with each increase in $i$, two vectors are added to 
$\mathbf{G}_i$, where one, namely 
$\imag( \mathbf{s}_{i-1}^\bot / \| \mathbf{s}_{i-1}^\bot \| )$
by \eqref{eq:lincomb},
is a linear combination of the vectors 
already in $\mathbf{G}_i$, and the other one, namely
$\real( \mathbf{s}_{i-1}^\bot / \| \mathbf{s}_{i-1}^\bot \| )$
by \eqref{eq:indcomb},
is a linear combination of the vectors already in $\mathbf{G}_i$
and the vector $\mathbf{s}_{i-1} / \| \mathbf{s}_{i-1}^\bot \|$.
We can therefore conclude that the 
$2(\vi-1)\times2(\vi-1)$ matrix $\mathbf{G}_{\vi}^T\mathbf{G}_{\vi}$ has at most 
$\vi$ nonzero eigenvalues.
The remaining $\vi-2$ eigenvalues are equal to zero.
By \eqref{eq:2kev} we must therefore have $\vi-2$ eigenvalues which are equal to 1.
Since we have $2\vi-2$ eigenvalues in total, it follows that
there is one pair of eigenvalues of the form
$\{ \lrv^{(\vi)}, 1 - \lrv^{(\vi)} \}$.
Finally, noting that $\lrv^{(\vi)}$ and $1-\lrv^{(\vi)}$ must be real-valued 
and positive, we can conclude that $\lrv^{(\vi)}\in[0,1]$.
\end{IEEEproof}
\end{lemma}

\begin{lemma}
\label{lemma:lemma3}
Given an orthonormal set of vectors 
$\mathbf{x}_1,\mathbf{x}_2,\ldots,\mathbf{x}_{\vn}\in \numID{C}^{N}$ with $\vn\le N$,
the eigenvalues of the $2\vn\times2\vn$ matrix
$\mathbf{Y}^T \mathbf{Y}$ with
\begin{align*}
\mathbf{Y} 
&= \big[ \, \real(\mathbf{X}) \,\;\, \imag(\mathbf{X}) \, \big]
\end{align*}
where
\begin{align*}
  \mathbf{X} 
&= 
  [ \, \mathbf{x}_1 \,\;\, \mathbf{x}_2 \,\;\, \cdots \,\;\, \mathbf{x}_{\vn} \, ]
\end{align*}
are given by
\begin{align}
\begin{split}
\bigg\{ \frac{1 + \mysqrt{\lambda_1(\mathbf{A})}}{2}, 
        \frac{1 - \mysqrt{\lambda_1(\mathbf{A})}}{2}, 
        \frac{1 + \mysqrt{\lambda_2(\mathbf{A})}}{2}, 
&       \frac{1 - \mysqrt{\lambda_2(\mathbf{A})}}{2}, \ldots,
\\
&       \frac{1 + \mysqrt{\lambda_\vn(\mathbf{A})}}{2}, 
        \frac{1 - \mysqrt{\lambda_\vn(\mathbf{A})}}{2} \bigg\}
\end{split}
\label{eq:2k_evs}
\end{align}
where 
$\mathbf{A} = \mathbf{X}^T \mathbf{X} \mathbf{X}^H \mathbf{X}^*$.

\begin{IEEEproof}
We start by noting that 
\begin{align*}
\mathbf{Y}' = \mysqrt{\frac{1}{2}} \, 
[ \, \mathbf{X}   \,\;\, 
     \mathbf{X}^* \, ]
\end{align*} 
satisfies
\begin{align*}
\mathbf{Y}' 
=
\mathbf{Y} \underbrace{ \left( 
\mysqrt{\frac{1}{2}} \; 
\left[ \begin{array}{cc}
1 &  1 \\
j & -j
\end{array} \right] \kron \mathbf{I}_{\vn} \right) }_{\mathbf{U}}
\end{align*}
where $\mathbf{U}$ is unitary. We can therefore conclude that
\begin{align*}
\lambda_{\vi}(\mathbf{Y}'^H \mathbf{Y}') 
= \frac{1}{2} \, 
\lambda_{\vi}\mo\left(
\left[ \begin{array}{cc} \mathbf{X}^H \\ \mathbf{X}^T \end{array} \right] 
     [ \, \mathbf{X} \,\;\, \mathbf{X}^* \, ] \right)
=
\lambda_{\vi}(\mathbf{U}^H \mathbf{Y}^H \mathbf{Y} \mathbf{U}) 
= 
\lambda_{\vi}(\mathbf{Y}^H \mathbf{Y}).
\end{align*}
Next, we note that $\mathbf{X}^H \mathbf{X} = \mathbf{X}^T \mathbf{X}^* = \mathbf{I}_{\vn}$
due to the orthonormality of the $\mathbf{x}_{\vn}$, and hence the eigenvalues of
$\mathbf{Y}^H \mathbf{Y}$ are given by the solutions of the characteristic equation
\begin{align*}
\ce(\mu) = 
\det\left( \left[ \begin{array}{cc}
(1-2\mu)\mathbf{I} & \mathbf{X}^H \mathbf{X}^* \\
\mathbf{X}^T \mathbf{X} & (1-2\mu)\mathbf{I} \end{array}
\right] \right). 
\end{align*}
For a Hermitian matrix $\mathbf{S}$ partitioned according to
\begin{align*}
\mathbf{S} = \left[ \begin{array}{cc}
\mathbf{A}   & \mathbf{B} \\
\mathbf{B}^* & \mathbf{C} \end{array} \right] 
\end{align*}
where $\mathbf{A}$ and $\mathbf{C}$ are square matrices, we have 
from the Schur complement formula \cite[Sec.~0.8.5]{HoJo85b}
\begin{align*}
\det(\mathbf{S}) = \det(\mathbf{A}) \, \det( \mathbf{C} - \mathbf{B}^* \mathbf{A}^{-1} \mathbf{B} ).
\end{align*}
It therefore follows that 
\begin{align}
p(\mu) = 
\det \big( (1-2\mu)^2 \mathbf{I} - \mathbf{X}^T \mathbf{X} \mathbf{X}^H \mathbf{X}^* \big).
\label{eq:eqR}
\end{align}
The solution of \eqref{eq:eqR} yields the eigenvalues of $\mathbf{Y}^H \mathbf{Y}$ as
\eqref{eq:2k_evs},
which completes the proof.
\end{IEEEproof}
\end{lemma}
Note that if $\vn\ge N/2$, $\mathbf{Y}$ must have $2\vn-N$ eigenvalues 
equal to 0 and therefore by \eqref{eq:2k_evs} 
also $2\vn-N$ eigenvalues equal to 1 so that
\eqref{eq:2k_evs} can be refined as
\begin{multline}
\bigg\{ 
        \frac{1 + \mysqrt{\lambda_1(\mathbf{A})}}{2}, 
        \frac{1 - \mysqrt{\lambda_1(\mathbf{A})}}{2}, 
        \frac{1 + \mysqrt{\lambda_2(\mathbf{A})}}{2}, 
        \frac{1 - \mysqrt{\lambda_2(\mathbf{A})}}{2}, 
        \ldots, 
\\
        \frac{1 + \mysqrt{\lambda_{N-\vn}(\mathbf{A})}}{2},
        \frac{1 - \mysqrt{\lambda_{N-\vn}(\mathbf{A})}}{2}
        \underbrace{ 1, \ldots, 1 }_{2\vn-N},
        \underbrace{ 0, \ldots, 0 }_{2\vn-N} \bigg\}.
\nonumber
\end{multline}

\begin{prop}
\label{prop:E_Z}
Given the conditional (on $\lrv$) \ac{RV}\footnote{For the sake of simplicity of notation, we committed an abuse of notation here, 
as in the main body (cf.~\propref{prop:dist_approx}) we used the symbol $Z(\lrv)$ to denote the unconditional \ac{RV} $Z(\lrv)=\sigma_{\PP}^2(\lrv X_{1}+(1-\lrv)X_{2})$.}
$Z(\lrv) \equidist \chi^2_{1,\sigma_{\PP}^2\lrv} + \chi^2_{1,\sigma_{\PP}^2(1-\lrv)}$,
where the two chi-square distributed terms are independent, 
and a constant $r\in\numID{R}^{+}$, we have
\begin{align}
  \expect\big\{ \log\big( r + Z(0) \big) \big\} 
\le 
  \expect\big\{ \log\big( r + Z(\lrv) \big) \big\}, 
\quad 
  0 \le \lrv \le 1. 
\label{eq:prop}
\end{align}

\begin{IEEEproof}
We start by noting that
$Z(\lrv) \equidist Z(1-\lrv)$ for $0\le\lrv\le1$ 
so that it suffices to prove \eqref{eq:prop}
for $0\le\lrv\le1/2$. We shall also need the properties 
$Z(0) \equidist Z(1) \equidist \chi^2_{1,\sigma_{\PP}^2}$ 
and $Z(1/2) \equidist \chi^2_{2,\sigma_{\PP}^2/2}$. 
Noting that
\begin{align*}
  \log\big( r + Z(\lrv) \big) 
&= 
  \log(r) + \log\big( 1 + Z(\lrv)/r \big) \\
&= 
  \log(r) + \log\big( 1 + \tilde{Z}(\lrv) \big) 
\end{align*}
where $\tilde{Z}(\lrv) \equidist \chi^2_{1,\tilde{\sigma}_{\PP}^2\lrv} 
                                  + \chi^2_{1,\tilde{\sigma}_{\PP}^2(1-\lrv)}$
and $\tilde{\sigma}_{\PP}^2 = \sigma_{\PP}^2/r$, 
it follows that proving \eqref{eq:prop} for $r=1$ is sufficient.

The 
\ac{CDF} of $Y(\xi)=\chi^2_{1,\xi}$ is 
\cite[Eq.\rsp26.4.19, Eq.\rsp6.5.16, Eq.\rsp6.1.8]{AbSt72b}
\begin{align}
\cdf_{Y(\xi)}(y) &= \erf\left( \mysqrt{ \frac{y}{2\xi} } \right)
\label{eq:cdfY}
\end{align}
where 
$\erf(x) = (2/\mysqrt{\pi})\int_0^x \exp(-t^2) \, \ud t$ 
denotes the error function.
The inverse function corresponding to \eqref{eq:cdfY} is given by
$\cdf^{-1}_{Y(\xi)}(x) = 2\xi\erfi^2(x)$,
where $\erfi(x)$ denotes the inverse error function.
Using the inverse method of generating random deviates \cite[Sec.\rsp26.8]{AbSt72b},
i.e., for a uniformly distributed \ac{RV} $U \in [0,1]$ we have
$\chi^2_{1,\xi} \equidist \cdf^{-1}_{Y(\xi)}(U)$,
together with the independence of the two terms in 
$Z(\lrv) \equidist \chi^2_{1,\sigma_{\PP}^2\lrv} + \chi^2_{1,\sigma_{\PP}^2(1-\lrv)}$,
we can now express $Z(\lrv)$ in terms of two
independent uniformly distributed \acp{RV} 
$U_1 \in [0,1]$ and $U_2 \in [0,1]$ as
\begin{align}
Z(\lrv) \equidist
  2\sigma^2 \lrv \, \erfi^2( U_1 ) 
+ 2\sigma^2 (1-\lrv) \, \erfi^2( U_2 ).
\label{eq:Zuni}
\end{align}

To prove the Theorem, it suffices to show that
\begin{align}
  \frac{\partial}{\partial \lrv} \, 
  \expect\big\{ \log\big( 1 + Z(\lrv) \big) \big\} 
&\ge 
  0, 
\quad 
  0 \le \lrv \le 1/2.
\label{eq:diffE}
\end{align}
Inserting \eqref{eq:Zuni} in \eqref{eq:diffE},
we get 
\begin{align}
  \frac{\partial}{\partial \lrv} \, 
  \expect\big\{ \log\big( 1 + Z(\lrv) \big) \big\} 
&=
  \frac{\partial}{\partial \lrv} \, 
  \int_{0}^{1}\int_{0}^{1} \log\big( 1 + 2\sigma^2\lrv\,\erfi^2(u_1) 
+ 2\sigma^2(1-\lrv)\,\erfi^2(u_2) \big) \, \ud u_1 \ud u_2
\nonumber
\\
&\stackrel{\text{(a)}}{=}
\log(e) 
\int_{0}^{1}\int_{0}^{1}
\frac{2\sigma^2 \big( \erfi^2(u_1) - \erfi^2(u_2) \big) }
     {1 + 2\sigma^2\lrv\,\erfi^2(u_1) + 2\sigma^2(1-\lrv)\,\erfi^2(u_2)} 
\, \ud u_1 \ud u_2
\nonumber
\\
&= \log(e) \int_{0}^{1}\int_{0}^{1}
f_1(u_1,u_2) f_2(u_1,u_2)
\, \ud u_1 \ud u_2
\label{eq:F1F2}
\end{align}
where in (a) we interchanged expectation and differentiation
(noting that the integrand is continuous for all $\eta\in[0,1]$), 
and we set
\begin{align*}
&
&
f_1(u_1,u_2) 
&= 
\erfi^2(u_1) - \erfi^2(u_2) 
\\
& 
&
f_2(u_1,u_2) 
&=
\left(1/(2\sigma^2)+\lrv\,\erfi^2(u_1)+(1-\lrv)\,\erfi^2(u_2)\right)^{-1}.
\end{align*}
Note that $f_1(u_1,u_2)$ is negative symmetric, i.e.,  
$f_1(u_1,u_2) = -f_1(u_2,u_1)$. 
Furthermore, the monotonicity of $\erfi(u)$ implies that 
for $u_1\ge u_2$ we have $f_1(u_1,u_2)\ge 0$, and
for $u_1\le u_2$ it holds that $f_1(u_1,u_2)\le 0$. 
We will now exploit these properties of $f_1(u_1,u_2)$ 
to complete the proof and start by rewriting the integral
in \eqref{eq:F1F2} according to
\begin{align}
& \int_{u_1=0}^1 \int_{u_2=0}^1
f_1(u_1,u_2) f_2(u_1,u_2)
\, \ud u_1 \, \ud u_2 
\nonumber
\\
& \qquad = \int_{v=0}^1 \left(
\int_{u_1=v}^1 f_1(u_1,v) f_2(u_1,v) \, \ud u_1 +
\int_{u_2=v}^1 f_1(v,u_2) f_2(v,u_2) \, \ud u_2 \right) \ud v 
\nonumber
\\
& \qquad = \int_{v=0}^1
\int_{u=v}^1 f_1(u,v) \big( f_2(u,v) - f_2(v,u) \big) \, \ud u \, \ud v
\label{eq:Fsymm}
\end{align}
where $f_1(u,v)\ge 0$ in the entire range of integration since $u\ge v$.
Finally, we will show that $f_2(u,v)-f_2(v,u) \ge 0$, 
which by \eqref{eq:Fsymm} then implies \eqref{eq:diffE}.
Straightforward manipulations reveal that the condition
$f_2(u,v)-f_2(v,u) \ge 0$ is equivalent to 
\begin{align*}
\lrv\,\erfi^2(u)+(1-\lrv)\,\erfi^2(v)
&\le
\lrv\,\erfi^2(v)+(1-\lrv)\,\erfi^2(u) 
\\
(2\lrv-1)\,\erfi^2(u)
&\le
(2\lrv-1)\,\erfi^2(v)
\end{align*}
and hence by $0\le\lrv\le 1/2$ to
\begin{align*}
\erfi^2(u) &\ge \erfi^2(v)
\end{align*}
which is satisfied for $u\ge v$ (i.e., over the entire range of integration)
because $\erfi(x)$ is nondecreasing.
\end{IEEEproof}
\end{prop}

\section*{Acknowledgment}
We would like to thank Vinko Erceg and Pieter van Rooyen 
for helping both plan and making possible the measurement campaign 
that led to the discovery of the phase noise problem
in \ac{TDMS}-based \ac{MIMO} channel sounding.

\setcounter{totalnumber}{3}
\newpage
\setstretch{1.0}

\clearpage
\setstretch{1.8}

\bibliographystyle{IEEEtran}
\bibliography{IEEEabrv,phase_noise}

\begin{thebibliography}{10}
\providecommand{\url}[1]{#1}
\csname url@samestyle\endcsname
\providecommand{\newblock}{\relax}
\providecommand{\bibinfo}[2]{#2}
\providecommand{\BIBentrySTDinterwordspacing}{\spaceskip=0pt\relax}
\providecommand{\BIBentryALTinterwordstretchfactor}{4}
\providecommand{\BIBentryALTinterwordspacing}{\spaceskip=\fontdimen2\font plus
\BIBentryALTinterwordstretchfactor\fontdimen3\font minus
  \fontdimen4\font\relax}
\providecommand{\BIBforeignlanguage}[2]{{%
\expandafter\ifx\csname l@#1\endcsname\relax
\typeout{** WARNING: IEEEtran.bst: No hyphenation pattern has been}%
\typeout{** loaded for the language `#1'. Using the pattern for}%
\typeout{** the default language instead.}%
\else
\language=\csname l@#1\endcsname
\fi
#2}}
\providecommand{\BIBdecl}{\relax}
\BIBdecl

\bibitem{MMHS02j}
G.~Matz, A.~F. Molisch, F.~Hlawatsch, M.~Steinbauer, and I.~Gaspard, ``On the
  systematic measurement error of correlative mobile radio channel sounders,''
  \emph{{IEEE} Trans. Commun.}, vol.~50, no.~5, pp. 808--821, May 2002.

\bibitem{GBGP02}
D.~Gesbert, H.~B{\"o}lcskei, D.~A. Gore, and A.~J. Paulraj, ``Outdoor {MIMO}
  wireless channels: {M}odels and performance prediction,'' \emph{{IEEE} Trans.
  Commun.}, vol.~50, no.~12, pp. 1926--1934, Dec. 2002.

\bibitem{CFGV02}
D.~Chizhik, G.~J. Foschini, M.~J. Gans, and R.~A. Valenzuela, ``Keyholes,
  correlations, and capacities of multielement transmit and receive antennas,''
  \emph{{IEEE} Trans. Wireless Commun.}, vol.~1, no.~2, pp. 361--368, Apr.
  2002.

\bibitem{KiVa97}
J.~Kivinen and P.~Vainikainen, ``Phase noise in a direct sequence based channel
  sounder,'' in \emph{Proc.\ IEEE Int.\ Symp.\ on Personal, Indoor \& Mobile
  Radio Commun.\ (PIMRC)}, vol.~3, Helsinki, Finland, Sep. 1997, pp.
  1115--1119.

\bibitem{KiVa00}
------, ``Calibration scheme for synthesizer phase fluctuations in virtual
  antenna array measurements,'' \emph{Microwave \& Optical Technol.\ Lett.},
  vol.~26, no.~3, pp. 183--187, Jun. 2000.

\bibitem{FeHe94j}
J.~A. Fessler and A.~O. Hero, ``Space-alternating generalized
  expectation-maximization algorithm,'' \emph{{IEEE} Trans. Signal Processing},
  vol.~42, no.~10, pp. 2664--2677, Oct. 1994.

\bibitem{AWTM05c}
P.~Almers, S.~Wyne, F.~Tufvesson, and A.~F. Molisch, ``Effect of random walk
  phase noise on {MIMO} measurements,'' in \emph{Proc.\ IEEE Veh.\ Technol.\
  Conf.\ (VTC) Spring}, vol.~1, Stockholm, Sweden, May 2005, pp. 141--145.

\bibitem{Gans02}
M.~J. Gans, N.~Amitay, Y.~S. Yeh, H.~Xu, T.~C. Damen, R.~A. Valenzuela,
  T.~Sizer, R.~Storz, D.~Taylor, W.~M. MacDonald, C.~Tran, and A.~Adamiecki,
  ``Outdoor {BLAST} measurement system at 2.44\unit{GHz}: {C}alibration and
  initial results,'' \emph{{IEEE} J. Select. Areas Commun.}, vol.~20, no.~3,
  pp. 570--583, Apr. 2002.

\bibitem{Kyritsi02}
P.~Kyritsi, R.~A. Valenzuela, and D.~C. Cox, ``Channel and capacity estimation
  errors,'' \emph{{IEEE} Commun. Lett.}, vol.~6, no.~12, pp. 517--519, Dec.
  2002.

\bibitem{LoKo02j}
S.~Loyka and A.~Kouki, ``On {MIMO} channel capacity, correlations, and
  keyholes: {A}nalysis of degenerate channels,'' \emph{{IEEE} Trans. Commun.},
  vol.~50, no.~12, pp. 1886--1888, Dec. 2002.

\bibitem{AlTM03}
P.~Almers, F.~Tufvesson, and A.~F. Molisch, ``Measurement of keyhole effect in
  a wireless multiple-input multiple-output ({MIMO}) channel,'' \emph{{IEEE}
  Commun. Lett.}, vol.~7, no.~8, pp. 373--375, Aug. 2003.

\bibitem{MaNe88b}
J.~R. Magnus and H.~Neudecker, \emph{Matrix Differential Calculus with
  Applications in Statistics and Econometrics}, 2nd~ed., ser. Wiley Series in
  Probability \& Statistics.\hskip 1em plus 0.5em minus 0.4em\relax Chichester,
  UK: Wiley-VCH, 1999.

\bibitem{NeMa93j}
F.~D. Neeser and J.~L. Massey, ``Proper complex random processes with
  applications to information theory,'' \emph{{IEEE} Trans. Inform. Theory},
  vol.~39, no.~4, pp. 1293--1302, Jul. 1993.

\bibitem{FaKN90b}
K.-T. Fang, S.~Kotz, and K.-W. Ng, \emph{Symmetric Multivariate and Related
  Distributions}, ser. Monographs on Statistics and Applied Probability.\hskip
  1em plus 0.5em minus 0.4em\relax London, UK and New York, NY: Chapman and
  Hall, 1990.

\bibitem{AbSt72b}
M.~Abramowitz and I.~A. Stegun, Eds., \emph{Handbook of Mathematical Functions
  with Formulas, Graphs, and Mathematical Tables}, 9th~ed.\hskip 1em plus 0.5em
  minus 0.4em\relax New York, NY: Dover, 1972.

\bibitem{MMSH99c}
G.~Matz, A.~F. Molisch, M.~Steinbauer, F.~Hlawatsch, I.~Gaspard, and H.~Artés,
  ``Bounds on the systematic measurement error of channel sounders for
  time-varying mobile radio channels,'' in \emph{Proc.\ IEEE Veh.\ Technol.\
  Conf.\ (VTC) Fall}, vol.~3, Amsterdam, The Netherlands, Sep. 1999, pp.
  1465--1470.

\bibitem{FlJo96c}
P.~G. Flikkema and S.~G. Johnson, ``A comparison of time- and frequency-domain
  wireless channel sounding techniques,'' in \emph{Proc.\ IEEE SoutheastCon},
  Tampa, FL, Apr. 1996, pp. 488--491.

\bibitem{CuFM93j}
P.~J. Cullen, P.~C. Fannin, and A.~Molina, ``Wide-band measurement and analysis
  techniques for the mobile radio channel,'' \emph{{IEEE} Trans. Veh.
  Technol.}, vol.~42, no.~4, pp. 589--603, Nov. 1993.

\bibitem{PaDT91c}
J.~D. Parsons, D.~A. Demery, and A.~M.~D. Turkmani, ``Sounding techniques for
  wideband mobile radio channels: {A} review,'' \emph{Proc.\ IEE Part I:
  Commun., Speech \& Vision}, vol. 138, no.~5, pp. 437--446, Oct. 1991.

\bibitem{Rutm78}
J.~Rutman, ``Characterization of phase and frequency instabilities in precision
  frequency sources: {F}ifteen years of progress,'' \emph{Proc.\ IEEE},
  vol.~66, no.~9, pp. 1048--1075, Sep. 1978.

\bibitem{Rutm91j}
J.~Rutman and F.~L. Walls, ``Characterization of frequency stability in
  precision frequency sources,'' \emph{Proc.\ IEEE}, vol.~79, no.~7, pp.
  952--960, Jul. 1991.

\bibitem{GeBa85}
E.~A. Gerber and A.~Ballato, Eds., \emph{Precision Frequency Control: Acoustic
  Resonators and Filters (Volume 1), Oscillators and Standards (Volume
  2)}.\hskip 1em plus 0.5em minus 0.4em\relax New York, NY: Academic Press,
  1985.

\bibitem{DeMR00j}
A.~Demir, A.~Mehrotra, and J.~Roychowdhury, ``Phase noise in oscillators: {A}
  unifying theory and numerical methods for characterization,'' \emph{{IEEE}
  Trans. Circuits Syst.}, vol.~47, no.~5, pp. 655--674, May 2000.

\bibitem{Demi02j}
A.~Demir, ``Phase noise and timing jitter in oscillators with colored-noise
  sources,'' \emph{{IEEE} Trans. Circuits Syst.}, vol.~49, no.~12, pp.
  1782--1791, Dec. 2002.

\bibitem{Mehr02j}
A.~Mehrotra, ``Noise analysis of phase-locked loops,'' \emph{{IEEE} Trans.
  Circuits Syst.}, vol.~49, no.~9, pp. 1309--1316, Sep. 2002.

\bibitem{UFFCe}
\BIBentryALTinterwordspacing
(2006) {IEEE} ultrasonics, ferroelectrics, and frequency control society.
  [Online]. Available: \url{http://www.ieee-uffc.org/}
\BIBentrySTDinterwordspacing

\bibitem{Wood53b}
P.~M. Woodward, \emph{Probability and Information Theory with Application to
  Radar}.\hskip 1em plus 0.5em minus 0.4em\relax London, UK: Pergamon Press,
  1953.

\bibitem{SRTW02c}
G.~Sommerkorn, A.~Richter, R.~S. Thom{\"a}, and W.~Wirnitzer, ``Antenna
  multiplexing and time alignment for {MIMO} channel sounding,'' in
  \emph{Proc.\ Int.\ Union of Radio Science (URSI) General Assembly (GA)},
  Maastricht, The Netherlands, Aug. 2002, {CP.P.28}, paper no.~1879.

\bibitem{HaLe99b}
A.~Hajimiri and T.~H. Lee, \emph{The Design of Low Noise Oscillators}.\hskip
  1em plus 0.5em minus 0.4em\relax Dordrecht, The Netherlands: Kluwer Academic
  Publishers, 1999.

\bibitem{Wood56j}
R.~A. Wooding, ``The multivariate distribution of complex normal variables,''
  \emph{Biometrika}, vol.~43, no. 1/2, pp. 212--215, Jun. 1956.

\bibitem{Tela99j}
I.~E. Telatar, ``Capacity of multi-antenna {G}aussian channels,'' \emph{Eur.\
  Trans.\ Telecommun.}, vol.~10, no.~6, pp. 585--595, Nov. 1999.

\bibitem{SFGK00j}
D.-S. Shiu, G.~J. Foschini, M.~J. Gans, and J.~M. Kahn, ``Fading correlation
  and its effect on the capacity of multielement antenna systems,''
  \emph{{IEEE} Trans. Commun.}, vol.~48, no.~3, pp. 502--513, Mar. 2000.

\bibitem{ONBP03j}
{\"O}.~Oyman, R.~U. Nabar, H.~B{\"o}lcskei, and A.~J. Paulraj, ``Characterizing
  the statistical properties of mutual information in {MIMO} channels,''
  \emph{{IEEE} Trans. Signal Processing}, vol.~51, no.~11, pp. 2784--2795, Nov.
  2003.

\bibitem{ALTV04c}
G.~Alfano, A.~Lozano, A.~M. Tulino, and S.~Verd\'{u}, ``Mutual information and
  eigenvalue distribution of {MIMO} {R}icean channels,'' in \emph{Proc.\ IEEE
  Int.\ Symp.\ on Inform.\ Theory \& Appl. (ISITA)}, Parma, Italy, Oct. 2004,
  pp. 1040--1045.

\bibitem{ZhCL05j}
Q.~T. Zhang, X.~W. Cui, and X.~M. Li, ``Very tight capacity bounds for
  {MIMO}-correlated {Rayleigh}-fading channels,'' \emph{{IEEE} Trans. Wireless
  Commun.}, vol.~4, no.~2, pp. 681-- 688, Mar. 2005.

\bibitem{TuLV05j}
A.~M. Tulino, A.~Lozano, and S.~Verd\'{u}, ``Impact of antenna correlation on
  the capacity of multiantenna channels,'' \emph{{IEEE} Trans. Inform. Theory},
  vol.~51, no.~7, pp. 2491--2509, Jul. 2005.

\bibitem{KaAl06j-Rayl}
M.~Kang and M.~S. Alouini, ``Capacity of correlated {MIMO} {Rayleigh}
  channels,'' \emph{{IEEE} Trans. Wireless Commun.}, vol.~5, no.~1, pp.
  143--155, Jan. 2006.

\bibitem{Jake91j}
W.~C. Jakes, Jr., ``A comparison of specific space diversity techniques for
  reduction of fast fading in {UHF} mobile radio systems,'' \emph{{IEEE} Trans.
  Veh. Technol.}, vol.~20, no.~4, pp. 81--92, Nov. 1971.

\bibitem{Lee73j}
W.~Lee, ``Effects on correlation between two mobile radio base-station
  antennas,'' \emph{{IEEE} Trans. Commun.}, vol.~21, no.~11, pp. 1214--1224,
  Nov. 1973.

\bibitem{Kreu05}
\BIBentryALTinterwordspacing
K.~Kreutz-Delgado, ``The complex gradient operator and the
  {$\numID{C}\numID{R}$}-calculus,'' Lecture Supplement to ECE275A,
  Ver.~ECE275CG-F05v1.3c, University of California, San Diego, USA, 2005.
  [Online]. Available: \url{http://dsp.ucsd.edu/~kreutz/PEI05.html}
\BIBentrySTDinterwordspacing

\bibitem{MaPr92b}
A.~M. Mathai and S.~B. Provost, \emph{Quadratic Forms in Random Variables:
  Theory and Applications}.\hskip 1em plus 0.5em minus 0.4em\relax New York,
  NY: M.~Dekker, 1992.

\bibitem{Tork01j}
M.~Torki, ``Second-order directional derivatives of all eigenvalues of a
  symmetric matrix,'' \emph{Nonlinear Anal.: Theory, Methods \& Applications},
  vol.~46, no.~8, pp. 1133--1150, Dec. 2001.

\bibitem{HoJo85b}
R.~A. Horn and C.~R. Johnson, \emph{Matrix Analysis}.\hskip 1em plus 0.5em
  minus 0.4em\relax Cambridge, UK: Cambridge University Press, 1985.

\bibitem{EaPe73j}
M.~L. Eaton and M.~D. Perlman, ``The non-singularity of generalized sample
  covariance matrices,'' \emph{Ann.\ Stat.}, vol.~1, no.~4, pp. 710--717, Jul.
  1973.

\bibitem{CaKo99b}
M.~Capi\'nski and P.~E. Kopp, \emph{Measure, Integral and Probability},
  2nd~ed.\hskip 1em plus 0.5em minus 0.4em\relax London, UK: Springer, 1999.

\bibitem{McAL96j}
T.~McKelvey, H.~Ak\c{c}ay, and L.~Ljung, ``Subspace-based multivariable system
  identification from frequency response data,'' \emph{{IEEE} Trans. Automat.
  Contr.}, vol.~41, no.~7, pp. 960--979, Jul. 1996.

\bibitem{Ande03b}
T.~W. Anderson, \emph{An Introduction to Multivariate Statistical Analysis},
  3rd~ed.\hskip 1em plus 0.5em minus 0.4em\relax John Wiley \& Sons, 2003.

\bibitem{Math99b}
A.~M. Mathai, \emph{An Introduction to Geometrical Probability: Distributional
  Aspects with Applications}.\hskip 1em plus 0.5em minus 0.4em\relax Newark,
  NJ: Gordon and Breach, 1999.

\bibitem{Edel89t}
A.~Edelman, ``Eigenvalues and condition numbers of random matrices,'' Ph.D.
  dissertation, MIT, Cambridge, MA, 1989.

\bibitem{BaBo04c}
D.~S. Baum and H.~B{\"o}lcskei, ``Impact of phase noise on {MIMO} channel
  measurement accuracy,'' in \emph{Proc.\ IEEE Veh.\ Technol.\ Conf.\ (VTC)
  Fall}, vol.~3, Los Angeles, CA, Sep. 2004, pp. 1614--1618.

\bibitem{Simo02b}
M.~K. Simon, \emph{Probability Distributions Involving {Gaussian} Random
  Variables: {A} Handbook for Engineers and Scientists}.\hskip 1em plus 0.5em
  minus 0.4em\relax Kluwer Academic Publishers, 2002.

\bibitem{HaBo04c}
J.~C. Hansen and H.~B{\"o}lcskei, ``A geometrical investigation of the rank-1
  {Ricean} {MIMO} channel at high {SNR},'' in \emph{Proc.\ IEEE Int.\ Symp.\ on
  Inform.\ Theory (ISIT)}, Chicago, IL, Jun. 2004, p.~64.

\bibitem{Stew80j}
G.~W. Stewart, ``The efficient generation of random orthogonal matrices with an
  application to condition estimators,'' \emph{SIAM J.\ Numerical Anal.},
  vol.~17, no.~3, pp. 403--409, Jun. 1980.

\bibitem{Lee97b}
P.~M. Lee, \emph{Bayesian Statistics: An Introduction}, 2nd~ed.\hskip 1em plus
  0.5em minus 0.4em\relax London, UK: Arnold, New York, NY: Wiley, 1997.

\bibitem{HoMT04j}
B.~M. Hochwald, T.~L. Marzetta, and V.~Tarokh, ``Multiple-antenna channel
  hardening and its implications for rate feedback and scheduling,''
  \emph{{IEEE} Trans. Inform. Theory}, vol.~50, no.~9, pp. 1893--1909, Sep.
  2004.

\bibitem{MiEG99c}
D.~G. Michelson, V.~Erceg, and L.~J. Greenstein, ``Modeling diversity reception
  over narrowband fixed wireless channels,'' in \emph{Proc.\ IEEE MTT-S Symp.\
  on Technol.\ for Wireless Appl.}, Vancouver, BC, Canada, Feb. 1999, pp.
  95--100.

\bibitem{diss_baum}
D.~S. Baum, ``Information-theoretic analysis of a class of {MIMO} channel
  measurement devices,'' Ph.D. dissertation, ETH Zurich, Z{\"u}rich,
  Switzerland, 2007, {N}o.~17395.

\end{thebibliography}

\end{document}